%% file: phdthesis.tex
\documentclass[12pt,a4paper]{uq-thesis}

\usepackage[utf8]{inputenc}
\usepackage{rotating}
\usepackage{graphicx}
\usepackage{amsbsy}
\usepackage{amsmath}
\usepackage{amssymb}
\usepackage{color}
\usepackage{setspace}
\usepackage{amsfonts}
\usepackage[numbers]{natbib}
\usepackage{psfrag}
\usepackage{import}
\usepackage{bbold}
\usepackage{booktabs}
\usepackage{comment}
\usepackage{bibentry}
\definecolor{linkcol}{rgb}{0,0,0.4}
\usepackage[hyperfootnotes=false,colorlinks=true,linkcolor=linkcol,citecolor=linkcol,urlcolor=linkcol,bookmarks=false]{hyperref}

\newcommand{\Bkt}[1]{\left( #1 \right)}
\newcommand{\bkt}[2][]{#1( #2 #1)}
\newcommand{\Bktcl}[1]{\left\{ #1 \right\}}
\newcommand{\bktcl}[2][]{#1\{ #2 #1\}}
\newcommand{\Bktsq}[1]{\left[ #1 \right]}
\newcommand{\bktsq}[2][]{#1[ #2 #1]}

\newcommand{\Norm}[1]{\left \lVert #1 \right \rVert}
\newcommand{\norm}[2][]{#1 \lVert #2 #1 \rVert}
\newcommand{\Abs}[1]{\left \lvert #1 \right \rvert}
\newcommand{\abs}[2][]{#1 \lvert #2 #1 \rvert}

\newcommand{\bra}[2][]{#1\langle #2 #1\rvert}
\newcommand{\ket}[2][]{#1\lvert #2 #1\rangle}
\newcommand{\braket}[3][]{#1\langle #2 #1| #3 #1\rangle}
\newcommand{\braoket}[4][]{#1\langle #2 #1| #3 #1| #4 #1\rangle} 
\newcommand{\ketbra}[3][]{#1\lvert #2 #1\rangle #1\langle #3 #1\rvert}

\newcommand{\deriv}[3][]{\frac{d^{#1} #2}{d #3^{#1}}}
\newcommand{\fderiv}[2]{\frac{\delta #1}{\delta #2}}
\newcommand{\pderiv}[3][]{\frac{\partial^{#1} #2}{\partial #3^{#1}}}

\renewcommand{\vec}[1]{\mathbf{#1}}

\newcommand{\del}[0]{\nabla}
\newcommand{\grad}[0]{\del}

\newcommand{\Exval}[1]{\left\langle #1 \right\rangle}
\newcommand{\exval}[2][]{#1\langle #2 #1\rangle}
\DeclareMathOperator{\Tr}{Tr}

\newcommand{\Zfield}{\mathbb{Z}}
\newcommand{\Rfield}{\mathbb{R}}
\newcommand{\Cfield}{\mathbb{C}}

\newcommand{\Zcal}{\mathcal{Z}}
\newcommand{\Lcal}{\mathcal{L}}
\newcommand{\Hcal}{\mathcal{H}}
\newcommand{\Kcal}{\mathcal{K}}
\newcommand{\Ncal}{\mathcal{N}}
\newcommand{\Mcal}{\mathcal{M}}
\newcommand{\Pcal}{\mathcal{P}}
\newcommand{\Fcal}{\mathcal{F}}

\renewcommand{\v}[1]{\mathbf{#1}}

\newcommand{\Tc}{T_c}
\newcommand{\Nex}{N_\text{ex}}
\newcommand{\ao}{\hat{a}}
\newcommand{\aodag}{\hat{a}^\dagger}
\newcommand{\bo}{\hat{b}}
\newcommand{\bodag}{\hat{b}^\dagger}
\newcommand{\fo}{\hat{\psi}}
\newcommand{\fodag}{\hat{\psi}^\dagger}
\newcommand{\identop}{\hat{\mathbb{1}}}
\newcommand{\Hsp}{H_{\text{sp}}}

\newcommand{\vx}{\v{x}}

\newcommand{\Braoket}[3]{\left\langle #1 \left|#2\vphantom{#1 #3}%
                         \right| #3 \right\rangle}

\newcommand{\Evalat}[2]{\left.#1\right|_{#2}\!}

\renewcommand{\Re}{\operatorname{Re}}
\renewcommand{\Im}{\operatorname{Im}}

\newcommand{\rC}{\text{\bf{C}}} 
\newcommand{\rI}{\text{\bf{I}}} 
\newcommand{\PC}{\mathcal{P}_{\rC}}
\newcommand{\bfI}{\fo_{\rI}}
\newcommand{\cf}{\psi_{\rC}} 
\newcommand{\Tkt}{T_\text{KT}}
\newcommand{\md}{\hat{\v{p}}} 
\newcommand{\punpair}{p_u} 

\newcommand{\foC}{\fo_{\rC}}
\newcommand{\foI}{\fo_{\rI}}
\newcommand{\foCdag}{\foC^\dagger}
\newcommand{\foIdag}{\foI^\dagger}
\newcommand{\PI}{\Pcal_{\rI}}

\title{Physics of Low-Dimensional Ultracold Bose Gases}
\author{Christopher James Foster}
\authordegrees{BSc(Hons), BMath}
\school{School of Mathematics and Physics}
\thesistype{Doctor of Philosophy}
\submitdate{2011}

\begin{document}

\input{frontmatter}
\subimport{introduction/}{introduction}
\subimport{background/}{background}

\subimport{bkt/}{bkt}

\subimport{sfrac/}{sfrac}

\subimport{var1d/}{var1d}

\input{conclusion}

\bibliographystyle{plain}
\addcontentsline{toc}{chapter}{References}
\renewcommand{\k}[1]{#1} 

\input{phdthesis2.bbl}

\appendix
\input{funcderivs}

\subimport{bkt/}{bkt_appendix}

\subimport{var1d/}{var1d_appendix}

\input{other_papers}

\end{document}

%% file: frontmatter.tex
\PublicationsDuringCandidature{
Peer reviewed papers:
\begin{itemize}
\item[\cite{Foster2010}]
C.~J. Foster, P.~B. Blakie, and M.~J. Davis.
\newblock Vortex pairing in two-dimensional {B}ose gases.
\newblock {\em Phys. Rev. A}, 81:023623, 2010.

\item[\cite{Cavalcanti2007}]
E.~G. Cavalcanti, C.~J. Foster, M.~D. Reid, and P.~D. Drummond.
\newblock Bell inequalities for continuous-variable correlations.
\newblock {\em Phys. Rev. Lett.}, 99:210405, 2007.

\item[\cite{Mendonca2008}]
P.~E. M.~F. Mendon\c{c}a, R.~d.~J. Napolitano, M.~A. Marchiolli, C.~J. Foster,
  and Y.-C. Liang.
\newblock Alternative fidelity measure between quantum states.
\newblock {\em Phys. Rev. A}, 78:052330, 2008.

\end{itemize}
}

\ContributionsToJointWorksStatement{
\hspace{\parindent}
Ref.~\cite{Foster2010} is incorporated as chapters \ref{bkt_chapter},
\ref{sfrac_chapter} and appendix \ref{bkt_appendix} of the thesis.
I was responsible for writing the majority of this paper and performed all
technical calculations.  P. Blair Blakie assisted in the writing of the
introductory and concluding parts of the paper.  Matthew J. Davis guided the
research and helped with interpretation of results.  

Ref.~\cite{Cavalcanti2007} is reproduced verbatim in appendix
\ref{additional_work_appendix}.
Eric G. Cavalcanti was responsible for writing most of the paper and performing
most of the technical calculations.
I contributed the no-go proof on pages three and four in collaboration with
Eric, and helped with the final stages of writing.
Margaret D. Reid analysed the effect of detector inefficiencies.
Peter D. Drummond conceived and guided the research.

Ref.~\cite{Mendonca2008} is also reproduced verbatim in appendix
\ref{additional_work_appendix}.
Paulo E. M. F. Mendonça wrote the initial draft of the manuscript
containing several of the presented scientific results.
Reginaldo d. J. Napolitano obtained the results of section III and the proof of
proposition III.2.
Marcelo A. Marchiolli contributed with discussions and improvements of the
text, and collaborated with Paulo in conjecturing inequality (29) of the paper.
I contributed the numerical results and written description of section IV; I
also assisted in revising the manuscript as a whole.
Yeong-Cherng Liang conceived section IV including initial Matlab code, and
contributed to the revision of the manuscript as a whole.
}

\ContributionsByOthersStatement{Simon A. Haine provided initial ideas and
Matthew J. Davis provided initial ideas and general guidance for the work
described in chapter \ref{var1d_chapter}.  Vivien J. Challis, S. A. Haine and
M. J. Davis proofread the thesis and provided suggestions.
}

\OtherDegreeStatement{
The two works incorporated verbatim in appendix \ref{additional_work_appendix}
were previously incorporated into PhD theses by their respective primary
authors:
Ref.~\cite{Cavalcanti2007} was submitted by fellow author Eric G. Cavalcanti
as part of his PhD at the University of Queensland, degree awarded 19 May
2008;
Ref.~\cite{Mendonca2008} was submitted by fellow author Paulo E. M. F.
Mendonça as part of his PhD at the University of Queensland, degree awarded
11 August 2009.
}

\acknowledgments{
There are many people who contributed positively to this thesis in one way or
another.  The most obvious of course is my supervisor Matthew Davis.  Matt has
been involved with the project from the beginning, and I am deeply grateful
for his unwavering support and good humoured patience.

Thanks are also due to Blair Blakie, Ashton Bradley, Joel Corney, Peter Drummond,
Simon Haine and Tod Wright, all of whom were involved in my supervision at some
stage, either officially or unofficially.  I am particularly grateful to Blair
for providing the final impetus to publish the work of chapter
\ref{bkt_chapter}, and to Simon for finding the time to proofread much of this
thesis and provide thoughtful suggestions.

My PhD experience would have been vastly inferior without the friendship and
camaraderie of my fellow students.  Most obviously there were ``the usual
suspects'': my office mates Eric Cavalcanti, Andy Ferris, Geoff Lee,
Yeong-Cherng Liang, Terry McRae and Paulo Mendonça, who together made the
office a lively and intellectually stimulating environment.  Memories which
will not be easily forgotten include the animated discussions with Eric on the
nature of consciousness, coding and juggling adventures with Andy, and many
fascinating fragments of the home cultures of Eric, Paulo, and Yeong-Cherng.  I
also enjoyed interacting with other students within and outside the department,
including Andrew Sykes, Tim Vaughan, Rodney Polkinghorne, Mark de Burgh and
many more.  Most of these friends have already moved on from UQ, so I can say
with certainty that they are sorely missed.  

The research group at the UQ node of the Australian Research Council Centre of
Excellence for Quantum-Atom Optics (ACQAO) also deserves my thanks for hosting
me during my PhD.  The members of the group have always provided a good
environment to learn about theoretical physics, but also about the
\emph{process} of doing theoretical physics.  In addition, ACQAO as an
organisation provided funding for both part of my scholarship and travel to
various conferences.  I am particularly grateful for the opportunity to travel
to conferences in France and Germany during early 2005.

Finally, I thank my family and particularly my wife Vivien for emotional and
practical support over these last several years.  To Vivien I would like to say
--- ``Thank you for your love, thank you for your support, and thank you for
being my last line of defence against sloppy definitions and awkward
sentences!''  It is quite possible that this thesis would not exist without
Vivien, and with that in mind, I dedicate it to her.
}

\abstract{
In this thesis we investigate the properties of ultracold Bose gases in one and
two dimensions.  Low-dimensional systems have several striking differences from
their three-dimensional counterparts, most notably the absence of Bose-Einstein
condensation (BEC) in the infinite homogeneous case.  Recent experimental
progress has brought low-dimensional systems within reach in the laboratory, and
we provide numerical simulations with experimentally relevant parameters.  We
present simulations of the Berezinskii-Kosterlitz-Thouless (BKT) phase
transition in two dimensions and investigate equations of motion for a Bose gas
constrained to one dimension.

Recent experiments have attributed superfluidity in two-dimensional systems to
the BKT phase transition.  We perform classical field simulations using the
projected Gross-Pitaevskii equation (PGPE) formalism to model the 
two-dimensional Bose gas at finite temperature.  We confirm the presence of the
BKT phase via the observation of two unique features: algebraic decay of the
first-order correlation function; and vortex pair unbinding.  Unbinding of vortex
pairs at the BKT transition is clearly demonstrated via a coarse-graining
procedure which reveals unpaired vortices.  The BKT transition temperature
identified via correlations and vortex behaviour agrees well with the
temperature deduced from a calculation of the superfluid fraction.
Surprisingly, we observed no separation between the temperature of the BEC
transition --- which is present due to the finite simulation size --- and the
BKT transition.  We relate our results to experimental observations and show
that an interpretation based on BKT physics is justified.

In investigating the two-dimensional system we found it necessary to compute
the superfluid fraction.  Calculating the superfluid fraction is a delicate
procedure because it involves connecting the macroscopic quantum phenomenon of
superfluidity to a detailed microscopic simulation.  We present an efficient
method which overcomes this difficulty using a tensor decomposition of the
momentum density autocorrelations.  Our method gives results consistent with
the other physical properties of the BKT phase in two-dimensional systems.

One-dimensional effective equations allow for efficient simulation of very
elongated systems of ultracold Bose gases.  We use a Gaussian ansatz and
Lagrangian approach to derive an effective equation for a Bose gas constrained
to one dimension at zero temperature.  In some respects our method outperforms
alternative one-dimensional equations such as the non-polynomial Schrödinger
equation.  However, our scheme is found to be inherently unstable in a
wide range of cases.  We analyse the instability and find that it is an inherent
feature of a whole class of nonlinear ansätze.
}


\AnzsrClassification{ANZSRC code: 020601, Degenerate Quantum Gases and Atom
Optics, 100\%}

\FoRClassification{FoR code: 0206, Quantum Physics, 100\%}

\newacronym{BEC}{BEC}{Bose-Einstein condensation}
\newacronym{BKT}{BKT}{Berezinskii-Kosterlitz-Thouless}
\newacronym{GPE}{GPE}{Gross-Pitaevskii equation}
\newacronym{HFB}{HFB}{Hartree-Fock-Bogoliubov}
\newacronym{LDA}{LDA}{local density approximation}
\newacronym{MDE}{MDE}{Muñoz Mateo-Delgado equation}
\newacronym{NPSE}{NPSE}{non-polynomial Schrödinger equation}
\newacronym{PGPE}{PGPE}{projected Gross-Pitaevskii equation}
\newacronym{SPGPE}{SPGPE}{stochastic projected Gross-Pitaevskii equation}
\newacronym{1D}{1D}{one-dimensional}
\newacronym{2D}{2D}{two-dimensional}
\newacronym{3D}{3D}{three-dimensional}
\glsaddall

\makefrontmatter

%% file: introduction/introduction.tex
\chapter{Introduction} \label{chap:introduction}

In this thesis we investigate the behaviour of dilute, degenerate Bose gases
confined to one and two dimensions.  This introductory chapter is mostly
dedicated to explaining the physics contained in the previous sentence: The
quantum statistics of bosons, the nature of degeneracy, and the physical
differences between three and lower dimensions.

It is a surprising fact that elementary particles of the same type are not
distinguishable, even in principle.  The first hints of the importance of this
fact appeared even before the advent of quantum mechanics as a resolution
to the Gibbs paradox of classical statistical mechanics: The calculated entropy
of an ideal gas is not extensive\footnote{A measured quantity is
\emph{extensive} if it is directly proportional to the size of the system;
for a new system composed of two exact copies of the system joined together the 
extensive quantity is doubled.
} unless the atoms are treated
as indistinguishable.  As it turns out, there are exactly two consistent
statistical behaviours for indistinguishable quantum particles in three
dimensions\footnote{In two dimensions there is in fact a continuum of
possibilities known as anyonic statistics.  The systems considered in this
thesis are embedded in real three-dimensional space however, so anyonic
statistics are irrelevant at the level of the constituent particles.}.  Bose
discovered the first of these in 1924 \cite{Bose1924} in a successful attempt to
derive Planck's blackbody radiation formula from first principles.  Bose had
invented a new way to count multi-photon states, which Einstein immediately
applied to atoms to predict the equation of state \cite{Einstein1924}, and
in 1925 the ``condensation'' \cite{Einstein1925} of what became known as the
ideal Bose gas.  Bose-Einstein condensation (BEC) was invoked in the following
decades to explain various condensed matter phenomena including superfluidity
in liquid helium and low temperature superconductivity.  In 1926 Fermi
\cite{Fermi1926} and Dirac \cite{Dirac1926} discovered the second type of
quantum statistics, motivated partly by the behaviour of electrons as newly
articulated in the Pauli exclusion principle.  Particles obeying Bose-Einstein
and Fermi-Dirac statistics are now known as \emph{bosons} and \emph{fermions}
respectively.

We reflect briefly on what exactly we mean by ``particle statistics''.
Statistical mechanics starts from a probability distribution over the system
microstates\footnote{A microstate is a full microscopic description of the
details of a physical system; a state of maximal information.
}
and uses this distribution to calculate expected properties of the system.  
The distribution is specified by some macroscopic variables (for example the
temperature) but before the distribution can be given, the possible microstates 
must be described.  Bosons, fermions and classical particles differ in
exactly which microstates are possible.  For a noninteracting system such as
the ideal gas, the state of the full system may be given by specifying which
single-particle states, or \emph{modes}\footnote{
It is easy to confuse the terminology when referring to ``single-particle states''
versus ``states of the system as a whole''.  We hope to avoid this problem by
following a common practice from optics where the single-particle states are
referred to as \emph{modes}.
},
are occupied.  Here is where the difference arises --- any number of bosons may
occupy a given mode, while the occupation is constrained to zero or one
particle for fermions.  Labelling the states of the full system using only the
occupation of modes incorporates particle indistinguishability.  For classical
distinguishable particles, we would also need to keep track of \emph{which}
particles are in each mode, in addition to how many.

Quantum statistical behaviour is connected to intrinsic angular momentum by the
celebrated spin statistics theorem: particles of half-integer spin (in
multiples of $\hbar$) are fermions while particles with integer spin are
bosons.  Although the elementary constituents of normal matter are all fermions,
bosonic statistics may be observed in composite particles --- atoms for example
--- composed of even numbers of fermions.  For this to be true, it is
sufficient that the interactions between composites be at much lower energy
than the internal energy levels \cite{Ehrenfest1931}.

In order to observe the consequences of quantum statistics a system must be
\emph{degenerate}.  In this context ``degenerate'' means that the number of
energetically accessible modes is similar to or smaller than
the total number of particles; it is only with this kind of crowding of the
available states that the differences in counting microstates become apparent.
Degeneracy may be reached by either increasing the density, which increases the
number of particles per state, or reducing the temperature which reduces the
number of accessible states.  For the particular case of an ideal gas, it is
convenient to discuss degeneracy in terms of the de~Broglie wavelength:
At a given temperature $T$, the expected energy of a classical ideal gas
particle is $\frac{1}{2}k_B T$ for each degree of freedom, according to the
equipartition theorem, where $k_B$ is Boltzmann's constant.  In three 
dimensions we therefore have $\exval{p^2/2m} = \frac{3}{2}k_B T$, and using
de~Broglie's formula $\lambda=h/p$ gives a value
$\lambda = h / \sqrt{3 m k_B T}$ for the typical quantum wavelength.  It is
conventional\footnote{The convention for $\lambda_\text{dB}$ is selected to
absorb the constants in the expression for the ideal gas partition function.
}
to define the \emph{thermal de~Broglie wavelength} as
\begin{equation} \label{lambda_dB}
    \lambda_\text{dB} = \sqrt{\frac{2\pi\hbar^2}{m k_B T}}.
\end{equation}
With this definition the number of accessible states can be estimated as
$Z=V/\lambda_\text{dB}^3$ where $V$ is the volume of the system\footnote{
Here $Z$ is actually the canonical partition function; if the energy scale is
adjusted such that the lowest energy state has energy zero, $Z$ counts the
accessible states.  (See, for example, \cite[\S{}6.1]{Schroeder2000}.)
}.  Quantum degeneracy occurs when the number 
of particles is greater than or approximately equal to the number of accessible 
states, that is, $N \gtrsim Z$.  Rearranging, we arrive at the degeneracy
condition
\begin{equation}
    n \gtrsim \lambda_\text{dB}^{-3}
\end{equation}
for the density $n = N/V$, which has a nice physical interpretation: Degeneracy 
occurs when the size of the quantum wavepacket is comparable to or larger than
the interparticle spacing.

Apart from Bose's work, the first historical uses of quantum statistics were to
compute properties of the ideal gas in the quantum degenerate regime.
Nevertheless, there was no experimentally accessible system with both weak
interactions and strong quantum statistical behaviour until 1995 when the first
degenerate, dilute Bose gases were created in the laboratory
\cite{Anderson1995,Davis1995}.  This feat was followed shortly afterwards with
the creation of a degenerate Fermi gas in 1999 \cite{Demarco1999}.  There were 
several good reasons why it took seventy years for experimental systems to
catch up with the theory.  On the one hand, quantum degeneracy occurs in many
sorts of condensed matter systems, including striking examples such as
superfluidity in liquid $^4$He.  However, these systems have strong
interactions between particles which renders the simplest theories useless from
a quantitative perspective.  On the other hand, gases with weak interactions
are nowhere near degeneracy at easily accessible temperatures, and tend to form
liquids and solids when cooled.  At low densities the formation of liquids or
solids from atomic gases is suppressed, and cooling to extremely low
temperatures --- of the order of 100~nK --- allows experiments to reach the
degenerate regime.  Perfecting the experimental tools necessary to trap and
cool atomic clouds to such extraordinarily low temperatures was a major
undertaking.

From now on we focus exclusively on bosons which are the topic of this thesis.
The typical ultracold boson experiment begins with a gas of neutral atoms ---
$^{87}$Rb for example --- in a vacuum chamber.  In the first stage the atoms
are captured in a magneto-optical trap and cooled via laser cooling to
temperatures of order 100~$\mu$K \cite{Phillips1998}.  Atoms are then loaded
into a tighter magnetic or optical dipole trap and further cooled by
successively removing the most energetic atoms --- a process known as
evaporative cooling.  During this process the gas undergoes condensation
to form a BEC at temperatures of order 1~$\mu$K to 100~nK, with total number of
atoms ranging from $10^8$ to $10^3$ \cite[\S 1.1]{Pethick2008}.  The
trapping potential is usually well approximated by a parabola, with the spatial
extent of the atomic cloud on the order of 10--100~$\mu$m, depending on the
trap anisotropy \cite{Anglin2002}.

The precision and control available to experiments has continued to improve 
over the last 15 years.  Experiments can now control the interaction strength
via Feshbach resonances, and detailed control of the trapping potential is
available using combinations of lasers and magnetic fields.  Ultimately, the
experimental accessibility of weakly interacting and precisely controllable
quantum gases has been a great resource for basic quantum physics: It has
allowed rigorous comparison and evaluation of first principles approaches to
quantum field theory.

\section{Bose statistics and condensation}

The phenomenon of BEC is a phase transition, with a normal gaseous phase at
higher temperature, and a fluid with macroscopic quantum behaviour at low
temperature.  In the low temperature phase, a macroscopic number of particles 
all occupy the same quantum state, in a sense to be made more precise below.  
For the particular case of the noninteracting homogeneous gas in three
dimensions, the expected occupation of the ground state is
\begin{equation}
    N_0 = N \bktsq[\bigg]{1 - \Bkt{\frac{T}{\Tc}}^{3/2}},
\end{equation}
where $\Tc$ is the transition temperature.  We discuss the value of $\Tc$ and
provide a derivation of this basic relation in section
\ref{ideal_gas_condensation}.

Unlike most phase transitions, BEC occurs even in systems without interparticle
interactions: it is a purely statistical effect.  To understand qualitatively
how this happens, we consider the effect of Bose statistics on a two mode
toy system containing exactly $N$ particles.  For definiteness, let us think of
these modes as a pair of ``left'' and ``right'' potential wells.

First let us consider the case where each well has the same energy, and count 
the number of states available to the system as a whole.  For classical
distinguishable particles, each particle may be in the left or right well 
independently of the rest, so adding a particle to the system multiplies the
number of available states by two; the total number of states for $N$ particles
is then $2^N$.  On the other hand, for indistinguishable particles a state of
the system as a whole is fully specified by listing the number of particles in
each well.  The number of particles in the left well, $N_L$, is between $0$ and
$N$ particles, and $N_R$ is always equal to $N-N_L$, so there are only $N+1$
states.

\begin{figure}[phtb]
  \begin{center}
    \includegraphics[width=14cm]{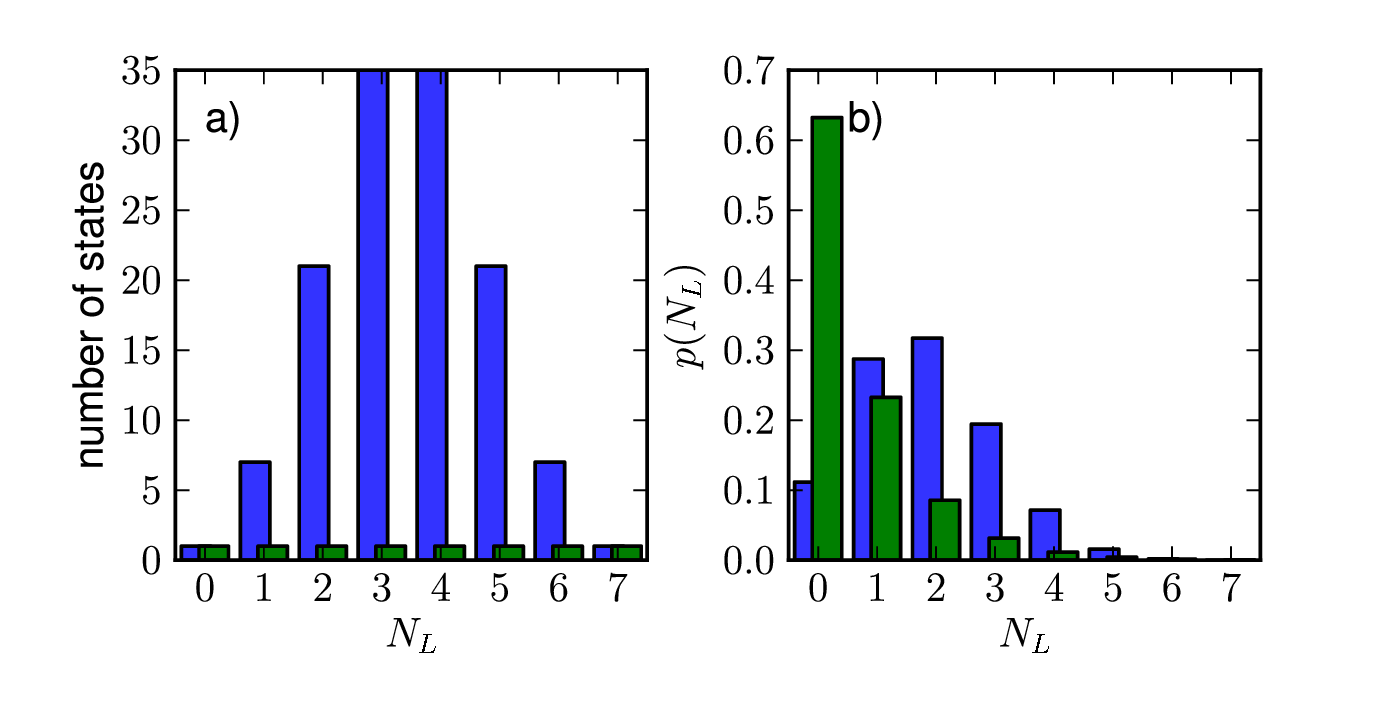}
    \includegraphics[width=14cm]{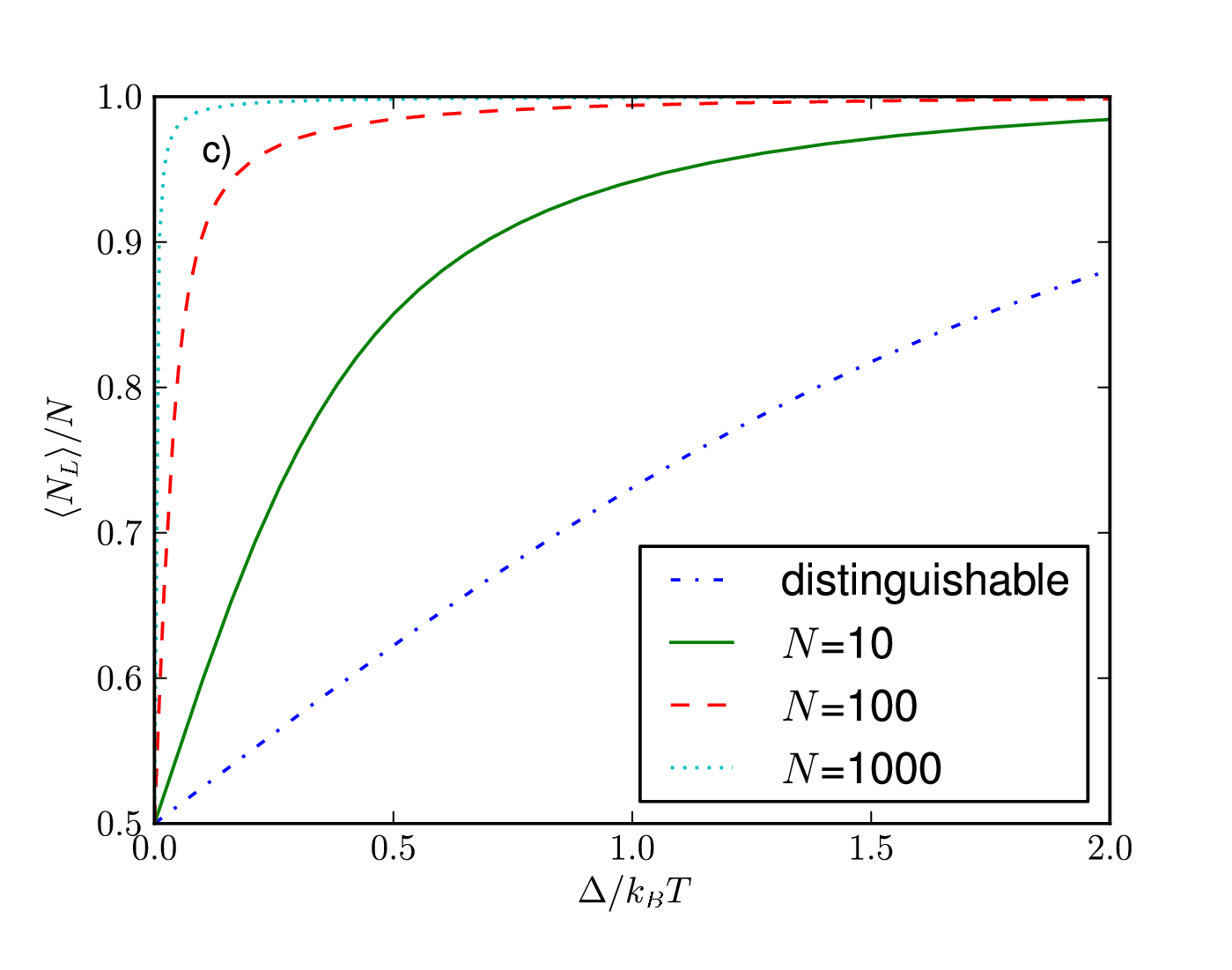}
  \end{center}
  \caption[Bose statistics versus distinguishable statistics in a two mode system]
  {\label{bose_stats}
    Bose statistics versus classical distinguishable statistics in a two mode
    toy system.
    (a) Count of system states with a given number of particles in the left well 
    for the distinguishable case (blue) and bosonic case (green).  This is
    proportional to the probability density $p(N_L)$ when $\Delta = 0$.
    (b) Probability density $p(N_L)$ for finding the system with $N_L$ particles 
    in the left well, when $\Delta/k_B T = 1$ [colours as in (a)].
    Both (a) and (b) show statistics for a system of $N=7$ atoms.  
    (c) Expected fraction of particles in the left well as a function of 
    $\Delta/k_B T$ for the bosonic case with three different atom numbers.  The 
    distinguishable case is included for reference and is the same for 
    \emph{any} $N$.
  }
\end{figure}

If both modes have the same energy then all states of the system have the same
energy and are therefore equally likely\footnote{The assumption that states 
with equal energy are equally probable is the founding assumption of
statistical mechanics, see, for example, Ref.~\cite[\S 7.1]{Schroeder2000}.
}.  The probability of having all $N$
particles in the left well is then $1/2^N$ for the distinguishable case, but
$1/(N+1)$ for indistinguishable particles.  From this simple example, we can
already see how Bose statistics exponentially enhances the probability of
finding the system with all particles in a single well, simply because of the
way states are counted.  One way to visualise this is to plot the number of
states with a particular $N_L$ as in Fig.~\ref{bose_stats}(a) --- even for the
very small value of $N=7$ shown, there are many more distinguishable states
where the particles are evenly distributed between the wells.

If we now let the right well have energy $\Delta$ greater than the left, the 
probability distribution $p(N_L)$ is skewed proportionally to
$e^{-N_R \Delta/k_B T}$, as shown in Fig.~\ref{bose_stats}(b)\footnote{
The Boltzmann factor $e^{-N_R \Delta/k_B T}$ arises from analysing the system 
in the canonical ensemble, see, for example, Ref.~\cite[Ch.~6]{Schroeder2000}.
}.
Nevertheless, the distinguishable case may still have a peak at nonzero
$N_L$ due to the statistical pressure toward equal numbers; this effect becomes
much more pronounced for larger, more realistic values of $N$.  Taking the
expected value of $N_L$ from the probability distribution, we can also
calculate\footnote{The calculation of $\exval{N_L}/N$ is immediate in the
canonical ensemble via direct summation over states.
}
a ``condensate fraction'' --- the expected fraction of atoms in the lowest
energy state, $\exval{N_L}/N$.  Figure~\ref{bose_stats}(c) shows the condensate
fraction $\exval{N_L}/N$, as a function of $\Delta/k_B T$, for various total
number $N$.  It is clear that for bosons the condensate fraction rapidly converges
to 1 as $N$ increases, regardless of the size of the energy gap between the
wells.  On the other hand, for distinguishable particles the fraction in the
left well is independent of $N$.  It is worth noting that in this toy system
--- or any system with a finite number of modes --- there must always be a mode
with an extensive population, so strictly speaking the existence of
condensation is a trivial fact in this case.  Even so, it is instructive to 
observe the differences between the bosonic and distinguishable cases.

An infinite homogeneous gas has an infinite number of modes, so the existence of
a nonzero condensate fraction in the lowest energy mode is a nontrivial fact in
this case (see section \ref{ideal_gas_condensation}).  The existence of BEC in
a system with infinitely many modes turns out to depend on the density of
states at low energies.  In the three-dimensional homogeneous case, there are 
``sufficiently few'' low-energy modes meaning that particles are forced to gather in
the lowest energy mode and BEC occurs.  However, in one and two dimensions the
density of low-energy modes is larger and BEC is forbidden, at least in the
homogeneous case.  We emphasise that the addition of a trapping potential
modifies the density of states, which can lead to BEC even in low dimensions.

%

\section{Low-dimensional systems}
\label{lowdim_intro}

The physical behaviour of one- and two-dimensional systems can be qualitatively
different from a three-dimensional (3D) system containing the same constituent
particles.  A dramatic example is the homogeneous ideal Bose gas which undergoes 
the BEC phase transition in 3D, but never condenses in lower dimensions at any 
nonzero temperature.  This absence of BEC is a specific case of the more
general Mermin-Wagner-Hohenberg theorem \cite{Mermin1966,Hohenberg1967}, which
forbids long-range order in a whole class of one- and two-dimensional systems.

The qualitative differences between three and lower dimensions are due to both
geometric and topological effects.  For example, the existence of BEC
depends critically on the density of states at low energy, which for the
homogeneous case scales as the surface area of a $(d-1)$-sphere in momentum
space, where $d$ is the dimension; this is a geometric effect (see section
\ref{ideal_gas_condensation}).  On the other hand, the nature and existence of
elementary topological excitations such as vortices also depends on the
dimensionality \cite[Ch.~9]{Chaikin1995} and these excitations have a 
significant effect on the thermodynamics of the system.

At first sight one might question the relevance of low-dimensional systems,
given that the world is really three-dimensional.  However, many 3D systems can
be made to act as if they are low-dimensional by tight confinement.  When the
confinement is sufficiently strong and the temperature low enough, the
transverse degrees of freedom are effectively ``frozen out'' and all dynamics
occurs within the plane or line.  More precisely, when interactions are
unimportant it is clear that freeze out occurs for $\hbar\omega \gg k_B T$, 
where we have compared one quantum of transverse excitation energy $\hbar\omega$ 
with the characteristic magnitude $k_B T$ of the thermal fluctuations.  This
estimate holds even when interactions are important, as discussed in
Ref.~\cite[Ch.~15]{Pethick2008}.

Ref.~\cite{Gorlitz2001} presents a useful characterisation of
low-dimensional regimes in terms of length scales: Let us consider a BEC in a
cylindrically symmetric system, where the extent of the trapped cloud is given
by radii $R_\perp$ and $R_z$ in the radial and longitudinal directions
respectively.  As discussed in Ref.~\cite{Gorlitz2001}, there are two
additional length scales relevant in this problem, the healing length\footnote{
The healing length $\xi = 1/\sqrt{8\pi n a_s}$ is the length scale on which a
condensate recovers (``heals'') from the effects of a local perturbation, where
$n$ is the density.  $\xi$ may be estimated as the length scale on which the 
kinetic term in the in the Gross-Pitaevskii equation (described in section
\ref{GPE_derivation}) is equal to the size of the interaction term far from 
the perturbation.  For details see Ref.~\cite[\S 6.4]{Pethick2008}.
}
$\xi$ and the scattering length $a_s$.  In many experiments, we have the
situation $R_\perp, R_z \gg \xi \gg a_s$, in which case the dynamics are fully
three-dimensional.  In contrast, the 2D and 1D cases correspond to the regimes
$R_\perp \gg \xi > R_z$ and $R_z \gg \xi > R_\perp$, where the BEC is
conventionally called ``pancake-shaped'' and ``cigar-shaped'' respectively.  In
all these cases we assume that $R_\perp, R_z \gg a_s$ so that the scattering
remains three-dimensional, even when the dynamics is two- or one-dimensional.
This is called the \emph{quasi}-low-dimensional regime \cite{Gorlitz2001}.

We briefly discuss some unique properties of two-dimensional systems.  As noted 
above, condensation does not happen in a homogeneous 2D system.  However, the 2D
Bose gas supports topological defects in the form of vortices, and in the
presence of interactions can instead undergo a Berezinskii-Kosterlitz-Thouless
(BKT) \cite{Berezinskii1971,Kosterlitz1973,Posazhennikova2006} transition to a 
quasi-coherent superfluid state.  By quasi-coherent we mean a state with
algebraic\footnote{
``Algebraic'' decay is the name conventionally used in this context for a
simple power law decay.
}
decay of spatial correlations, rather than the traditional off
diagonal long-range order of a coherent condensate \cite{Yang1962}, or the
short range exponential decay characteristic of a high temperature phase.  This 
is one of several unique features of the BKT phase.

The most striking feature of the BKT transition is the pairing of topological
defects in the low temperature phase.  In the superfluid case this corresponds
to pairing of quantised vortices of opposite rotation, which begin to unbind as
the temperature is increased through the transition; the high temperature phase
is a plasma of bound and unbound vortices \cite{Kosterlitz1973}.  

A simple argument shows why vortex unbinding is related to the destruction of
superfluidity above the BKT transition.  Consider a channel filled with a
superfluid, and transport a single unpaired vortex from one boundary of the
channel to the other.  This imparts a single quantum of velocity to the fluid
within the channel\footnote{It is easiest to see how the single quantum is
imparted by considering a toroidal channel where the superflow represents
angular momentum, and transporting a vortex from the centre to the outside.
},
and provides a mechanism for the decay of supercurrents.  On the other hand, 
transport of \emph{pairs} of vortices does not impart any net momentum to the
channel.  We defer further presentation of the aspects of the BKT transition
relevant to 2D Bose gas experiments until chapter \ref{bkt_chapter}.


In addition to 2D BKT physics, we also consider BECs in 1D systems in
this thesis.  In contrast to the 2D case, we make no particular attempt to
investigate features unique to one dimension, since our focus is on developing a
numerical approximation technique.  However, we do rely on one uniquely 1D
feature for testing our numerical method: In 1D there are stable soliton 
solutions with well understood behaviour; in higher dimensions such solitons 
would be unstable, and eventually decay via the ``snake instability''
\cite{Anderson2001}.

Experiments in cold gases have been able to access the low-dimensional regime
since 2001 when Görlitz \emph{et~al.} created both 1D and 2D systems using 
magnetic and optical traps respectively \cite{Gorlitz2001}.  The BKT transition
was first observed in liquid helium thin films \cite{Bishop1978}, but observing
aspects of BKT physics in an ultracold Bose gas proved difficult until a series
of experiments by the
Dalibard group at ENS in Paris \cite{Stock2005,Hadzibabic2006,Kruger2007}.  In
the Paris experiments, an optical lattice was used to create two parallel
pancake-shaped BECs.  Using these as mutual phase references allowed unpaired
vortices to be detected in an interference experiment \cite{Stock2005}.
Various aspects of BKT physics including the algebraic decay of correlations
were subsequently observed in Refs.~\cite{Hadzibabic2006,Kruger2007}.  Further
experimental work by other groups has followed
\cite{Schweikhard2007a,Clade2008}, further adding to the evidence of BKT
physics in the 2D cold gas system.

%
%


\section{Outline of the thesis} 

After the current introductory chapter, some background theory is introduced in
chapter \ref{background_chapter}, including the two main pieces of formalism
used later in the thesis:  The time dependent Gross-Pitaevskii equation (to be
used in chapter \ref{var1d_chapter}) is introduced in section
\ref{GPE_derivation}; the classical field methods for finite temperature 
calculations (in particular the projected Gross-Pitaevskii equation, to be used 
in chapter \ref{bkt_chapter}) are introduced in section \ref{PGPE_derivation}.
An attempt has been made to keep the background self-contained,
starting from the level of an advanced undergraduate.  For this reason, a
section on basic Bose statistics is included, along with a discussion of how
nonrelativistic quantum field theory arises as a generalisation of
first-quantised many-body quantum mechanics.

The two-dimensional experiments discussed in the previous section present a
challenge for theory.  One particular issue is to understand the relation
between the BKT phase and BEC, which exists in real experiments due to the
harmonic trapping potential.  In chapter \ref{bkt_chapter} we make use of
classical field methods to simulate the finite temperature physics of 2D
systems, with a view to understanding this question and to understand which
aspects of BKT physics may be observed in current experiments.

As part of this work, we developed a method for calculating the superfluid 
fraction from a classical field simulation.  This applicability of this method is
more general than the specific use to which it is put in chapter
\ref{bkt_chapter}, so we describe the derivation in isolation as chapter
\ref{sfrac_chapter}.

We change focus in chapter \ref{var1d_chapter} to examine effective
equations for 1D systems.  Motivated by the desire to study dispersive shock
waves in high resolution and perform expansions of suddenly untrapped atomic
clouds, we consider a variational ansatz for the quasi-1D time evolution.  We 
implement the method numerically and characterise the behaviour using several 
test problems.

The thesis concludes with a summary in chapter \ref{conclusion_chapter}.
Several appendices are included: appendix \ref{funcderivs_appendix} describes 
subtleties surrounding differentiation of Lagrangian functionals with respect 
to complex fields.  Appendices \ref{bkt_appendix} and \ref{var1d_appendix} 
describe additional details pertaining to chapters \ref{bkt_chapter} and 
\ref{var1d_chapter} respectively.  Finally, appendix
\ref{additional_work_appendix} contains verbatim copies of
Refs.~\cite{Cavalcanti2007,Mendonca2008}.  These two papers were produced
during the author's PhD candidature, but were not related to the main theme of
the thesis.  

%% file: background/background.tex
\chapter{Background Theory}
\label{background_chapter}

\begin{chap_desc}
    In this chapter we describe some basic theoretical background followed by
    the specific theoretical methods used in the thesis.  We start with some
    elementary remarks about condensation in Bose gases and then show how
    nonrelativistic bosonic quantum field theory arises from single-particle
    quantum mechanics.  We next derive the Gross-Pitaevskii equation from the
    variational point of view.  The projected Gross-Pitaevskii equation is
    covered in some detail, including a discussion of the motivation and
    procedure for taking the classical limit.
\end{chap_desc}

\section{Bose statistics and BEC in noninteracting systems}

Bose-Einstein condensation is a unique phase transition because it exists even 
in the absence of interactions between particles: it is a purely statistical
effect.  Without interactions, the behaviour of a Bose gas can readily be
analysed using the methods of elementary statistical mechanics.  The following
outlines the argument, showing that the occupation of the lowest energy state
is extensive below some transition temperature, which we then compute.

%
%
%
%

\subsection{The Bose distribution}

In this section we derive the Bose distribution from basic statistical 
mechanical considerations.  Before we start it is worth noting that the only
quantum mechanical ingredient required is indistinguishability of particles.  
When we talk about microstates of the system, we may think about either
classical states with a particular total energy and number of particles, or
quantum eigenstates of the Hamiltonian.

In the grand canonical ensemble, the probability of observing a given
microstate $s$ of a system is proportional to the Gibbs factor:
\begin{equation}
    P(s) \propto e^{-\beta[E(s) - \mu N(s)]}.
\end{equation}
Here $E(s)$ is the energy, $N(s)$ is the number of particles in the microstate
and $\beta = 1/k_BT$ is the inverse temperature.  The Gibbs
factor comes about by considering a system that can exchange energy and
particles with a much larger reservoir, and applying the fundamental assumption
of statistical mechanics --- that all microstates in a closed system are
equally likely --- to the closed combination of system and reservoir (see, for 
example, \cite[\S 7.1]{Schroeder2000}).  The normalisation constant for $P$
is the sum of the Gibbs factors for all states, known as the partition function
\begin{equation}
    \Zcal = \sum_s e^{-\beta[E(s) - \mu N(s)]};
\end{equation}
with this definition we have $P(s) = \Zcal^{-1} e^{-\beta[E(s) - \mu N(s)]}$.

We consider a system with a number of discrete single-particle quantum states
or \emph{modes}, $\bktcl{\ket{0}, \ket{1}, \ket{2}, \dotsc}$.  Letting the 
single-particle energies for the modes be $\epsilon_i$ and considering the case 
where particles have no interaction energy, the total energy is
$E(s) = \sum_i \epsilon_i n_i(s)$, with corresponding total number
$N(s) = \sum_i n_i(s)$ where $n_i$ is the occupation for the $i$th mode.  With
these expressions, the partition function for a noninteracting system is
\begin{align}
    \Zcal &= \sum_s e^{-\beta\sum_i (\epsilon_i - \mu) n_i(s)} \\
          &= \sum_s \prod_i e^{-\beta(\epsilon_i - \mu) n_i(s)}
\end{align}
where the product ranges over all modes.

To proceed further, we need to make use of Bose statistics to specify over which
states the sum $\sum_s$ occurs.  Because bosons are indistinguishable, a state
is uniquely defined by the number of particles in each mode,
$s = [n_1(s), n_2(s), \dotsc]$.  Therefore, after some
thought\footnote{
To make this transformation slightly less mysterious, consider only two modes.
Enumerating the possible states as $[0,0], [1,0], [0,1], [2,0], [1,1], [0,2],
[3,0], [2,1], [1,2], [0,3], \dotsc$, and abbreviating
$a = e^{-\beta(\epsilon_1 - \mu)}$, $b = e^{-\beta(\epsilon_2 - \mu)}$,
we see that we're dealing with the expression
\begin{align*}
\Zcal = \sum_s a^{n_1(s)} b^{n_2(s)}
    &= 1 + a + b + a^2 + ab + b^2 + a^3 + a^2b + ab^2 + b^3 + \dotsc \\
    &= (1 + a + a^2 + a^3 + \dotsc)(1 + b + b^2 + b^3 + \dotsc)
    = \sum_{n=0}^\infty a^n \sum_{n=0}^\infty b^n.
\end{align*}
}
$\Zcal$ can be rewritten in the simpler form
\begin{align}
    \label{ideal_Zcal_factorised}
    \Zcal &= \prod_i \sum_{n=0}^\infty e^{-\beta(\epsilon_i-\mu)n} \\
          &\equiv \prod_i \Zcal_i,
\end{align}
where we have defined the single-mode partition function
$\Zcal_i \equiv \sum_{n=0}^\infty e^{-\beta(\epsilon_i-\mu)n}$.
We note that the simplicity of Eq.~\eqref{ideal_Zcal_factorised} is what makes 
the analytical calculation tractable, and is peculiar to the grand canonical
ensemble.

An explicit formula for $\Zcal_i$ is easily obtained by summing the geometric
series:
\begin{equation}
    \Zcal_i = \sum_{n=0}^\infty e^{-\beta(\epsilon_i - \mu)n}
            = \sum_{n=0}^\infty \Bkt{e^{-\beta(\epsilon_i - \mu)}}^n
            = \frac{1}{1 - e^{-\beta(\epsilon_i - \mu)}}.
\end{equation}
This puts us in a position to calculate the expected occupation of the $j$th
mode.  Because of the factorised form of Eq.~\eqref{ideal_Zcal_factorised}, we
can show that functions of $n_j$ are normalised using only $\Zcal_j$ rather than
$\Zcal$,
\begin{equation}
    \exval{f(n_j)}
        = \sum_s f(n_j(s)) P(s) 
        = \frac{1}{\Zcal_j} \sum_{n=0}^\infty f(n) e^{-\beta(\epsilon_j - \mu) n}
\end{equation}
for any function $f$.  For the particular case of $f(n) = n$ we can
evaluate the sum\footnote{These kinds of sums may be reduced to the 
sum of a geometric series using the common derivative trick:
\begin{equation*}
    \sum_{n=0}^\infty n x^n
        = \sum_{n=0}^\infty n e^{\ln(x)n}
        = \pderiv{}{y} \sum_{n=0}^\infty e^{yn}\bigg|_{y=\ln{x}}
        = \pderiv{}{y} \frac{1}{1 - e^y} \bigg|_{y=\ln{x}}
        = \frac{x}{(1-x)^2}.
\end{equation*}
}, leading to the famous Bose distribution function
\begin{equation}
    \exval{n_j} = \frac{1}{e^{\beta(\epsilon_j - \mu)} - 1}.
\end{equation}

\subsection{Condensation in the 3D homogeneous Bose gas}
\label{ideal_gas_condensation}

Having derived the Bose distribution, we turn our attention to the statistical
mechanics of Bose condensation in a 3D gas.  The expected total number of
atoms in the system is given by the sum over modes
\begin{equation}
    N = \sum_i \exval{n_i} = \sum_i \frac{1}{e^{\beta(\epsilon_i - \mu)} - 1}.
\end{equation}
For a given system the modal energies $\{\epsilon_i\}$ are known, so at a
particular total number $N$ and inverse temperature $\beta$ we may in
principle use this equation to determine $\mu$.  However, solving this equation
is not possible except in the simplest cases, and we therefore turn to
approximations.

The general procedure is to split the atoms into two groups: the expected
numbers in the ground and excited states, $N_0$ and $\Nex$ respectively.
The original argument due to Einstein is that in the 3D homogeneous gas there is
an upper bound on the number of excited state atoms $\Nex$ at any given
temperature.  For any system with more atoms than this bound, the remaining
atoms must be found condensed in the ground state \cite{Einstein1925}:

\begin{quote}
    \it
    [...] something similar happens as when isothermally compressing a vapour
    beyond the volume of saturation.  A separation occurs; a part
    ``condenses'', the rest remains a ``saturated ideal gas''.
\end{quote}

The expected number of excited state atoms is the sum over excited state modal occupations,
\begin{equation}
    \Nex = \sum_{i\ne 0} \frac{1}{e^{\beta(\epsilon_i - \mu)} - 1}.
\end{equation}
To estimate $\Nex$ analytically we need a few transformations and
approximations, starting by rewriting the sum as a sum over the mode energies 
rather than state indices.
Defining the discrete density of states $g_\text{dsc}(\epsilon)$ to be the
number of modes at energy $\epsilon$, we have
\begin{equation}
    \Nex = \sum_\epsilon g_\text{dsc}(\epsilon) \frac{1}{e^{\beta(\epsilon - \mu)} - 1}.
\end{equation}
Next, we recognise that $N_0 > 0$ implies $\mu < \epsilon_0$, which in turn implies
$1/\bkt{e^{\beta(\epsilon_i - \mu)} - 1} < 1/\bkt{e^{\beta(\epsilon_i - \epsilon_0)} - 1}$
so that we can replace $\mu$ with $\epsilon_0$ to bound $\Nex$ from above:
\begin{equation}
    \Nex < \sum_\epsilon g_\text{dsc}(\epsilon) \frac{1}{e^{\beta(\epsilon - \epsilon_0)} - 1}.
\end{equation}
In the particular case of the ideal gas, $\epsilon_0 = 0$ and we have
\begin{equation}
    \Nex < \sum_\epsilon g_\text{dsc}(\epsilon) \frac{1}{e^{\beta\epsilon} - 1}.
\end{equation}

Unfortunately evaluating this sum is not straightforward, so to progress further we
transform it into an integral via an approximation.  We have
\begin{equation} \label{Nex_integral}
    \Nex \lesssim \int d\epsilon\; g(\epsilon) \frac{1}{e^{\beta(\epsilon - \mu)} - 1},
\end{equation}
where $g(\epsilon)$ is the density of states that will correspond to a
smoothed version of $g_\text{dsc}(\epsilon)$.


The most elementary method for computing an appropriately smoothed $g$ is to
first evaluate the total number of states $G(\epsilon)$ with energy less than
$\epsilon$ (see, for example, \cite{Pethick2008}).  With the simplest 
approximate evaluation, $G(\epsilon)$ turns out to be smooth and we may define
a smooth $g$ using $g(\epsilon) \equiv \deriv{G}{\epsilon}$.
For the 3D gas in a finite periodic box of size $L\times L\times L$, the
possible modes are indexed by the integer wavenumber $\v{q} \in \Zfield^3$ and
have energy
\begin{equation}
    E(\v{q}) = \frac{2\hbar^2\pi^2\v{q}^2}{mL^2}.
\end{equation}
The number of modes with energy less than $\epsilon$ is then approximately
the volume of the sphere of radius $\sqrt{mL^2\epsilon/2\hbar^2\pi^2}$
containing the set $\bktcl{\v{q}\in\Rfield^3 \colon E(\v{q}) < \epsilon}$, that
is,
\begin{equation}
    G(\epsilon) = \frac{4\pi}{3} \Bkt{\frac{mL^2\epsilon}{2\hbar^2\pi^2}}^{3/2}.
\end{equation}
Differentiating $G$ leads to the density of states
\begin{equation}
    g(\epsilon) = \deriv{G}{\epsilon}
        = \frac{4\sqrt{2}\pi L^3 m^{3/2}}{h^3} \epsilon^{1/2},
\end{equation}
which we see varies as the square root of $\epsilon$.

Putting this into Eq.~\eqref{Nex_integral} and performing the 
integral\footnote{The integral may be evaluated by expanding in powers of 
$e^{-x}$, followed by a change of variables and recognising the Riemann zeta 
function $\zeta$ and gamma function $\Gamma$:
\begin{equation*}
    \int_0^\infty dx\; \frac{x^{1/2}}{e^x-1} = 
    \sum_{n=1}^\infty \int_0^\infty dx\; x^{1/2}e^{-xn} =
    \sum_{n=1}^\infty n^{-3/2} \int_0^\infty dx\; x^{1/2}e^{-x} =
    \zeta\Bkt{\tfrac{3}{2}} \Gamma\Bkt{\tfrac{3}{2}}.
\end{equation*}
},
we finally arrive at
\begin{equation}
    \Nex \lesssim \Bkt{\frac{2\pi m L^2 k_B}{h^2}}^{3/2}
        \zeta\Bkt{\tfrac{3}{2}} T^{3/2}
\end{equation}
where $\zeta(n) = \sum_{k=1}^\infty 1/k^n$ is the Riemann zeta function, with 
$\zeta(\frac{3}{2}) \approx 2.61$.  This expression tells us that there must
be some critical temperature $\Tc$ below which $N$ is strictly greater than
$\Nex$, and the remaining $N-\Nex$ atoms must go into the ground state.  At
the critical temperature, $N = \Nex(\Tc)$ which implies
\begin{equation}
    \Tc = \frac{h^2 N^{2/3}}{2\pi m L^2 k_B \zeta\Bkt{\tfrac{3}{2}}^{2/3}}.
\end{equation}
The fraction of atoms in the ground state as a function of temperature is
\begin{equation} \label{3d_condensate_fraction}
    f_c = \frac{N-\Nex}{N} = 1 - \Bkt{\frac{T}{\Tc}}^{3/2}.
\end{equation}

Determining the condensate fraction using the standard derivation presented
above depends critically on the functional form of the density of states.  In
turn, the density of states depends on the dimensionality and any trapping
potential, both of which modify the possible single-particle quantum states.
Indeed, in the two-dimensional homogeneous gas the integral
in Eq.~\eqref{Nex_integral} fails to converge, suggesting --- though not by itself
proving --- an absence of condensation.

\section{Field theory and interacting Bose gases}
\label{field_theory}


In this section we develop some basics of quantum field theory in a form that
applies to cold gases of atomic bosons.  In particular, the theory developed
below is strictly nonrelativistic, and assumes bosonic symmetry for the 
many-particle quantum state.

\subsection{Second quantisation}
\label{second_quantisation}

The formalism of \emph{second quantisation}\footnote{
The name ``second quantisation'' is somewhat obscure.  Historically,  one
method of deriving a non-interacting quantum field theory was to apply the
procedure of canonical quantisation to the Schrödinger field (see 
\cite[page~81]{Teller1995} for a discussion).  Insofar as one step of
quantisation has already been used to obtain the Schrödinger equation, it
appeared that this procedure constituted quantising a second time.
} is an elegant and compact way to
express the quantum mechanics of identical particles.  Second quantisation
succeeds by incorporating the symmetries of the many-particle state into the
theory from the start.
At heart, it may be viewed as a convenient reformulation
of basic many-particle quantum mechanics, extended to allow for superpositions
of states with different particle number.  There are many good explanations of
second quantisation in the literature; for additional exposition we refer the
reader to Refs.~\cite{Altland2006,Teller1995}.


\subsubsection{Many-particle symmetries and the number state basis}

To describe second quantisation we start with the quantum mechanics of an $N$
particle system.  In the general case, the Hilbert space for a system of $N$
distinguishable particles is the tensor product of $N$ one-particle
Hilbert spaces (see, for example, \cite[\S 2.2.8]{Nielsen2000}):
\begin{equation} \label{general_quantum_state_space}
    \Hcal = \Hcal_1 \otimes \Hcal_2 \otimes \dotsb \otimes \Hcal_N.
\end{equation}
Accordingly, the possible states of $N$ \emph{identical} particles live in the
Hilbert space
\begin{equation} \label{N_particle_tensor_product_space}
    \Hcal = \underbrace{\Hcal_1 \otimes \Hcal_1 \otimes \dotsb \otimes 
            \Hcal_1}_{N \text{ times}}
        \equiv \Hcal_1^N.
\end{equation}
States from the natural tensor product basis for $\Hcal_1^N$ have the
form\footnote{
We omit the tensor product symbols between kets for brevity;
$\ket{a}\ket{b} \equiv \ket{a}\otimes\ket{b}$.
}
$\ket{j_1}\ket{j_2}\dotsm\ket{j_N}$, where the $j_i$ are positive integers
and we use $\bktcl{\ket{1}, \ket{2},\dotsc}$ as a basis for $\Hcal_1$.

It is possible to solve problems directly in $\Hcal_1^N$.  However, the
indistinguishability of particles implies a strict symmetry requirement on
which states of $\Hcal_1^N$ correspond to realisable physical states.  In the
case of bosons, the state must remain unchanged under an interchange of
particle indices.  For example, a two-particle state $\ket{\psi} = \sum_{i,j}
c_{ij} \ket{i}\ket{j}$ must have the property $c_{ij} = c_{ji}$; in a
continuous basis this takes the more familiar form $\psi(x_1,x_2) =
\psi(x_2,x_1)$.  We note that the correct derivation of the allowable boson
and fermion symmetry rules is considerably more subtle than the common argument
based on swapping labels in the many-particle wavefunction \cite{Leinaas1977}.

Taking symmetry into account, the general state for $N$ bosons may be written
\begin{equation} \label{explicitly_symmetrised_wavefunction}
\ket{j_1,\dotsc,j_N}_\text{sym} =
    \frac{1}{\sqrt{N! \prod_{i=1}^\infty n_i!}} \sum_P \ket{j_{P_1}} \ket{j_{P_2}} \dotsm \ket{j_{P_N}}
\end{equation}
where the sum is over all permutations $P$ of the integers $\{1,\dotsc,N\}$ and
$n_i$ is the number of particles in mode $\ket{i}$ as before.  Such
wavefunctions are usually extremely inconvenient to use for computational
purposes due to the large number $\prod_{i=1}^{\infty}n_i!/N!$ of unique terms.


At this stage it is useful to introduce the number state notation.  To ensure 
uniqueness, the indices in the state $\ket{j_1,\dotsc,j_N}_\text{sym}$ must be 
ordered --- for example, $\ket{1,1,1,2}_\text{sym}$ and $\ket{1,1,2,1}_\text{sym}$
are the same state due to the sum over all permutations.  This 
redundancy can be removed by instead listing the occupations of all modes.  We
have
\begin{equation} \label{number_state_definition}
    \ket{n_1,n_2,\dotsc} \equiv
        \ket{\underbrace{1,\dotsc,1}_{n_1\text{ times}},
        \underbrace{2,\dotsc,2}_{n_2\text{ times}}, \dotsc}_\text{sym},
\end{equation}
so that, for example, $\ket{1,1,1,2}_\text{sym} = \ket{3,1,0,0,\dotsc}$.  
The state $\ket{n_1,n_2,\dotsc}$ has a well-defined number of bosons in each 
mode and is therefore called a \emph{number state}.

\subsubsection{Fock space and the creation and annihilation operators}
%

Let $\Fcal^N$ be the subspace of $\Hcal_1^N$ spanned by all the correctly
symmetrised number states.  $\Fcal^N$ contains all the possible states of the
$N$ bosons, and therefore avoids the redundancy of nonphysical states contained 
in $\Hcal_1^N$.  The \emph{Fock space} is constructed via the direct sum\footnote{
The direct sum of two Hilbert spaces, $\Hcal_1\oplus\Hcal_2$, is the space of 
ordered pairs $\ket{(\psi,\phi)} \in \Hcal_1\times\Hcal_2$, with inner product
defined by
\begin{equation*}
    \braket{(\psi_1, \phi_1)}{(\psi_2, \phi_2)} = \braket{\psi_1}{\psi_2} + \braket{\phi_1}{\phi_2}.
\end{equation*}
}
of all the $\Fcal^N$ with varying $N$:
\begin{equation}
    \Fcal = \bigoplus_{N=0}^\infty \Fcal^N.
\end{equation}
Note the presence of the Hilbert space $\Fcal^0$ of zero particles which has a
single basis vector known as the \emph{vacuum state} and conventionally written
$\ket{0}$.

The efficiency of second quantisation arises because we can perform
calculations entirely within $\Fcal$ and the symmetry of the number states is
built in from the start.  The formalism of \emph{creation} and
\emph{annihilation operators} is the tool used for this task.  We define the
creation operator for the $i$th mode by its action on the number basis:
\begin{equation}\label{creation_op_definition}
    \aodag_i\ket{n_1,\dotsc,n_i,\dotsc} = (n_i+1)^{1/2}\ket{n_1,\dotsc,n_i+1,\dotsc}.
\end{equation}
That is, the action of the creation operator $\aodag_i$ is to create a 
particle in the $i$th mode.  Note that this is a complete specification of
$\aodag_i$, because we have defined its action for every number state and the 
number states are a basis for $\Fcal$.  From the definition we have
$\braoket{m+1}{\aodag}{n} = (n+1)^{1/2}\delta_{mn}$; taking the conjugate and 
rearranging shows that the adjoint $\ao_i$ is the annihilation operator for
mode $i$ because it reduces the occupation by one:
\begin{equation}\label{annihilation_op_definition}
    \ao_i\ket{n_1,\dotsc,n_i,\dotsc} = n_i^{1/2}\ket{n_1,\dotsc,n_i-1,\dotsc}.
\end{equation}

A basic and important property of the creation and annihilation operators is
their commutation relations.  These can be computed directly from
Eqs.~\eqref{creation_op_definition} and \eqref{annihilation_op_definition}
yielding
\begin{equation}
    \bktsq[\big]{\ao_i,\aodag_j} = \delta_{ij},
    \qquad
    \bktsq[\big]{\aodag_i,\aodag_j} = 0,
    \qquad
    \bktsq[\big]{\ao_i,\ao_j} = 0.
\end{equation}
The commutation relations are vital for calculations, but may also be used as an 
alternate starting point to Eq.~\eqref{creation_op_definition}, resulting
in the same operators.

The operators $\aodag_i$ are defined with respect to a particular
single-particle basis $\{\ket{i}\}$; a different set of operators $\bodag_\kappa$ arises 
if we consider a different basis $\{\ket{\kappa}\}$.  (Note that we are abusing
the notation somewhat and treating kets with Greek indices as a distinct
basis from those with Latin indices.)  The change of basis law
may be derived by considering the action of $\aodag_i$ on the vacuum state, and
using the resolution of identity $\identop=\sum_\kappa\ketbra{\kappa}{\kappa}$
in the one-particle Hilbert space:
\begin{gather}
    \aodag_i \ket{0} = \ket{i} =
        \Bkt{\sum_\kappa \ketbra{\kappa}{\kappa}} \ket{i} =
        \sum_\kappa \braket{\kappa}{i} \; \ket{\kappa} = 
        \sum_\kappa \braket{\kappa}{i} \; \bodag_\kappa \ket{0}.
\end{gather}
The change of basis is therefore given by
\begin{equation} \label{ao_aodag_basis_change}
    \aodag_i = \sum_\kappa \braket{\kappa}{i} \; \bodag_\kappa
    \quad\text{and}\quad
    \ao_i = \sum_\kappa \braket{i}{\kappa} \; \bo_\kappa.
\end{equation}

Using the analogous transformation law for the continuous position
basis
$\{\ket{\vx}\}$ gives us the powerful concept of the \emph{boson field operator}
\begin{equation}
    \fo(\vx) = \sum_i \braket{\vx}{i} \; \ao_i
              = \sum_i \psi_i(\vx) \ao_i
\end{equation}
where $\psi_i(\vx) \equiv \braket{\vx}{i}$ is the shape of mode $i$ in position space.
Note that we use a Greek letter for the field operator by convention; we could
equally well have used $\aodag(\vx)$ to emphasise the similarity with the
operators $\aodag_i$ for the discrete basis.  We interpret the operation of the
object $\fodag(\vx)$ on a quantum state as the creation of a particle at position
$\vx$.  Alternatively, it may be thought of as a field that associates an
operator on $\Fcal$ to each point of space.  The commutation relations for the
continuous case are analogous to the discrete case:
\begin{equation}
    \bktsq[\big]{\fo(\vx),\fodag(\v{y})} = \delta(\vx-\v{y}),
    \qquad
    \bktsq[\big]{\fodag(\vx),\fodag(\v{y})} = 0,
    \qquad
    \bktsq[\big]{\fo(\vx),\fo(\v{y})} = 0.
\end{equation}
Various expressions are shorter and more familiar when written in the position
basis using $\fo$, as will become clear in the following sections.

\subsubsection{Representation of operators on Fock space}
%

The creation and annihilation operators are useful because of two important
properties: First, they act in a particularly straightforward way on the number
state basis of $\Fcal$, as per the definition.  Second, the operators of
interest on $\Hcal_1^N$ can be written simply in terms of $\aodag_i$ and
$\ao_i$.  Once the necessary operators are expressed in the natural basis for
$\Fcal$ there is no longer any need to consider explicitly symmetrised states
defined on $\Hcal_1^N$.  This section examines the representation of one and 
two-particle operators on Fock space, largely following the development in
Ref.~\cite{Altland2006}.

We examine the case of single-particle operators first.  Consider an operator
$\hat{q}$ acting on $\Hcal_1$.  $\hat{q}$ may be extended to act on the space
of $N$ particles as the sum
\begin{equation}
    \hat{Q} = \sum_{i=1}^N \hat{Q}_i
\end{equation}
where $\hat{Q}_i$ represents $\hat{q}$ operating on the $i$th particle\footnote{
$\hat{Q}_i$ may be written explicitly as
$\hat{Q}_i = \identop^{\otimes(N-i)} \otimes \hat{q} \otimes \identop^{\otimes i}$
where 
$\identop^{\otimes i} \equiv \underbrace{\identop \otimes \dotsm \otimes \identop}_{i \:\text{times}}$
is the identity on $\Hcal_1^i$, constructed from $i$ copies of the identity 
operator $\identop$ on $\Hcal_1$.
}.  Suppose now that
$\hat{q}$ is diagonal in some basis $\{\ket{\kappa}\}$ of $\Hcal_1$ so that
$\hat{Q} = \sum_\kappa q_{\kappa\kappa} \ketbra{\kappa}{\kappa}$.  Using the
explicitly symmetrised wavefunction in Eq.~\eqref{explicitly_symmetrised_wavefunction}
the action of $\hat{Q}$ on a number state may be computed.  After some algebra,
we find
\begin{align}
    \hat{Q}\ket{n_{\kappa 1}, n_{\kappa 2},\dotsc}
        &= \bkt[\Big]{\sum_\kappa q_{\kappa\kappa} n_\kappa}
            \ket{n_{\kappa 1}, n_{\kappa 2},\dotsc} \\
        &= \bkt[\Big]{\sum_\kappa q_{\kappa\kappa} \bodag_\kappa\bo_\kappa}
            \ket{n_{\kappa 1}, n_{\kappa 2},\dotsc} \\
    \implies \hat{Q} &= \sum_\kappa q_{\kappa\kappa} \bodag_\kappa\bo_\kappa
                 = \sum_\kappa \braoket{\kappa}{\hat{q}}{\kappa} \bodag_\kappa\bo_\kappa
\end{align}
In the second line we have used the fact that the \emph{occupation number
operator} for the $\kappa$th mode is $\hat{n}_\kappa = \bodag_\kappa
\bo_\kappa$.  Changing to the non-diagonal basis using
Eq.~\eqref{ao_aodag_basis_change} gives the general expression for any
one-particle operator
\begin{equation} \label{fock_space_one_particle_operator}
    \hat{Q} = \sum_{i,j} \braoket{i}{\hat{q}}{j} \aodag_i\ao_j.
\end{equation}

A related procedure can be carried out for multi-particle operators, with
similar results.  If $\hat{q}'$ is a two-particle operator acting on the space
$\Hcal_1^2$, an extension to $\Hcal_1^N$ is
\begin{equation} \label{two_particle_operator_extension}
    \hat{Q}' = \sum_{\substack{i,j=1 \\ i\ne j}}^N \hat{Q}'_{ij}
\end{equation}
where $\hat{Q}'_{ij}$ is $\hat{q}'$ acting on the subspace of the $i$th and
$j$th particles.  The operator $\hat{Q}'$ can be expressed in terms of the
creation and annihilation operators.  The procedure is similar to the 
derivation for the single-particle operators above, with some care required in 
handling the sum so that $i \ne j$.  The general result is
\begin{equation}
    \hat{Q}' = \sum_{i,j,k,l} \bkt[\big]{\bra{i}\otimes\bra{j}}\hat{q}'\bkt[\big]{\ket{k}\otimes\ket{l}}
                \; \aodag_i\aodag_j \ao_k \ao_l.
\end{equation}
Note that including both orderings $(i,j)$ and $(j,i)$ in
Eq.~\eqref{two_particle_operator_extension} is redundant because $\hat{q}'$ is 
symmetric for identical particles so $\hat{Q}'_{ij} = \hat{Q}'_{ji}$.

The second-quantised operators we will be interested in are conveniently
compact and familiar in the position basis, as exemplified by the Hamiltonian.
The single-particle part of the Hamiltonian\footnote{
It is conventional to omit the hat from $\Hsp$ and other operators acting on 
the single-particle configuration space.  This is somewhat inconsistent, but
serves as a useful reminder of the subspace on which these operators act when 
dealing with the full Fock space.
},
\begin{equation}\label{Hsp_definition}
    \Hsp = -\frac{\hbar^2}{2m} \del_\vx^2 + V(\vx),
\end{equation}
extended to Fock space is
\begin{equation}
    \hat{H}_1 = \int d\vx\; \fodag(\vx) \Bktsq{-\frac{\hbar^2}{2m} \del_\vx^2 + V(\vx)} \fo(\vx).
\end{equation}
It is notable that this equation has only a single integral over the
variable $\vx$ rather than the double integral --- or sum in the discrete case
--- which would be expected in general (cf.\
Eq.~\eqref{fock_space_one_particle_operator}).  This is because $\Hsp$ acts only
locally in the position basis, although it is not quite diagonal.

\newcommand{\Vint}[1]{V_\text{int}\bkt[\big]{#1}}

If the interaction potential between a pair of particles located at $\vx$ and
$\vx'$ is given by $\Vint{\vx-\vx'}$, the interaction term in the $N$
particle Hamiltonian is
\begin{equation}
    \sum_{\substack{i,j=1 \\ i<j}}^N \Vint{\vx_i-\vx_j}
        = \frac{1}{2} \sum_{\substack{i,j=1 \\ i\ne j}}^N \Vint{\vx_i-\vx_j}.
\end{equation}
This is diagonal in position space, so writing it in second-quantised form is
simple:
\begin{equation}
    \hat{H}_2 = \frac{1}{2}
        \iint d\vx\; d\vx'\; \fodag(\vx)\fodag(\vx') \Vint{\vx-\vx'} \fo(\vx') \fo(\vx).
\end{equation}
To summarise, the second-quantised Hamiltonian for the dilute Bose gas is given
by
\begin{align} \label{full_bose_gas_Hamiltonian}
    \hat{H} = \int d\vx\; \fodag(\vx) \Hsp \fo(\vx) + 
        \frac{1}{2} \iint d\vx\; d\vx'\; \fodag(\vx)\fodag(\vx') \Vint{\vx-\vx'} \fo(\vx') \fo(\vx).
\end{align}
This Hamiltonian is the starting point for every theoretical analysis of the 
single species Bose gas \cite{Proukakis2008}; it contains a complete
description of the physics in the low-energy regime.

\subsection{The low-energy Hamiltonian}
\label{s_wave_hamiltonian_approx}

Experiments with ultracold gases often take place in the dilute regime --- the
typical interatomic distance is much larger than the range of the potential
$V_\text{int}$.
As a consequence, the interactions may be treated as two-body scattering events
to a high degree of accuracy.  Solving the two-body scattering problem involves
expanding the wavefunction in a series of partial waves.  At sufficiently low
energies only the $s$ partial wave is important and the scattering may be
characterised by a single parameter $a_s$ called the \emph{s-wave scattering length}.
The scattering length is on the order of $100a_0$ for alkali atoms, where $a_0$ 
is the Bohr radius, though there is considerable experimental flexibility in
tuning the value using Feshbach resonances.  For a detailed account of the
scattering theory and related issues, we direct the reader to 
Ref.~\cite[Ch.~5]{Pethick2008}.

In practice, the dilute and low energy nature of the system implies that the 
theory can be greatly simplified by formally replacing the full interaction
potential by a delta function
\begin{equation}\label{s_wave_V_replacement}
    \Vint{\vx-\vx'} \to U_0\delta(\vx-\vx')
\end{equation}
in the Hamiltonian, where $U_0 = 4\pi\hbar^2 a_s / m$.  Strictly speaking, this
replacement arises by integrating out the high energy states --- resulting in a
two-body T-matrix description --- followed by taking the zero energy limit
(see, for example, \cite[\S 2.1]{Blakie2008}).  The resulting theory necessarily
contains a high energy cutoff that prevents the delta potential from
unphysically scattering waves of arbitrarily high momentum.
Nevertheless, we ignore this complication in what follows because the main
approximations used later in the thesis are not sensitive to it.

Performing the replacement of Eq.~\eqref{s_wave_V_replacement} leads to the
approximate low-energy Hamiltonian:
\begin{equation} \label{swave_bose_gas_Hamiltonian}
    \hat{H} = \int d\vx\; \fodag(\vx) \Hsp \fo(\vx)
        + \frac{U_0}{2}\int d\vx\; \fodag(\vx)\fodag(\vx) \fo(\vx)\fo(\vx).
\end{equation}
A theory built from this Hamiltonian retains all the smooth long wavelength
behaviour, while all short range correlations between particles are discarded.


\subsection{Operator equations of motion}

In section \ref{second_quantisation} we wrote operators and states for $N$
particle quantum mechanics in the convenient notation of second quantisation 
but we said nothing about time evolution.  Clearly we may use the
\emph{Schrödinger picture} in which the quantum state on Fock space evolves
according to the usual Schrödinger equation.  Equivalently we may use the 
\emph{Heisenberg picture} of time evolution where the operators evolve and the 
quantum state is fixed.  The Heisenberg equation of motion for an operator
$\hat{A}_H$ is
\begin{equation} \label{general_heisenberg_eoms}
    i\hbar \pderiv{\hat{A}_H(t)}{t} = \bktsq{\hat{A}_H(t), \hat{H}},
\end{equation}
where we have assumed that the corresponding Schrödinger picture operator 
$\hat{A}_S = \hat{A}_H(0)$ is time independent.  Note that this equation is linear --- in
the usual sense that linear combinations of solutions are new solutions ---
because the so-called superoperator $[\cdot, \hat{H}]$ is linear.

We can in principle write any desired Heisenberg picture observable in terms of
the Heisenberg picture field operator $\fo_H(\vx,t)$.  Therefore the full
dynamics of the system are encoded in the behaviour of $\fo_H(\vx,t)$, the
general evolution of which is given by
\begin{equation}
    i\hbar \pderiv{\fo_H(\vx,t)}{t} = \bktsq{\fo_H(\vx,t), \hat{H}}.
\end{equation}
We now drop the subscripts for brevity, assuming that from now on we will work
in the Heisenberg picture unless otherwise specified.  Using the Hamiltonian
in Eq.~$\eqref{swave_bose_gas_Hamiltonian}$ along with the commutation relations for
$\fodag$ and $\fo$, we obtain the operator equation of motion for $\fo$ (see, 
for example, \cite{Proukakis2008}):
\begin{equation} \label{field_heisenberg_eoms}
    i\hbar \pderiv{\fo(\vx,t)}{t} =
        \Hsp \fo(\vx,t) + U_0 \fodag(\vx,t)\fo(\vx,t)\fo(\vx,t).
\end{equation}
The apparently nonlinear form of this equation presents a puzzle when compared
to the manifestly linear form of the equation of motion
\eqref{general_heisenberg_eoms} for a general operator.  What we have here is a 
tradeoff: We have removed the explicit dependence on the Hamiltonian operator
so that the equation is expressed purely in terms of $\fo$; the price is that
to actually evaluate the time evolution requires computing a nonlinear function
of field operators.  Nevertheless, the underlying linearity of the solutions 
is preserved\footnote{For a very simple example, consider a single-mode
system with Hamiltonian $\hat{H}=\frac{1}{2}\aodag\aodag\ao\ao$ representing a two-particle
interaction.  The Heisenberg equation of motion for $\ao$ is
\begin{equation*}
    i\hbar \pderiv{\ao}{t} = \aodag\ao\ao,
\end{equation*}
which appears to be nonlinear.  However, $\hat{H}$ may also be expressed in the
number basis as $\hat{H} = \frac{1}{2}\sum_n n(n-1) \ketbra{n}{n}$.  Writing
$\ao(t) = \sum_{n,m} a_{nm}(t) \ketbra{n}{m}$, the evolution equation for
the components of $\ao(t)$ is clearly linear:
\begin{equation*}
    i\hbar \pderiv{a_{nm}}{t} = \frac{1}{2}\bktsq[\big]{m(m-1) - n(n-1)} a_{nm}.
\end{equation*}
}.

The Heisenberg equations of motion arising from very simple Hamiltonians can
sometimes be solved directly.  However, including interactions as in 
Eq.~\eqref{field_heisenberg_eoms} generally makes the problem analytically
intractable and attacking it directly using numerics is also out of the
question, due to the infinite number of degrees of freedom.  Nevertheless, the
Heisenberg equations of motion for the field operator give us a useful starting
point for certain types of approximations, including the projected
Gross-Pitaevskii equation discussed in section \ref{PGPE_derivation}.


\section{The Gross-Pitaevskii equation}
\label{GPE_derivation}

%

The Gross-Pitaevskii equation\footnote{The GPE is commonly known as the
(cubic) nonlinear Schrödinger equation (NLSE) in other areas of physics and
in mathematics.} (GPE) is a remarkably successful simplification 
of the full quantum field equations at the level of mean-field theory.  We have 
already seen that for the ideal Bose gas all particles condense into a single
quantum mode at zero temperature.  The central approximation of the 
Gross-Pitaevskii theory is to assume this wavefunction is also reasonable for
\emph{dynamics} with small but nonzero interactions and excitation energies.
That is, we assume the Schrödinger picture wavefunction is
\begin{equation} \label{TDHF_ansatz}
    \Phi\Bkt{\vx_1, \dotsc, \vx_N, t} \equiv \prod_{i=1}^N \varphi\Bkt{\vx_i, t},
\end{equation}
and compute an equation of motion for $\varphi$ so that the dynamics of $\Phi$
approximates the full quantum dynamics of $\Psi$ as closely as possible.  The
assumed form for the wavefunction given above is the \emph{time dependent
Hartree-Fock} ansatz.  (We note that $N$ bosons in a single mode is a
special case where the fully symmetrised Hartree-Fock ansatz is identical to
the simple product state, also known as the Hartree ansatz.)

There are numerous methods for deriving the equations of motion for $\varphi$.
In the cold atoms literature the most commonly used are geared toward computing
not only an equation for $\varphi$ but also the next order corrections, giving the
so-called Hartree-Fock-Bogoliubov theories.  These techniques may be broadly
grouped into the \emph{symmetry breaking} (see, for example, \cite{Dalfovo1999})
and \emph{number conserving} formalisms (see, for example,
\cite{Gardiner1997,Morgan1999,Gardiner2007}).  For a discussion of the 
difference between these techniques, see Refs.~\cite{Morgan1999,Proukakis2008}.

To derive only the lowest order theory, we choose a more elementary method
based on a time dependent variational principle.

\subsection{The Dirac-Frenkel time dependent variational principle}
\label{DF_variational_principle}

%

The Dirac-Frenkel time dependent variational principle \cite[\S II.1]{Lubich2008}
has been commonly used in quantum chemistry and nuclear theory in the context
of time dependent Hartree-Fock theory \cite{Jackiw1979}.  It is also an 
expedient method for deriving the time dependent GPE \cite[\S~7.1]{Pethick2008} 
and with that in mind we briefly outline some of its generic properties before
making use of it in the next section.  The Dirac-Frenkel variational principle
is characterised by stationarity of the quantum effective action\footnote{
In path integral quantisation, one builds the quantum theory out of a classical
action integrated over all paths.  This is distinct from the full quantum 
``effective action'' for the same system, which is a functional of the quantum
rather than classical state.
}
\begin{equation}
    S[\Psi] = \int_{t_0}^{t_1} dt\; \braoket[\big]{\Psi}{i\hbar \partial_t - \hat{H}}{\Psi}
            = \int_{t_0}^{t_1}dt\int d\vx\; \Psi^*\bkt[\big]{i\hbar \partial_t - \hat{H}}\Psi
\end{equation}
with respect to small variations in $\Psi^*$. For an unconstrained state $\Psi$
in the full Hilbert space $\Hcal$, the variational principle is equivalent to
\begin{equation} \label{FD_variational_principle_full_Psi}
    \fderiv{S}{\Psi^*} = 0,
\end{equation}
where $\delta S/\delta\Psi^*$ is the functional derivative of the
nonholomorphic function\footnote{A complex-valued function of complex arguments
is \emph{holomorphic} if it is complex differentiable according to the usual
limit-based definition of derivative in the complex plane.  For nonholomorphic 
functions we must use a modified definition of the derivative; see section
\ref{sec_wirtinger_calculus}.}
$S$ with respect to the complex variable $\Psi^*$.  For a detailed discussion
including how to compute with these derivatives in practice, see appendix
\ref{sec_functional_derivatives}.  Evaluating the functional derivative gives
\begin{equation}
    \fderiv{S}{\Psi^*} = \bkt[\big]{i\hbar\partial_t - \hat{H}}\Psi,
\end{equation}
which immediately yields the time dependent Schrödinger equation when set to
zero.

In contrast to the trivial case above, the variational principle is most useful 
when constraining $\Psi$ to some submanifold $\Mcal$ of $\Hcal$; in our
case $\Mcal$ will be the set of all states of the form given in Eq.~\eqref{TDHF_ansatz}.
In fact, the Dirac-Frenkel variational principle arises as a result of trying
to find the ``best possible'' approximation of this kind: Consider some 
\emph{approximate} evolution $\Psi(t)\in\Mcal$.  According to the Schrödinger
equation, the full dynamics attempts to evolve $\Psi(t)$ in the direction
$\hat{H}\Psi(t)$, but this evolution generally takes us out of the manifold $\Mcal$
and into the bulk of the higher dimensional space $\Hcal$.  To avoid this
we want an evolution direction $v$ constrained to the tangent space
$T_{\Psi(t)}\Mcal$ of $\Mcal$ at $\Psi(t)$ but chosen so that the error
$\norm{v - \hat{H}\Psi(t)}$ is as small as possible.  Evolving $\Psi$ according to
$v$ means $i\hbar\partial_t\Psi(t) = v$ by definition, so we want to minimise
$\norm{i\hbar\partial_t\Psi(t) - \hat{H}\Psi(t)}$.  This leads to the condition that 
the residual of the Schrödinger equation is perpendicular to the tangent 
space $T_{\Psi(t)}\Mcal$:
\begin{equation}
    \braoket[\big]{u}{i\hbar \partial_t - \hat{H}}{\Psi} = 0
        \qquad \forall\; u\in T_\Psi\Mcal.
\end{equation}
The action in the Dirac-Frenkel variational principle follows from the
orthogonality condition above as discussed in Ref.~\cite{Lubich2008}.  It is also
shown that the system of equations arising from the variational principle
automatically conserves the expected energy, $E[\Psi] =
\braoket{\Psi}{\hat{H}}{\Psi}$.  Also conserved is any operator $A$ that commutes
with the Hamiltonian and keeps $A\Psi$ within $T_\Psi \Mcal$:
\begin{equation}
    A\Psi \in T_\Psi\Mcal \qquad\forall\; \Psi \in \Mcal\cap D(A)
\end{equation}
where $D(A)$ is the domain of $A$.

\subsection{Deriving the GPE}

%
%

Deriving the GPE is a straightforward application of the variational principle
introduced in the previous section, with approximation manifold $\Mcal$ equal
to the set of all states of the form given in Eq.~\eqref{TDHF_ansatz}.  For convenience, 
we repeat the action
\begin{equation}
    S[\Psi] = \int_{t_0}^{t_1} dt\; \braoket[\big]{\Psi}{i\hbar \partial_t - \hat{H}}{\Psi}.
\end{equation}
To compute this for our particular assumed wavefunction $\Phi$, we need expressions for $\braoket{\Phi}{\hat{H}}{\Phi}$ and
$\braoket{\Phi}{i\hbar \partial_t}{\Phi}$.  For the first part, we note that 
our wavefunction at some time $t$ has the form $\Phi=\ket{N,0,\dotsc}$ in 
second-quantised notation, where the only occupied mode is $\varphi(t)$.  The 
Schrödinger picture field operators act on the state as
\begin{align}
    \fo\ket{N,0,\dotsc} &= \varphi(\vx,t)\sqrt{N}\ket{N-1,0,\dotsc}, \\
    \fodag\ket{N-1,0,\dotsc} &= \varphi^*(\vx,t)\sqrt{N}\ket{N,0,\dotsc}
\end{align}
and therefore the expectation value of the Hamiltonian
in Eq.~\eqref{swave_bose_gas_Hamiltonian} is
\begin{equation}
\begin{split}
    \bra{N,0,\dotsc}\hat{H}\ket{N,0,\dotsc} &=
        N \int d\vx\; \varphi^*(\vx,t) \Hsp \varphi(\vx,t) \\
        &+\frac{U_0}{2} N(N-1)\int d\vx\; \varphi^*(\vx,t)\varphi^*(\vx,t) \varphi(\vx,t)\varphi(\vx,t).
\end{split}
\end{equation}
Alternatively, this expression could be obtained by working directly with the
sums over all particles in the first-quantised Hamiltonian\footnote{
The first-quantised Hamiltonian is
\begin{equation*}
    H(\vx_1,\dotsc,\vx_N) = \sum_{i=1}^N \Hsp(\vx_i) +
        \frac{U_0}{2}\sum_{\substack{i,j=1 \\ i< j}}^N \delta(\vx_i-\vx_j),
\end{equation*}
which is equivalent to Eq.~\eqref{swave_bose_gas_Hamiltonian} when the number 
of particles is fixed at $N$.
}.

For the time derivative, it is simplest to work in the first-quantised position
representation; we see that
\begin{equation}
    i\hbar \partial_t\Phi
        = i\hbar\partial_t \prod_{i=1}^N \varphi(\vx_i, t)
        = i\hbar \sum_{i=1}^N \partial_t\varphi(\vx_i, t)  \prod_{\substack{j=1\\ j\ne i}}^N \varphi(\vx_j, t),
\end{equation}
so that
\begin{align}
    \Braoket{\Phi}{i\hbar \partial_t}{\Phi}
        &= i\hbar \int d\vx_1\dotsi\int d\vx_N \prod_{i=1}^N \varphi^*(\vx_i, t) 
            \partial_t \prod_{k=1}^N \varphi(\vx_k, t) \\
        &= N i\hbar \int d\vx\; \varphi^*(\vx, t) \partial_t \varphi(\vx,t).
\end{align}
Putting the parts of the action together, we have
\begin{equation}
    S[\varphi] = N \int_{t_0}^{t_1} dt \int d\vx\;
            \varphi^*(\vx,t)\bktsq[\big]{ i\hbar \partial_t - \Hsp }\varphi(\vx,t)
            - \frac{U_0}{2} (N-1)\abs{\varphi(\vx,t)}^4.
\end{equation}
To simplify further, we introduce the \emph{condensate wavefunction}\footnote{
Note that the condensate wavefunction is not a permissible single-particle 
wavefunction due to the normalisation convention.
},
\begin{equation}
    \phi(\vx,t) \equiv N^{1/2}\varphi(\vx,t),
\end{equation}
so that $\abs{\phi}^2$ is the particle density, and use the approximation
$(N-1)/N \approx 1$.
The action is then given by
\begin{equation}
    S[\phi] = \int_{t_0}^{t_1} dt \int d\vx\;
            \phi^*(\vx,t)\bktsq[\big]{ i\hbar \partial_t - \Hsp }\phi(\vx,t)
            - \frac{U_0}{2} \abs{\phi(\vx,t)}^4.
\end{equation}

As a final step we apply the variational principle $\delta S/\delta\phi^* = 0$.
Dropping the space and time indices for brevity, the functional derivative is
\begin{equation}
    \fderiv{S}{\phi^*} = \bktsq[\big]{ i\hbar \partial_t - \Hsp }\phi
            - U_0 \abs{\phi}^2\phi;
\end{equation}
we again direct the reader to appendix \ref{sec_functional_derivatives} for
details on computing functional derivatives.
Setting this to zero and using the usual form for the single-particle 
Hamiltonian as in Eq.~\eqref{Hsp_definition} yields the Gross-Pitaevskii
equation,
\begin{equation}\label{GPE}
    i\hbar\partial_t\phi =
        -\frac{\hbar^2}{2m}\del^2 \phi + V\phi + U_0\abs{\phi}^2\phi.
\end{equation}

\subsection{Validity of the GPE}
\label{GPE_validity}


The assumptions underlying the GPE are physically reasonable at very low 
energies\footnote{It is often said that the GPE is valid at ``zero 
temperature''.  Strictly speaking, the GPE does not describe a state with a
well-defined temperature because it is time dependent, but we may start from a 
true zero temperature state and apply a coherent excitation.
}
and when the scattering length $a_s$ is much less than the interparticle spacing.
This is borne out in the many successful applications to ultracold gas
experiments, where the GPE is an important theoretical tool for understanding
the dynamics of nearly pure condensates \cite{Pethick2008}.

Nevertheless, we emphasise that the derivation given above is not particularly
rigorous from a mathematical point of view: We
have said nothing about the size of the time dependent errors incurred by
making the product state ansatz of Eq.~\eqref{TDHF_ansatz}; neither have
we shown that the product state is a good initial state for the evolution.
Perhaps more severely, the effect of the replacement $V_\text{int}(\vx) =
\frac{U_0}{2}\delta(\vx)$ has also been glossed over, effectively neglecting the
short range correlations \cite{Erdos2007} that arise from the real potential
$V_\text{int}$.  Significant mathematical effort has been put into resolving 
these problems and convergence proofs are now available
\cite{Erdos2007,Pickl2010} for the limit $N\to\infty$ with $Na_s=\text{fixed}$,
known as the Gross-Pitaevskii or mean-field limit.

It is important to understand the nature of convergence in such proofs.  In
particular, the ansatz $\Phi$ does \emph{not} converge to the exact solution
$\Psi$ of the $N$ particle Schrödinger equation in the limit of large $N$,
\begin{equation}
    \norm[\big]{\Psi(\vx,t) - \Phi(\vx,t)} \nrightarrow 0 \qquad \text{as}\:N\to\infty.
\end{equation}
Instead, it is convergence of the one body reduced density matrix\footnote{
For a well-defined number of atoms $N$ in the pure quantum state $\Psi$, this
definition is equivalent to the definition
$\rho(\vx,\vx') = N\int\dotsi\int d\vx_2 \dotsm d\vx_N\; \Psi^*(\vx,\vx_2,\dotsc,\vx_N)\Psi(\vx',\vx_2,\dotsc,\vx_N)$,
so that the normalisation is $\int d\vx\; \rho(\vx,\vx) = N$
\cite[\S 13.5]{Pethick2008}.  This normalisation is conventional in ultracold 
atoms research so we use it here even though normalisation to the identity is
more convenient for the discussion of convergence.
}
$\rho(\vx,\vx') \equiv \braoket[\big]{\Psi}{\fodag(\vx)\fo(\vx')}{\Psi}$
to $\varphi^*(\vx)\varphi(\vx')$:
\begin{equation}\label{GPE_weak_convergence_eqn}
    \norm[\big]{\tfrac{1}{N}\rho(\vx,\vx') - \varphi^*(\vx)\varphi(\vx')} \to 0
        \qquad \text{as}\;N\to\infty
\end{equation}
where $\norm{\cdot}$ is the trace norm\footnote{The trace norm is defined by
$\norm{A} = \Tr\bktsq[\big]{(A^\dagger A)^{1/2}}$.
}.  In terms of the Penrose-Onsager criterion for condensation
\cite{Penrose1956}, this means that the system converges to a pure condensate 
in the mean-field limit.

To get a feeling for the importance of using the density matrix rather than
the state when proving convergence, consider the $N$ particle state
\begin{equation}\label{nearly_product_state}
    \Phi' \equiv
    \mathcal{S}\Bktcl{ \bktsq[\bigg]{\prod_{k=1}^{N-1} \varphi(\vx_k, t)} \varphi_\perp(\vx_N,t)},
\end{equation}
where $\varphi_\perp$ satisfies $\braket{\varphi_\perp}{\varphi} = 0$ and
$\mathcal{S}$ ensures correct Bose symmetrisation of the wavefunction.
Intuitively this state has ``nearly all particles in state $\varphi$'' and yet
it is orthogonal to the product $\Phi = \prod_{k=1}^{N} \varphi(\vx_k, t)$
so that $\norm{\Phi - \Phi'}$ is never small, even as $N\to\infty$.  On 
the other hand, we may show\footnote{
In second-quantised notation, $\ket{\Phi'} = \ket{N-1,1,0,\dotsc}$ in the basis
$\{\varphi,\varphi_\perp,\dotsc\}$, so we have
\begin{equation*}
    \fo(\vx)\ket{\Phi'} = (N-1)^{1/2}\varphi(\vx)\ket{N-2,1,0,\dotsc} +
        \varphi_\perp(\vx)\ket{N-1,0,\dotsc},
\end{equation*}
which allows $\rho$ to be computed immediately.  This is a small example of the 
great practical benefit of working in the second-quantised formalism over
having to deal with explicitly symmetrised wavefunctions as in 
Eq.~\eqref{nearly_product_state}.
} that the reduced density operator for $\Phi'$ is
\begin{equation}
    \braoket[\big]{\Phi'}{\fodag(\vx)\fo(\vx')}{\Phi'} =
        (N-1) \varphi^*(\vx)\varphi(\vx') + \varphi_\perp^*(\vx)\varphi_\perp(\vx')
\end{equation}
so that $\frac{1}{N}\rho(\vx,\vx')$ clearly converges to
$\phi^*(\vx)\phi(\vx')$ as $N\to\infty$.


\input{pgpe}


%
%

%% file: background/pgpe.tex
\section{The Projected Gross-Pitaevskii equation}
\label{PGPE_derivation}

The full quantum field evolution given in 
Eq.~\eqref{field_heisenberg_eoms} contains answers to any question we might ask
about cold Bose gases: Given a state $\ket{\Psi}$ and a \emph{solution}
$\fo(\vx,t)$ to the field equation, any observable $\hat{O}(t)$ can be written
as a function $\hat{O}(\fodag(\vx,t), \fo(\vx,t))$ of the field operator; the
expectation value of $\hat{O}$ at some time $t$ is
\begin{equation}
    \exval[\big]{\hat{O}(t)} \equiv
        \Braoket{\Psi}{\hat{O}\bkt[\big]{\fodag(\vx,t),\fo(\vx,t)}}{\Psi}.
\end{equation}
Unfortunately solving the operator equations of motion for $\fo$ directly is
intractable in all but the simplest of cases.  As a consequence there is an
extensive literature on methods for approximating the equations and/or
using an alternative but more tractable form.  Exact reformulations typically
recast the problem of computing expectation values as a problem of stochastic
sampling from some complicated probability distribution; this is the approach taken by the
various flavours of quantum Monte Carlo and the phase space methods commonly
used in quantum optics.  Approximation methods are many and varied, but usually
draw on physical insight about the kinds of quantum states which are of
interest to reduce the complexity of a solution.  In the following section we
describe the classical field approximation in the particular form of the
projected Gross-Pitaevskii equation (PGPE), originally developed in
Refs.~\cite{Davis2001a,Davis2001b}.

\subsection{Conceptual introduction}
\label{PGPE_conceptual_intro}

Under certain conditions a system governed by a quantum field theory can
instead be well approximated using a classical field.  This depends on both the state
of the system and the kinds of measurements we wish to make.  The prototypical
example is the electromagnetic field: while the underlying theory is quantum
mechanical, the classical theory --- in the form of Maxwell's equations for the
classical electric and magnetic fields --- describes a huge range of wave
phenomena very successfully.

This begs the question: which states and types of measurements can be
adequately described using a classical field?  Qualitatively, we can say that
classical fields describe only the collective wave-like behaviour of systems
that are in reality made up of quanta, and that we should have an
appropriately large number of those quanta so that the system is approximately
continuous rather than discrete.  More formally, this is a difficult question
to answer because taking the classical limit of a quantum theory is a subtle
business.  The standard quantum states that lead to classical wave-like
behaviour are the Glauber coherent states \cite{Glauber1963a}\footnote{
A single-mode coherent state (conventionally written $\ket{\alpha}$) is defined
to be an eigenvector of the annihilation operator, such that $\ao\ket{\alpha} =
\alpha \ket{\alpha}$ where $\alpha$ is a complex number.  In a similar way,
there is a full multimode coherent state $\ket{\Psi}$ for each complex-valued
field $\Psi(\vx,t)$ such that $\fo \ket{\Psi} = \Psi(\vx,t) \ket{\Psi}$.
}.
Unfortunately, there are conceptual and practical reasons that make it hard to
accept coherent states as a good representation of the actual state found in a
cold gas experiment.  Conceptually, a coherent state is a superposition of
states with different numbers of quanta, which has led to an ongoing debate
about whether such states can even be generated in principle\footnote{In
fact, the possibility of creating coherent states has been
called into doubt even in optics \cite{Molmer1997} where the photon number is 
not conserved.  We direct the reader to Ref.~\cite{Bartlett2006} for a presentation 
of both sides of the debate about the reality of coherent states, and a
possible resolution.
}
\cite{Bartlett2006}.  From a practical point of view, the number statistics of
a real system is likely to include additional classical noise which would be 
inconsistent with the Poissonian number statistics of a coherent state
\cite{Haine2009}.  These objections can be avoided if we treat the state as a
classical ensemble of coherent states, as discussed in more detail at the end
of the next section.

The electromagnetic field is the most obvious example of a classical field 
arising out of an underlying quantum theory, but the same kind of approximation
can be made for field theories describing ultracold gases of atoms:  While the
details of the equations of motion and the Hamiltonian differ, both theories
ultimately describe the same kinds of objects --- indistinguishable bosonic 
quanta having both wave-like and particle-like properties.  For an excellent
discussion of the nature of quanta, we direct the reader to Ref.~\cite{Teller1995}.

One might worry that we now have applied the label ``classical'' to two
mutually contradictory theories, both ostensibly describing massive particles:
On the one hand, we have classical Newtonian particle mechanics, and on the 
other a classical field theory.  We note that this is nothing other than a
manifestation of wave particle duality and the fact that different classical
limits are applicable to different situations \cite{Polkovnikov2010}.

To summarise, gases of cold bosons are not well described by the familiar
limit of classical \emph{particle} mechanics.  As a result, we are led to the
quantum theory of many particles in the form of a quantum field theory.  Being
difficult to solve, this theory is approximated --- but not by the classical
theory that we had to begin with.  Instead the appropriate limit is a
classical \emph{field} theory emphasising the collective wave-like aspects of 
the atomic ensemble.

\subsection{Derivation of the PGPE}
\label{PGPE_derivation_details}

Having said a few general things about the use of classical field theories to 
describe systems of massive particles, we turn to describing the PGPE that
will be used in chapter \ref{bkt_chapter} of this thesis.  We stated earlier
that classical fields are a good approximation when the number of particles is
large.  More precisely, it is the number of particles per mode $\exval{n_i}$
which is of importance because the errors are of order $1/\exval{n_i}$ as argued
in Ref.~\cite{Kagan1997}.  As a result, we want each mode to be highly occupied,
that is, $\exval{n_i} \gg 1$ for all $i$.  A system obeying this condition is
necessarily highly Bose degenerate (see the start of chapter
\ref{chap:introduction}).  We note that the number of particles per
mode is the same as the density of particles in phase space when the modes in
question are the momentum eigenstates --- that is, for a homogeneous system.
For this reason the terms Bose degeneracy and phase space density are often
used interchangeably.

In a spatially finite system we have an infinite number of field modes, so only
a finite number of those modes can be highly Bose degenerate as required for 
the classical field approximation.  To deal with this issue, the PGPE 
formalism splits the set of modes into two subsets, now conventionally labelled
the $\rC$ and $\rI$ regions \cite{Blakie2008}.  The $\rC$ or \emph{c-field}
region\footnote{The $\rC$ region has also been called the \emph{coherent}
region in earlier papers.} is chosen to contain the highly occupied modes, and
is simulated using a classical field.  The $\rI$ or \emph{incoherent} region
contains the remaining modes and is assumed to be thermalised.  In thermal 
equilibrium the occupation depends on the mode energy, so the split between
$\rC$ and $\rI$ regions is conveniently implemented using an appropriate energy
cutoff $\epsilon_\text{cut}$ to be defined more precisely later.  We emphasise
that the choice of splitting depends on the total number of atoms, so any given
splitting is specialised for a subset of the possible quantum states.

%


There are two well known methods for deriving the PGPE.  The first of these is 
a heuristic approach using what we would like to call ``dequantisation'' ---
the assertion that one may simply replace $\fo(\vx)$ by a classical field
$\cf(\vx)$ in the operator equations of motion.  This is the approach taken,
for example, in Ref.~\cite{Davis2001a} and will be discussed further below due to
its intuitive appeal.

The second approach uses the truncated Wigner function formalism to derive the 
PGPE as originally described in Ref.~\cite{Gardiner2003} and reviewed in depth
in Ref.~\cite{Blakie2008}.
The Wigner function method is attractive because it puts the classical field
approximation on firmer mathematical ground and allows for systematic treatment
of additional quantum behaviour as well as interactions with the $\rI$ region.  
Nevertheless, we will not need these additional features and there is a
significant cost in mathematical machinery, so we will not delve into the
details here.  We instead direct the reader to Ref.~\cite[\S 2]{Blakie2008} for
a clear and detailed explanation.

For a given partitioning of the modes into sets $\rC$ and $\rI$, we define a
pair of projection operators onto the subspaces spanned by the sets of mode
functions $\{\phi_i\colon i\in\rC\}$ and $\{\phi_i\colon i\in\rI\}$, respectively:
\begin{align}
    \PC\Bktcl{f}(\vx) &=
        \sum_{i\in\rC} \phi_i(\vx) \int d\vx'\; \phi^*_i(\vx') f(\vx'), \\
    \PI\Bktcl{f}(\vx) &=
        \sum_{i\in\rI} \phi_i(\vx) \int d\vx'\; \phi^*_i(\vx') f(\vx').
\end{align}
Note that the projectors act on \emph{spatial} functions $f(\vx)$, as
distinct from the creation and annihilation operators that act on the Fock
space.  The field $\fo$ may then be split according to
\begin{equation}\label{PGPE_fo_decomposition}
    \fo(\vx,t) = \foC(\vx,t) + \foI(\vx,t),
\end{equation}
where $\foC\equiv\PC\bktcl{\fo}$ and $\foI\equiv\PI\bktcl{\fo}$.  At this stage
no approximation has been made, and one can easily derive a coupled pair of
equations of motion for $\foC$ and $\foI$ that taken together are equivalent
to the full evolution.  Applying the projector $\PC$ to the field equation
\eqref{field_heisenberg_eoms} gives
\begin{equation}\label{foC_full_evolution}
    \begin{split}
    i\hbar \pderiv{\foC}{t} =
        \Hsp \foC +
        U_0 \PC\bktcl[\Big]{&\foCdag\foC\foC + \\
            &\foIdag\foC\foC + 2\foCdag\foI\foC + \\
            &\foCdag\foI\foI + 2\foIdag\foI\foC + \\
            &\foIdag\foI\foI
        }
    \end{split}
\end{equation}
where it is assumed that the basis $\phi_i$ is an eigenbasis of the 
single-particle Hamiltonian so that $\PC$ commutes with $\Hsp$ giving $\PC\{\Hsp \fo\} =
\Hsp\PC\{\fo\} = \Hsp\foC$.

We now proceed by the rather gross approximation of discarding the terms
coupling the $\rI$ region to the evolution of $\foC$ to obtain
\begin{equation}\label{projected_foC_evolution}
    i\hbar \pderiv{\foC}{t} =
        \Hsp \foC + U_0 \PC\bktcl[\Big]{\foCdag\foC\foC}.
\end{equation}
This has the great advantage of providing a closed system which makes numerical
work much simpler, but is hard to justify in general (see \cite{Davis2001b} for
further comments).  Nevertheless, numerical experience shows that the isolated
$\rC$ region as described by the PGPE evolves to thermal equilibrium
\cite{Davis2001a}, and a range of useful results have been obtained (see
\cite[\S 3.3--4]{Blakie2008} for several examples).  We note that the step
of discarding all interactions with the $\rI$ region is a defining feature of
the PGPE regardless of which derivation is used.  Retaining these terms is
necessary in many cases, in particular for realistic simulations of condensate
formation.  In such cases it is necessary to use a more powerful alternative, a
role filled by the so-called stochastic PGPE (SPGPE) \cite{Gardiner2003}.

The final step in the derivation is to make the classical field approximation, 
that is, to take the classical limit of the field.  On a purely formal level
this is achieved by simply making the replacement $\foC(\vx,t) \to \cf(\vx,t)$
in the equations of motion\footnote{Equivalently, this is a replacement of each
mode operator $\ao_i$ with a complex amplitude $c_i$ for all $i\in\rC$.}, where
$\cf$ is a complex-valued classical field.  We then obtain the PGPE,
\begin{equation}\label{PGPE}
    i\hbar \pderiv{\cf}{t} = \Hsp \cf + U_0 \PC\bktcl[\Big]{\cf^*\cf\cf},
\end{equation}
which describes classical evolution of the highly occupied field modes.  While
this derivation is expedient and intuitive, it leaves much to be desired from a
mathematical standpoint.

The most basic physical justification for the replacement $\foC\to\cf$ seems
to be to regard it as a ``dequantisation'' or the opposite of canonical
quantisation\footnote{Canonical quantisation has a somewhat unfortunate name: It
is not the canonical method of quantisation, but a method of quantisation
making use of classical canonical coordinates.}.  The recipe of canonical
quantisation produces a quantum theory from a classical one by replacing the
classical canonical position and momentum variables with operators that satisfy
the usual commutation relations (see, for example, \cite[\S 1.4]{Ficek2004}).
We imagine dequantisation as the opposite process --- forming a classical limit
by replacing the quantum field with a classical one.

One conceptual problem with this story is that the nature of the quantum and
classical fields are completely different.  As discussed in
Ref.~\cite[Ch.~5]{Teller1995}, a classical field theory uses field equations to
describe the
state of the system but the quantum field is not a state-like object.  Instead
the quantum field is more akin to a Green's function: a solution $\fo(\vx,t)$
encodes all possible evolutions independently of any particular quantum state.
In light of this, it is confusing to imagine replacing or somehow approximating
the field operator by a classical field.  Clearly some further mathematical
justification is required.

The classic mathematical procedure for taking the classical limit is to argue
that the class of ``classical-like'' quantum states of interest are well
approximated by coherent states \cite{Glauber1963a}.  From this viewpoint the
replacement $\foC\to\cf$ is implemented by assuming the state is coherent and
taking the expectation value of both sides of Eq.~\eqref{projected_foC_evolution}.
For example, on the left hand side we have
\begin{equation}
    \braoket[\bigg]{\Psi}{i\hbar \pderiv{\foC(\vx,t)}{t}}{\Psi} =
    i\hbar \pderiv{\braoket{\Psi}{\foC(\vx,t)}{\Psi}}{t} =
    i\hbar \pderiv{\cf(\vx,t)\braket{\Psi}{\Psi}}{t} =
    i\hbar \pderiv{\cf(\vx,t)}{t},
\end{equation}
where $\foC(\vx,t)\ket{\Psi} = \cf(\vx,t)\ket{\Psi}$ because $\ket{\Psi}$ is a
coherent state\footnote{The standard notation $\foC$ and $\cf$ is somewhat
unfortunate here, since it suggests that the state-like object $\cf$ arises as
an approximation to the quantum field $\foC$.  This is not the case: it is more
accurate to view $\cf$ as a representation of the state $\ket{\Psi}$.  A less
confusing notation might be to write $\Psi(\vx,t)$ in place of $\cf(\vx,t)$ so
that $\foC\ket{\Psi} = \Psi(\vx,t)\ket{\Psi}$.}.  This approach is known as the 
\emph{broken symmetry} or \emph{mean-field} approach since the average
of the quantum field $\exval{\foC} = \cf$ is nonzero for a coherent state.



The use of coherent states is attractively simple and provides a
straightforward way to take the classical limit.  On the other hand, the
coherent states are not obviously adequate for cold atom experiments, for the 
reasons noted in Sec.~\ref{PGPE_conceptual_intro}.  Luckily, we can avoid those
particular objections if we generalise the state to a statistical mixture of 
coherent states as in Ref.~\cite{Kagan1997}.  The density matrix is then
expressed as
\begin{equation}\label{P_representation}
    \hat{\rho} = \int \mathcal{D}\alpha\; P(\alpha)\, \ketbra{\alpha}{\alpha},
\end{equation}
where $P(\alpha)$ is a distribution\footnote{
As written here, $P$ is actually the density in the Glauber-Sudarshan
$P$ representation \cite{Glauber1963a}, and as a result is much more general than 
one would guess at first sight.  Not all density matrices have a well behaved
$P$ representation (in general $P$ can be negative and extremely singular
\cite{Bonifacio1966}), but this is not a problem for the states of interest
since --- roughly speaking --- the more classical a state is, the better
behaved is the associated $P$ distribution.  We note that the Wigner
function mentioned above is equal to the $P$ distribution smoothed by a
Gaussian convolution.
}
over the set of classical fields $\alpha$.  With this representation at hand,
computing the dynamics involves evolving each coherent state independently
according to the classical field approximation \cite{Kagan1997}.  Observables 
become ensemble averages at time $t$ over the ensemble described by $P$.

The procedure outlined above is not a particularly satisfying
\emph{mathematical} account of the PGPE theory, but it provides a
flavour for the kinds of physical arguments that have been used in the field.  
With that in mind, we again direct the reader to Ref.~\cite{Blakie2008} for a 
description of the Wigner function formalism that puts the method on more 
solid theoretical ground: It shows us more clearly which ensemble we should be
using, which terms must be neglected in forming the equations of motion, and 
directions for extending the theory to deal with physical situations where
the PGPE is not adequate.

\subsection{Ergodicity and thermal averages}
\label{computing_PGPE_observables}

We are often interested in the properties of the system at a particular
temperature --- that is, the expectation values of observations of the
\emph{thermal ensemble}.  To
compute such thermal expectation values it is sufficient to sample any one of
the standard statistical ensembles at the desired temperature\footnote{The various 
statistical ensembles are equivalent in the thermodynamic limit.  It is worth
keeping in mind that they have different fluctuation properties for the
mesoscopic numbers of particles we are dealing with here.}.  The microcanonical
ensemble is simplest to deal with in this case --- microcanonical averages
should be taken over the hypersurface of constant energy, and the PGPE is energy
conserving.  If the classical energy is given by
\begin{equation}\label{classical_Hamiltonian_Hc}
    H_\rC[\cf] = \int d\vx\; \bkt[\Big]{\cf^*\Hsp\cf + \frac{U_0}{2} \abs{\cf}^4},
\end{equation}
then microcanonical averages are taken over all $\cf$ with respect to the phase
space density \cite{Wright2011}
\begin{equation} \label{ergodic_phase_space_density}
    P[\cf;E] =
    \begin{cases}
        \text{const.} \quad  &\text{where } H_\rC[\cf] = E \text{ and other 
        macroscopic constraints are satisfied} \\
        0                    &\text{elsewhere}.
    \end{cases}
\end{equation}
We mention other macroscopic constraints because there may be additional
constants of motion and we want any averages to take these into account. For
example, the momentum is conserved in a homogeneous system in which case the
ensemble of interest includes only microstates with the system at rest.

The crucial step in computing thermal averages is to assume that the PGPE is
\emph{ergodic} \cite{Davis2001a} which allows us to convert phase space
averages into time averages:
\begin{align}\label{ergodic_average_limit}
    \exval{F} &= \int \mathcal{D}\cf\; P[\cf;E] F[\cf] \\
              &= \lim_{\tau\to\infty} \frac{1}{\tau} \int_0^\tau dt\; F[\cf(t)]
\end{align}
where $F$ is some functional of the classical field representing an observable.
In practice the time average is implemented numerically by sampling the motion
at discrete time intervals $\{t_j\}$ and forming the sum
\begin{equation}
    \exval{F} \approx \frac{1}{M} \sum_{j=1}^{M} F[\cf(t_j)]
\end{equation}
for large $M$.  This recipe provides a convenient and efficient way to sample
states from the microcanonical ensemble.


Not all quantities may be easily written as functionals of the field $\cf$.
In particular we note that derivatives of entropy such as the temperature ($T$)
and chemical potential ($\mu_{\rC}$) are calculated by time-averaging
appropriate quantities constructed from the Hamiltonian
in Eq.~\eqref{classical_Hamiltonian_Hc} using the Rugh approach \cite{Rugh1997a}.  The
detailed implementation of the Rugh formalism for the PGPE is rather technical
and we refer the reader to Refs.~\cite{Davis2003,Davis2005} for additional
details of this procedure.

It is instructive to connect the discussion of ergodic averaging to the Wigner
function version of the PGPE derivation.  For a thermal state the initial Wigner
function is very delocalised, in contrast to the near delta function required 
for a straightforward single-trajectory interpretation.  Sampling directly from
such a distribution is difficult except in the limit of very low or high
temperatures where the Hamiltonian can be approximately diagonalised
\cite{Blakie2008}.  Happily, the PGPE with ergodic averaging avoids this 
problem by relying on the \emph{dynamics} to sample the distribution correctly.

\subsection{Treatment of the $\rI$ region}
\label{I_region_exvals}

At temperatures near the BEC transition the number of atoms in the $\rI$ region
is significant, as observed in chapter \ref{bkt_chapter} (see also 
\cite{Blakie2007} for three-dimensional systems).  For realistic comparisons 
with experiment these need to be taken into account.

A simple way to deal with the $\rI$ region atoms is to assume a gas of
uncorrelated semiclassical bosons in thermal equilibrium, interacting only 
indirectly via the average particle density \cite{Bezett2008,Davis2006}.  Under 
these assumptions the $\rI$ region can be described by an approximate single-particle 
Wigner function $W_\rI$.  (Note that this is a \emph{single}-particle 
Wigner function, distinct from the multi-particle Wigner function
referred to previously in section \ref{PGPE_derivation_details}.)  The
appropriate single-particle Wigner function is positive and can be interpreted
as a classical probability distribution over phase space:
\begin{equation}\label{I_region_wigner_fxn}
    W_\rI(\v{k},\v{x})
        = \frac{1}{(2\pi)^D} \frac{1}{e^{[E_\text{HF}(\v{k},\vx) - \mu]/k_B T} - 1}.
\end{equation}
In this equation $D$ is the number of dimensions, $\v{k}$ is the wavevector and
$\mu = \mu_\rC + 2U_0 n_\rI$ is the chemical potential.  The Hartree-Fock
energy in this expression is given by 
\begin{equation}
    E_\text{HF}(\v{k},\vx) =
        \frac{\hbar^2\v{k}^2}{2m} + V(\vx) + 2 U_0 \Bktsq{n_\rC(\vx) + n_\rI(\vx)},
\end{equation}
where $n_\rC$ and $n_\rI$ are the densities of the $\rC$ and $\rI$ region atoms
\cite{Bezett2008}; the temperature and chemical potential are calculated from
the $\rC$ region.  When the potential $V$ is not constant the unknown density
$n_\rI$ is spatially varying and must be calculated self-consistently
\cite{Davis2006} using
\begin{equation}
    n_\rI(\vx) = \int_{E_\text{HF}(\v{k},\vx) > \epsilon_\text{cut}} d\v{k}\; W_\rI(\v{k},\vx).
\end{equation}
This complication disappears in the homogeneous case relevant to the work in 
chapter \ref{bkt_chapter}.

With the semiclassical Wigner function at hand, any observable $F(\v{k},\vx)$
may be calculated using a phase space average
\begin{equation}
    \exval{F}_\rI = \int_{E_\text{HF}(\v{k},\vx) > \epsilon_\text{cut}}
        d\vx\; d\v{k}\; W_\rI(\v{k},\vx) F(\v{k},\vx).
\end{equation}
Note that the nontrivial region of integration satisfying
$E_\text{HF}(\v{k},\vx) > \epsilon_\text{cut}$
is chosen to avoid counting atoms that have already been taken into account 
via the $\rC$ region simulation.

\subsection{Comparison with the GPE}
\label{PGPE_vs_GPE}


We briefly contrast the PGPE formalism with the GPE, as the relationship 
between these is a recurring source of confusion: Given such apparently similar
equations, why is it claimed that the PGPE describes all modes at temperatures
up to the order of the transition temperature, while the GPE describes only the
condensate mode at zero temperature?  To further emphasise the point of
similarity, recall that there are various ways to derive the GPE with subtly
different interpretations as to the exact state described.  In
Sec.~\eqref{GPE_derivation} we presented the GPE as arising from the product
state ansatz of Eq.~\eqref{TDHF_ansatz}, but it can just as easily be derived
by assuming a time dependent coherent state\footnote{
The sense in which the GPE approximates the true state of the system is 
addressed in Sec.~\ref{GPE_validity}.  We point out here that it is consistent
for both the coherent \emph{and} product state ansätze to approximate the true
many particle state in the relatively weak sense of
Eq.~\eqref{GPE_weak_convergence_eqn}.

Furthermore, distinguishing between these apparently very different states in
an experiment can be surprisingly difficult.  This is nicely demonstrated in
Ref.~\cite{Yoo1997} that analyses an interference experiment,
finding that the difference in interference fringe contrast is of order $1/N$,
and very difficult to measure for even moderate numbers of particles $N$.
This is reminiscent of the approximation $N/(1+N) \approx 1$ that must be made
in deriving the GPE via the product state ansatz.
}.
This second approach is confusingly similar to the ``dequantisation'' step in
our derivation of the PGPE when interpreted in terms of coherent states.

The real heart of the difference between these formalisms is the ensemble
average that must be taken when using the PGPE to describe finite temperature
states of the field.  While this is implemented as an ergodic average, 
certain non-thermal states could also in principle be simulated by considering
a full set of trajectories drawn from an initial non-thermal distribution.
Averages would then be taken over this set of trajectories at some
time\footnote{Indeed, this is the approach generally taken by the whole group
of powerful methods arising from the truncated Wigner function formalism.}.
At finite temperature there must be some portion of non-condensed atoms, and 
indeed the PGPE is able to simulate these: The ensemble average results in a
single particle density matrix with a largest eigenvalue significantly less
than one; this is a non-pure condensate in the Penrose-Onsager sense.

In contrast, the GPE simulates a single trajectory representing the motion of
the condensate: The single particle density matrix becomes arbitrarily close to
pure in the GP limit, as made precise in Sec.~\ref{GPE_validity}.  The types of
states that the GPE simulates are thus small perturbations of the zero
temperature stationary solution\footnote{Non-stationary solutions are not,
strictly speaking, at zero temperature since simple definitions of thermal
equilibrium imply a stationary state.  A dynamical definition of temperature
such as given in Ref.~\cite{Rugh1997a} would presumably assign a small but
nonzero temperature to such states.}.

While the projection operator is required to provide control over the set of
modes to be approximated, the arguments above show that it is not at the heart
of the conceptual differences between the GPE and PGPE.  Nevertheless, we
emphasise that a projector is essential to clearly define the set of modes in
the $\rI$ region.  Without careful treatment of these high energy modes, one
can only hope to achieve qualitative agreement with experimental observations.

%% file: bkt/bkt.tex
\chapter{Vortex pairing in two-dimensional Bose gases}
\label{bkt_chapter}

\begin{chap_desc}
    In this chapter we investigate finite temperature 2D Bose gases using the
    PGPE, with a view to understanding the relation between BEC and BKT
    physics in finite-sized systems.  We calculate several physical properties
    including the amount of vortex pairing, the condensate and superfluid
    fractions, and the functional form of the spatial correlations.  We also
    relate our simulation to the experimental measurements described in
    Ref.~\cite{Hadzibabic2006}.
\end{chap_desc}

\section{Introduction}

As discussed in section \ref{lowdim_intro}, a two-dimensional homogeneous Bose 
gas does not undergo the BEC transition.  Nevertheless, 2D Bose gases do
display superfluid behaviour in the presence of interactions, due to a vortex
pairing phase transition known as the Berezinskii-Kosterlitz-Thouless (BKT)
transition.  Evidence for the BKT transition has been found in experimental
realisations of the 2D Bose gas in several studies
\cite{Stock2005,Hadzibabic2006,Kruger2007,Schweikhard2007a,Clade2008}.  We make 
particular note of the experiment described in Ref.~\cite{Hadzibabic2006}, which
was carried out at ENS in Paris, and which we will refer to when choosing
parameters for our study.

Experiments in the 2D regime present a new challenge for theory as  strong
fluctuations invalidate mean-field theories (see, for example,
\cite{Prokofev2001,Prokofev2002,Posazhennikova2006,Bisset2009d,Mathey2009a,Sato2009a,Gies2004a,Schumayer2007a}),
and only recently have quantum Monte Carlo \cite{Holzmann2008,Holzmann2009b}
and classical field (c-field)  \cite{Simula2006,Simula2008,Bisset2009} methods
been developed that are directly applicable to the experimental regime. 

In the current chapter we study a uniform Bose gas of finite spatial extent and
parameters corresponding to current experiments. To analyse this system we use
the PGPE, which is well suited to studying finite temperature Bose fields with
many highly occupied modes.
We examine two important applications: First, we
provide a quantitative validation of the interference technique used in the ENS
experiment to determine the nature of two-point correlation in the system. To
do this we simulate the interference pattern generated by allowing two
independent 2D systems to expand and interfere.  Applying the experimental
fitting procedure to analyse the interference pattern, we can extract the
inferred two-point correlations which we then compare against the
\textit{in situ} correlations that we calculate directly.  Second, we examine
the correlations between vortices and antivortices in the system to directly
quantify the emergence of vortex-antivortex pairing in the low temperature
phase. A similar study was made by Giorgetti \textit{et al.} using a
semiclassical field technique \cite{Giorgetti2007}. We find results for vortex
number and vortex pair distributions consistent with their results, and we show
how a coarse-graining procedure can be used to reveal the unpaired vortices in
the system.

\section{Formalism}\label{Sform}

Here we consider a dilute 2D Bose gas described by the Hamiltonian
\begin{equation}
\hat{H}=\int d^{2}\mathbf{x}\,\hat{\psi}^{\dagger}\left\{-\frac{\hbar^2\nabla^2_{\mathbf{x}}}{2m}\right\}\hat{\psi}+\frac{\hbar^2g}{2m}\int d^{2}\mathbf{x}\,\hat{\psi}^\dagger\hat{\psi}^{\dagger}\hat{\psi}\hat{\psi},
\end{equation}
where $\mathbf{x}=(x,y)$.  This is simply Eq.~\eqref{swave_bose_gas_Hamiltonian},
but specialised to the two-dimensional homogeneous case.
We take the two-dimensional geometry to be realised by tight confinement in the
$z$ direction that restricts atomic occupation to the lowest  $z$ mode.  The
dimensionless 2D coupling constant is
\begin{equation}
{g}=\frac{\sqrt{8\pi}a}{a_{z}},
\end{equation}
with $a_{z}$ the spatial extent of the $z$ mode\footnote{For example, for
tight harmonic confinement of frequency $\omega_z$ we have
$a_z=\sqrt{\hbar/m\omega_z}$.} and $a$ the s-wave scattering length. We will
assume that $a_{z}\gg a$ so that the scattering is approximately
three-dimensional \cite{Petrov2000}, a condition well-satisfied in the ENS and
NIST experiments \cite{Stock2005,Hadzibabic2006,Kruger2007,Clade2008}.  For
reference, the ENS experiment reported in Ref.~\cite{Hadzibabic2006} had
${g}\approx0.15$,  whereas in the NIST experiments ${g}\approx0.02$
\cite{Clade2008}.

In contrast to experiments we focus here on the uniform case; no trapping 
potential in the $xy$-plane is considered.  We perform finite-sized calculations
corresponding to a square system of size $L$ with periodic boundary conditions.
Working in the finite-size regime simplifies the simulations and is more 
representative of current experiments.  We note that the thermodynamic limit
corresponds to taking $L\to\infty$ while keeping the density,
$n=\langle\hat{\psi}^\dagger\hat{\psi}\rangle$, constant.

\subsection{Review of BKT physics}
\label{bkt_review_section}
The BKT superfluid phase has several distinctive characteristics, which we
briefly review.
\subsubsection{First-order correlations}
Below the BKT transition the first-order correlations decay according to an
inverse power law:
\begin{equation} \label{eqn:algebraic_decay}
g^{(1)}(\vec{x},\vec{x}') \propto \norm{\vec{x}-\vec{x}'}^{-\alpha}.
\end{equation}
Systems displaying such \emph{algebraic decay} are said to exhibit 
\emph{quasi-long-range order} \cite{Chaikin1995}.  This is in contrast
to both the high temperature (disordered phase) in which the correlations decay
exponentially, and long-range ordered case of the 3D Bose gas in which
$g^{(1)}\to\rm{const.}$ for $\norm{\vec{x}-\vec{x}'}\to\infty$.


\subsubsection{Superfluid density}
Nelson and Kosterlitz \cite{Nelson1977} found that the exponent of the
algebraic decay is related to the ratio of the superfluid density and
temperature.  To within logarithmic corrections
\begin{equation} \label{eqn:alpha_rhos_T}
	\alpha(T) = \frac{1}{\lambda_\text{dB}^2\rho_s(T)},
\end{equation}
where $\rho_s$ is the superfluid density and $\lambda_\text{dB}$ is 
the thermal de Broglie wavelength (Eq.~\eqref{lambda_dB}).  Furthermore,
Nelson and 
Kosterlitz showed that this ratio converges to a universal constant as
the transition temperature, $\Tkt$, is approached from below: $\lim_{T\to \Tkt^-}
\alpha(T) = 1/4$ (i.e.,~$\rho_s\lambda_\text{dB}^2=4$).  Thus, the superfluid fraction
undergoes a universal jump from  $\rho_s(\Tkt^+)=0$ to
$\rho_s(\Tkt^-)=4/\lambda_\text{dB}^2$ as the temperature decreases through $\Tkt$.

\subsubsection{Vortex binding transition}
Another important indicator of the BKT transition is the behaviour of
topological excitations, which are quantised vortices and antivortices in the
case of a Bose gas.  A single vortex has energy that scales with the logarithm
of the system size.  At low temperatures this means that the free energy for a
single vortex is infinite (in the thermodynamic limit), and vortices cannot
exist in isolation.  As originally argued in Ref.~\cite{Kosterlitz1973}, the
entropic contribution to the free energy also scales logarithmically with the
system size, and will dominate the free energy at high temperatures allowing
unbound vortices to proliferate.  This argument provides a simple estimate for
the BKT transition temperature.

Although unbound vortices are thermodynamically unfavoured at $T<\Tkt$, bound
pairs of counter-rotating vortices may exist because the total energy of such a
pair is finite\footnote{The vortex-antivortex pair energy depends on the pair
size rather than the system size.}.  This leads to a distinctive qualitative
characterisation of the BKT transition: as the temperature increases through
$\Tkt$ pairs of vortices unbind.

%
%



\subsubsection{Location of the BKT transition in the dilute Bose gas}
While the relation $\rho_s(\Tkt^-)=4/\lambda_\text{dB}^2$ between the superfluid density
and temperature at the transition is universal, the total density, $n$, at the
transition is not.  General arguments \cite{Popov,Kagan1987,Fisher1988} suggest
that the transition point for the dilute uniform 2D Bose gas is given by
\begin{equation}
(n\lambda_\text{dB}^2)_\text{KT} = \ln\left(\frac{\xi}{{g}}\right), \label{critpsd}
\end{equation}
where  $\xi$ is a constant.  Prokofév, Ruebenacker and Svistunov
\cite{Prokofev2001, Prokofev2002} studied the homogeneous Bose gas using Monte
Carlo simulations of an equivalent classical $\phi^4$ model on a lattice.
Using an extrapolation to the infinite-sized system, they computed a value for
the dimensionless constant, $\xi = 380 \pm 3$.  By inverting
Eq.~(\ref{critpsd}), we obtain the  BKT critical temperature for the infinite
system 
\begin{equation}
	\Tkt^\infty = \frac{2\pi \hbar^2 n}{m k_B \ln \Bkt{\xi \hbar^2 / m {g}}}.
\end{equation} 
We use the superscript $\infty$ to indicate that this result holds in the thermodynamic limit.

\section{Method}
\label{sec:method}

\subsection{c-field and incoherent regions} \label{secPGPEformalism}

We briefly outline some specifics regarding how the PGPE formalism described in
section \ref{PGPE_derivation} is applied to the two-dimensional homogeneous
problem.  In the homogeneous case, 
the fields $\cf$ and $\foI$ are defined as the low and
high energy projections of the full quantum field operator, separated by the
cutoff wave vector $K$. In our theory this cutoff is implemented in terms of
the plane wave eigenstates $\{\varphi_\v{n}(\mathbf{x})\}$ of the
time-independent single-particle Hamiltonian, that is,
\begin{align}
\varphi_\v{n}(\mathbf{x}) &= \frac{1}{{L}}e^{-i\mathbf{k}_\v{n}\cdot\mathbf{x}}, \\
\mathbf{k}_\v{n} &= \frac{\pi}{L}\v{n},
\end{align}
with $\v{n} = (n_x, n_y) \in\mathbb{Z}^2$.
The fields are thus defined by
\begin{align}
\cf(\mathbf{x}) &\equiv \sum_{\v{n}\in\rC}c_\v{n}\varphi_\v{n}(\mathbf{x}),\label{eqn:Cfield}\\
\bfI(\mathbf{x}) &\equiv \sum_{\v{n}\in\rI}\hat{a}_\v{n}\varphi_\v{n}(\mathbf{x}),
\end{align}
where the $\hat{a}_\v{n}$ are Bose annihilation operators, the $c_\v{n}$ are
complex amplitudes, and the sets of quantum numbers defining the regions are 
\begin{align}
\rC &= \{\v{n}\colon\norm{\mathbf{k}_\v{n}}\le K\},\\
\rI &= \{\v{n}\colon\norm{\mathbf{k}_\v{n}}> K\}. 
\end{align}

\subsubsection{Choice of $\rC$ region}\label{cregionchoice}
In general, the applicability of the PGPE approach to describing the finite
temperature gas relies on an appropriate choice for $K$, so that the modes at
the cutoff have an average occupation of order unity. In this work we choose an
average of five or more atoms per mode using a procedure discussed in appendix
\ref{sec:sim_details}.  This choice means that all the modes in $\rC$ are
appreciably occupied, justifying the classical field replacement
$\hat{a}_\v{n}\to c_\v{n}$. In contrast the $\rI$ region contains many sparsely
occupied modes that are particle-like and would be poorly described using a
classical field approximation.  Because our 2D system is critical over a wide
temperature range, additional care is needed in choosing $\rC$. Typically
strong fluctuations occur in the infrared modes up to the energy scale
$\hbar^2gn/m$. Above this energy scale the modes are well described by
mean-field theory (see, for example, the discussion in
\cite{Kashurnikov2001a,Prokofev2001}).  For the results we present here, we
have
\begin{equation}
\frac{\hbar^2K^2}{2m} \gtrsim \frac{\hbar^2g}{m}n\label{validitycond}
\end{equation}
for simulations around the transition region and at high temperature.  At
temperatures well below $\Tkt$, the requirement of large modal occupation near
the cutoff competes with this condition and we favour the former at the expense
of violating Eq.~\eqref{validitycond}.

\subsubsection{PGPE treatment of $\rC$ region}\label{SecformalismPGPE}

Specialising the PGPE (Eq.~\eqref{PGPE}) to 2D we have the equation of motion for
$\cf$
\begin{equation} \label{PGPE_2D}
i\hbar\frac{\partial \cf }{\partial t} = -\frac{\hbar^2\nabla^2_{\mathbf{x}}}{2m}\cf + \frac{\hbar^2g}{m} \PC\left\{ \abs{\cf}^2\cf\right\},
\end{equation}
where the projection operator 
\begin{equation}
\PC\{ F(\mathbf{x})\}\equiv\sum_{\v{n}\in\rC}\varphi_{\v{n}}(\mathbf{x})\int
d^2\mathbf{x}'\,\varphi_{\v{n}}^{*}(\mathbf{x}') F(\mathbf{x}'),\label{eq:projectorC}\\
\end{equation}
formalises our basis set restriction of $\cf$ to the $\rC$ region. The main
approximation used to arrive at the PGPE is to neglect dynamical couplings to
the incoherent region \cite{Davis2001b}.

We assume that the evolution under Eq.~(\ref{PGPE_2D}) is ergodic \cite{Davis2001a}, so that the 
microstates \{$\cf$\} generated through time evolution form an unbiased sample
of the equilibrium microstates.  Time-averaging can then be used to obtain
macroscopic equilibrium properties.  We generate the time evolution by solving
the PGPE with three adjustable parameters: (i) the  cutoff wave vector, $K$,
that defines the division between $\rC$ and $\rI$, and hence the number of
modes in the $\rC$ region; (ii) the number of $\rC$ region atoms, $N_{\rC}$;
(iii) the total energy of the $\rC$ region, $E_\rC$. The last two quantities,
defined as 
\begin{align}
E_{\rC} &= \int d^2\mathbf{x}\,\cf^*\left(-\frac{\hbar^2\nabla^2_{\mathbf{x}}}{2m} +  \frac{\hbar^2g}{2m} \abs{\cf}^2\right)\cf,\label{Ec}\\
N_{\rC} &= \int d^2\mathbf{x}\,\abs{\cf(\mathbf{x})}^2,
\end{align}
are important because they represent constants of motion of the PGPE
(Eq.~\eqref{PGPE_2D}), and thus control the thermodynamic equilibrium state of the
system.

\subsubsection{Obtaining equilibrium properties for the $\rC$ region}\label{PGPEeqprops}
To characterise the equilibrium state in the $\rC$ region it is necessary to
determine the average density, temperature and chemical potential, which in 
turn allow us to characterise the $\rI$ region (see section~\ref{sec:above_cutoff}).  
These and other $\rC$ region quantities can be computed by time-averaging as 
described in section \ref{computing_PGPE_observables}.  For example,
the average $\rC$ region density is given by
\begin{equation}
n_{\rC}(\mathbf{x})  \approx \frac{1}{M_s}\sum_{j=1}^{M_s}\Abs{\cf(\mathbf{x},t_j)}^2,\label{nc2}
\end{equation}
where $\{t_j\}$ is a set of $M_s$ times (after the system has been allowed to
relax to equilibrium) at which the field is sampled. We typically use 2000
samples from our simulation to perform such averages over a time of $\sim 16$ s.
Another quantity of interest here is the first-order correlation function,
which we calculate directly via the expression
\begin{equation}
G^{(1)}_{\rC}(\v{x},\v{x}')  \approx \frac{1}{M_s}\sum_{j=1}^{M_s}\cf^*(\mathbf{x},t_j)\cf(\mathbf{x}',t_j).\label{Gc1}
\end{equation}
The temperature ($T$) and chemical potential ($\mu_{\rC}$) are 
computed using the Rugh approach \cite{Rugh1997a} which was briefly touched 
upon in section \ref{computing_PGPE_observables}.

A major extension to the formalism of the PGPE made in this thesis is the
development of a method for extracting the superfluid fraction, $\rho_s$, from
these calculations.  For this we use linear response theory to relate the
superfluid fraction to the long wavelength limit of the second order momentum
density correlations.  An extensive discussion of this approach, and the
numerical methods used to implement it, are presented in chapter
\ref{sfrac_chapter}.

\subsection{Mean-field treatment of $\rI$ region} \label{sec:above_cutoff}
Occupation of the $\rI$ region modes, $N_\rI$, accounts for about 25\% of the total
number of atoms at temperatures near the phase transition.  We assume a time-independent 
state for the $\rI$ region atoms defined by a Wigner function
\cite{Naraschewski1999}, allowing us to calculate quantities of interest by
integrating over the above-cutoff momenta, $k > K$
\cite{Bezett2008, Davis2006}.

Our assumed Wigner function corresponds to the self-consistent Hartree-Fock
theory as applied in Ref.~\cite{Davis2006}.  Specialising Eq.~\eqref{I_region_wigner_fxn} to
two dimensions, this is
\begin{equation}\label{eqn:wignerFxn}
W_\rI(\v{k},\v{x})
  = \frac{1}{(2\pi)^2} \frac{1}{e^{(E_\text{HF}(\v{k}) - \mu)/k_B T} - 1 },
\end{equation}
where 
\begin{equation}
E_\text{HF}(\v{k}) =\frac{\hbar^2\v{k}^2}{2m} + \frac{2\hbar^2g}{m}(n_\rC + n_\rI),\label{EHF}
\end{equation}
is the Hartree-Fock energy,  $n_\rI$ is the $\rI$ region density, and $\mu = \mu_\rC + 2\hbar^2gn_\rI/m$ is the chemical potential (shifted by the mean-field interaction with the $\rI$ region atoms).  Note that the average densities are constant in the uniform system, so $W_\rI(\v{k},\v{x})$ has no explicit $\v{x}$ dependence, however, we include this variable for generality when defining the associated correlation function. 

The $\rI$ region density appearing  in Eq.~\eqref{EHF} is given by
\begin{equation}
n_\rI= \int_{\norm{\v{k}} \ge  K} d^2\v{k}\, W_\rI(\v{k},\v{x}),
\end{equation}
with corresponding atom number $N_\rI=n_\rI L^2$; total number is simply
\begin{equation}
N=N_\rC+N_\rI.
\end{equation}
An analytic expression for $n_\rI$ and simplified procedure for numerically calculating the first-order correlation function of the $\rI$ region atoms, $G^{(1)}_\rI$, can be obtained  by taking integrals over the phase space. These results are discussed in appendix \ref{sec:above_cutoff_integrals}.

\subsection{Equilibrium configurations with fixed $T$ and $N$} \label{sec:initial_conditions}

Generating equilibrium classical fields with given values of $E_\rC$ and $N_\rC$
is straightforward since the PGPE simulates a microcanonical system
(see appendix \ref{sec:initial_given_Ec_Nc}).  However, we wish to
simulate systems with a given temperature and total number.  As described in
the preceding two sections these can only be determined after a simulation has
been performed.  In appendix \ref{sec:sim_details} we outline a procedure for
estimating values of $E_\rC$ and $N_\rC$ for desired values of $N$ and $T$
based on a root finding scheme using a Hartree-Fock-Bogoliubov analysis for the
initial guess.

\section{Results}\label{sec:res}

We choose simulation parameters in analogy with the Paris experiment of
Hadzibabic \textit{et al.}~\cite{Hadzibabic2006}.  This experiment used an
elongated atomic cloud of approximately $10^5$ $^{87}$Rb atoms, with a spatial extent (Thomas-Fermi lengths) of 120 $\mu$m and 10 $\mu$m along the two loosely
trapped $x$ and $y$ directions. The tight confinement in the $z$ direction was provided by an optical lattice.

Although our simulation is for a uniform system, we have chosen 
similar parameters where possible.  Our primary simulations are for a system in a square box with $L=100$ $\mu$m, with  $4{\times}10^5$ $^{87}$Rb atoms. We also present results for systems with $L=50$ $\mu$m and $L=200$ $\mu$m at the same density in order
to better understand finite-size effects. All simulations are for the case of
$g=0.15$ corresponding to the experimental parameters reported in Ref.~\cite{Hadzibabic2006}.

The cutoff wave vector $K$ varied with temperature to ensure appropriate
occupation of the highest modes (see section~\ref{cregionchoice}). For the 100
$\mu$m system, the number of $\rC$ region modes ranged between 559 at low
temperatures to 11338 at the highest temperature studied.

\subsection{Simulation of expanded interference patterns between two systems}\label{siminterference}
In order to make a direct comparison with the experimental results of
Ref.~\cite{Hadzibabic2006}, we have generated synthetic interference patterns and 
implemented the experimental analysis technique.  Our simulated imaging
geometry is identical to that found in Ref.~\cite{Hadzibabic2006}, with expansion
occurring in the $z$-direction. The interference pattern is formed in the
$xz$-plane via integration of the density along the $y$-direction
(``absorption imaging'').

 Our algorithm for obtaining the interference pattern due to our classical field is 
very similar to that presented in Ref.~\cite{Hadzibabic2004}.  Our above-cutoff
thermal cloud is taken into account separately.
We consider a pair of fields $\psi_\rC^{(1)}(x,y), \psi_\rC^{(2)}(x,y)$ from different times
during the simulation, chosen such that the fields can be considered
independent.  The 3D wavefunction corresponding to each field is
reconstructed by assuming a harmonic oscillator ground state in the
tight-trapping direction.  These two reconstructed fields are spatially
separated by $\Delta=3$ $\mu$m, corresponding to the period of the optical lattice in
Ref.~\cite{Hadzibabic2004}.

Given this initial state, we neglect atomic interactions and only account for expansion
in the tightly-trapped direction.  This yields a simple analytical result for 
the full classical field $\psi_\rC(x,y,z,\tau)$ at later times.  The contribution of the
above-cutoff atoms is included by an incoherent addition of intensities.
The result is integrated along the $y$-direction to simulate the effect of
absorption imaging with a laser beam, that is,
\begin{align}
n_{\rm{im}}(x,z)
	&=\int_0^{L'} dy\,\bktsq[\Big]{\abs[\big]{\psi_\rC^{(T)}(x,y,z,\tau)}^2+n_\rI(x,y,z,\tau)}, \\
\psi_\rC^{(T)}
	&=\psi_\rC^{(1)}(x,y,z,\tau)+\psi_\rC^{(2)}(x\!-\!\Delta,y,z,\tau).
\end{align}
Rather than integrate the full field along the $y$-direction, we use only
a slice of length $L' = 10$ $\mu$m in keeping with the experimental geometry of
Ref.~\cite{Hadzibabic2006}.

The interference patterns, $n_{\rm{im}}(x,z)$, generated this way contained fine spatial detail not seen in the experimental images. To make a more useful comparison to experiment it is necessary to account for the finite optical imaging resolution  by applying a Gaussian convolution in the $xz$-plane with standard deviation 3 $\mu$m
\cite{Hadzibabic2007priv}.

\begin{figure}[htbp]
    \begin{center}
        \includegraphics[height=7.3cm]{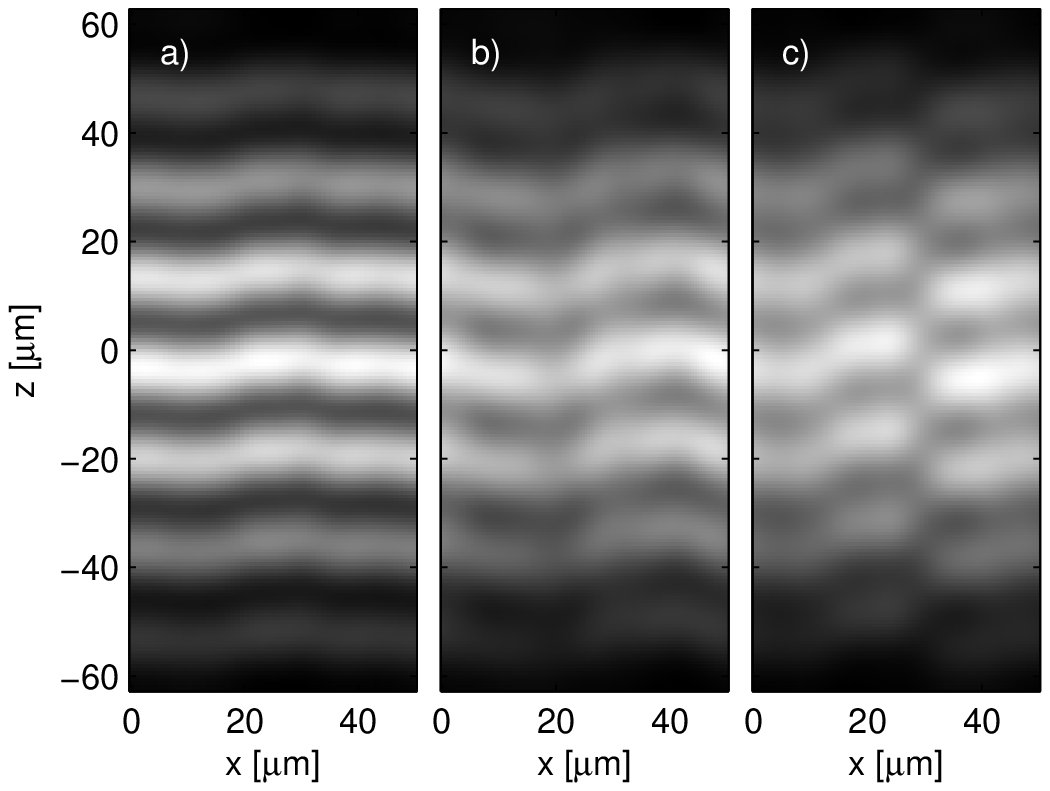}
        \includegraphics[height=7.3cm]{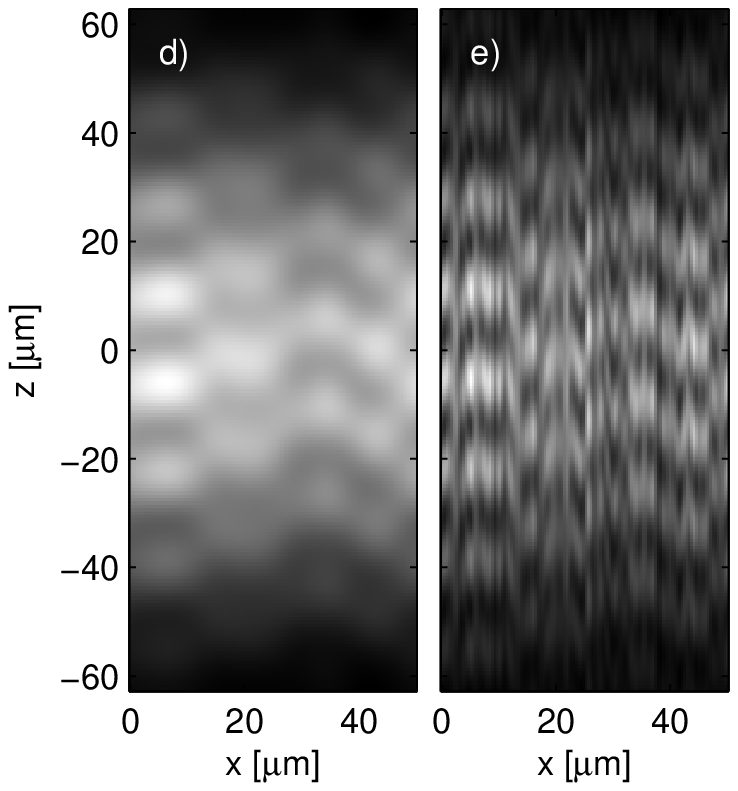}
    \end{center}
    \caption[Simulated interference fringes showing zipper patterns due to vortices]
    { \label{fig:interference_eg}
    Synthetic interference patterns generated from the 50 $\mu$m grid by simulation of the
    experimental procedure of Ref.~\cite{Hadzibabic2006}.  (a) At low temperatures,
    $T \approx 0.5\Tkt$, the interference fringes are straight.  (b) Just below 
    the transition temperature, $T \approx 0.95 \Tkt$, the fringes become wavy due 
    to decreased spatial phase coherence.  Phase dislocations become common at
    temperatures above the transition, (c) $T \approx 1.05\Tkt$, and (d)
    $T \approx 1.1\Tkt$.  These ``zipper patterns'' indicate the presence of free
    vortices.  (e) When simulation of the finite imaging resolution is disabled,
    the zipper patterns from the field in sub figure (d) are no longer clearly
    visible; the high frequency details obscure the phase information without
    providing obvious additional information about the existence of vortex pairs.
    }
\end{figure}

In accordance with the Paris experiment, we use a 22~ms expansion time to
generate interference patterns for quantitative analysis (see
section~\ref{G1interfere}). To obtain characteristic interference images for
display in Ref.~\cite{Hadzibabic2006}, the experiments used a shorter 11 ms expansion  \cite{Hadzibabic2007priv}.  We
exhibit examples of interference patterns at various temperatures in
Fig.~\ref{fig:interference_eg}, for this shorter expansion time. These images show a striking resemblance to 
the results presented in Ref.~\cite{Hadzibabic2006}.

\subsection{Condensate and superfluid fractions} \label{sec:fc_fs}
For a 2D Bose gas in a box we expect a nonzero condensate fraction due to the
finite spacing of low-energy modes.  A central question is whether we can 
observe a distinction between the crossover due to Bose condensation and that
due to BKT physics.  To address this question we have computed both the 
condensate and superfluid fractions from our dynamical simulations.

The condensate fraction in a homogeneous system is easily identified as the
average fractional occupation of the lowest momentum mode.  This is directly
available from our simulations as a time average of the $\v{k} = \v{0}$ mode of
the classical field,
\begin{equation}
    f_c = \Exval{c_\v{0}^* c_\v{0}^{\vphantom{*}}} / N.
\end{equation}

Superfluidity is the macroscopic tendency for some fraction $f_s$ of certain
fluids to flow without apparent viscosity.  The task of connecting this
phenomenology to the microscopic theory is not trivial, so extracting the
superfluid fraction from dynamical classical field simulations provides a more
difficult challenge.  Our approach relies on linear response theory to connect
the superfluid fraction with the long wavelength limit of the second order
momentum density correlations.  The details of this technique are presented
separately in chapter \ref{sfrac_chapter}.

\begin{figure}[htbp]
    \begin{center}
        \includegraphics[width=12cm]{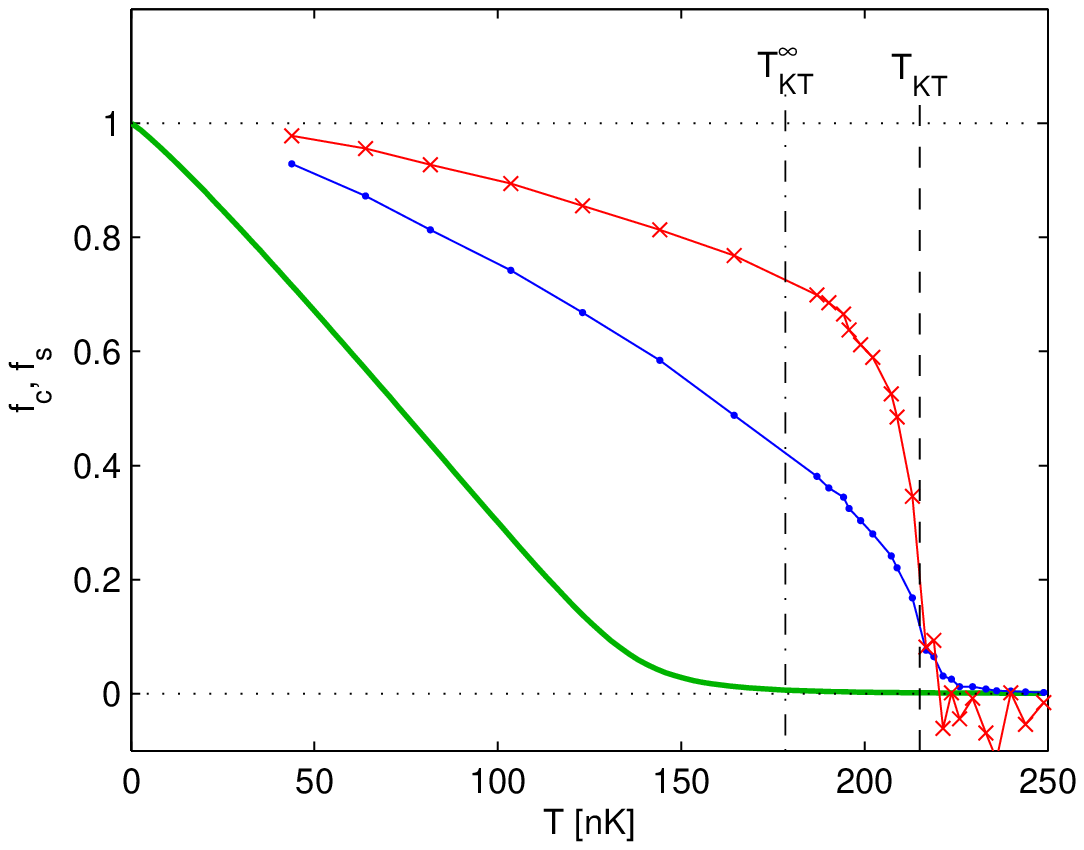}
    \end{center}
    \caption[Condensate and superfluid fractions as a function of temperature]
    { \label{fig:fc_fs_vs_T}
    Condensate fraction (solid dots) and superfluid fraction (crosses) as 
    functions of temperature for the 100 $\mu$m$^2$ grid.  The transition
    temperature in the thermodynamic limit, $\Tkt^\infty$ \cite{Prokofev2001},
    is shown as a vertical dot-dashed line.  The vertical dashed line shows our
    estimate for the transition temperature in the finite system.  The thick
    solid line is the condensate fraction for an ideal Bose gas in the grand 
    canonical ensemble with the same number of atoms and periodic spatial domain.
    The superfluid fraction becomes negative in places because the extrapolation
    of the momentum correlations to $\v{k} = 0$ is sensitive to statistical noise
    at high temperature (see section \ref{superfluid_fraction_numerics} for
    details).
    }
\end{figure}

\begin{figure}[htbp]
    \begin{center}
        \includegraphics[width=12cm]{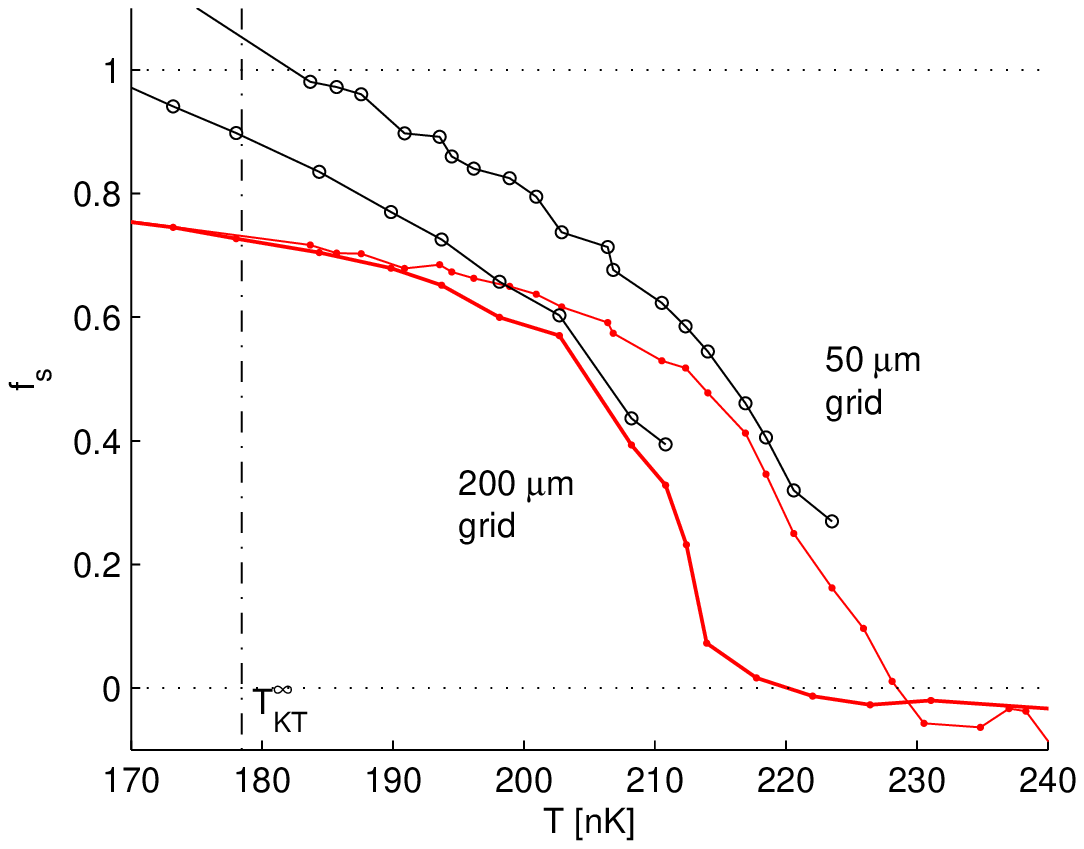}
    \end{center}
    \caption[Detail of the superfluid fraction near the transition temperature]
    { \label{fig:fs_all_vs_T}
    Detail of the superfluid fraction near the transition temperature.  Solid
    dots represent the calculation based on momentum correlations as
    described in chapter \ref{sfrac_chapter}.  Results for the
    largest and smallest grids are shown (left and right, respectively).  The
    data for the 100 $\mu$m grid is omitted for clarity, but lies between the
    curves shown as expected.  Open circles represent the calculation of
    the superfluid fraction from the associated fitted values for the decay
    coefficient $\alpha$, via Eq.~\eqref{eqn:alpha_rhos_T}.  The open circles 
    terminate where the power law fitting procedure fails.
    }
\end{figure}

Figure \ref{fig:fc_fs_vs_T} compares the results for the superfluid and
condensate fractions computed on the 100 $\mu$m grid.  These results are
qualitatively similar to the results for the larger and smaller grids.  In
particular, we note that there is no apparent separation between temperatures
at which the superfluid and condensate fractions fall to zero.  Also shown in
Fig.~\ref{fig:fc_fs_vs_T} is the condensate fraction for the ideal Bose gas
confined to an identical finite-size box in the grand canonical ensemble.  The large
shift between ideal and computed transition temperatures indicates the
effect of interactions in the 2D system. Because the average system density is uniform, this large shift is to   due to critical fluctuations (also see \cite{Kashurnikov2001a}).

In our calculations we identify the transition temperature, $\Tkt$, 
as where the superfluid fraction falls off most rapidly (i.e., the location of steepest slope on the $f_s$ versus $T$ graph; see Fig.~\ref{fig:fc_fs_vs_T}).  As the system
size increases, this transition temperature moves toward the value for an
infinite-sized system, $\Tkt^\infty$ \cite{Prokofev2001}.  This effect is
illustrated by the behaviour of the superfluid fraction in
Fig.~\ref{fig:fs_all_vs_T}.

Also shown in Fig.~\ref{fig:fs_all_vs_T} is an alternative calculation of the
superfluid fraction based on Eq.~\eqref{eqn:alpha_rhos_T}.  The two methods are
expected to match for temperatures at and slightly below the transition
temperature \cite{Nelson1977}.

\subsection{First-order correlations --- algebraic decay}
\label{sec:g1}
Algebraic decay of the first-order correlations, as described by
Eq.~\eqref{eqn:algebraic_decay}, is a characteristic feature of the BKT
phase.  Above the BKT transition, the first-order correlations should
revert to the exponential decay expected in a disordered phase.

The normalised first-order correlation function, $g^{(1)}$ is defined by
\begin{equation} \label{eqn:g1_definition}
    g^{(1)}(\vec{x},\vec{x}') = \frac{G^{(1)}(\vec{x},\vec{x}')}
    {\sqrt{n(\v{x})n(\v{x}')}},
\end{equation}
where $G^{(1)}(\vec{x},\vec{x}') =
\exval[\big]{\hat{\psi}^\dagger(\vec{x})\hat{\psi}(\vec{x'})}$ is the
unnormalised first-order correlation function \cite{Naraschewski1999}.
In a homogeneous isotropic system $g^{(1)}$ depends only the distance
$\norm{\vec{x} - \vec{x}'}$ and we may characterise the first-order
correlations by a function of one variable, $g^{(1)}(x) \equiv
g^{(1)}(\norm{\vec{x} - \vec{x}'}) = g^{(1)}(\vec{x},\vec{x}')$.

\subsubsection{Direct calculation of $g^{(1)}$} \label{sec:direct_calc_g1}
In the PGPE formalism the $\rC$ and $\rI$ contributions to the correlation function are additive \cite{Bezett2008}, that is,
\begin{equation}
    G^{(1)}(\vec{x},\vec{x}')=G^{(1)}_\rC(\vec{x},\vec{x}')+G^{(1)}_\rI(\vec{x},\vec{x}'),
\end{equation}
where $G^{(1)}_\rC$ and $G^{(1)}_\rI$ are defined in Eqs.~\eqref{Gc1} and \eqref{eqn:GI1}, respectively.
It is interesting to note that $G^{(1)}_\rC$ and $G^{(1)}_\rI$ individually
display an oscillatory decay behaviour --- originating from the cutoff --- an
effect which correctly cancels when the two are added together.

\begin{figure}[htbp]
    \begin{center}
        \includegraphics[width=12cm]{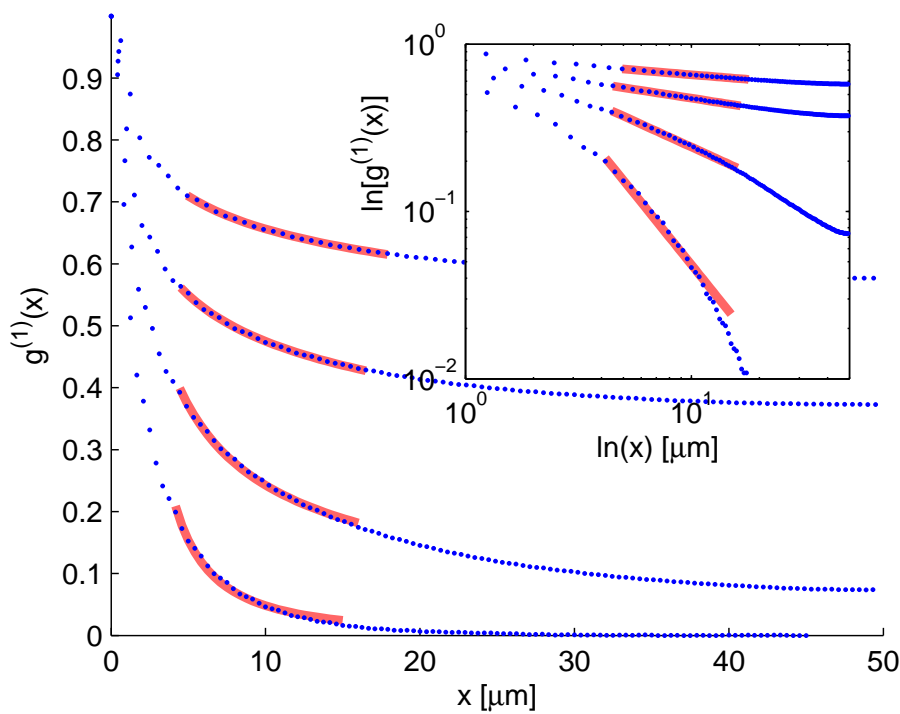}
  \end{center}
  \caption
  [Power law fits of $g^{(1)}$ in two dimensions]
  { \label{fig:alpha_fit_eg}
  Sample fits to the algebraic decay of $g^{(1)}$ at multiples
  $T \approx 0.77, 0.93, 1.01$ and $1.12$ of the transition temperature.
  High temperatures correspond to curves at the bottom of the figure which have
  rapid falloff of $g^{(1)}$ with distance.
  Fits are shown on a log-log scale in the inset to emphasise the failure of a
  power law in describing the behaviour of $g^{(1)}$ at high temperature.
  }
\end{figure}

Having calculated $g^{(1)}$, we obtain the coefficient $\alpha$ by fitting the
algebraic decay law, Eq.~\eqref{eqn:algebraic_decay}, using nonlinear least
squares; sample fits are shown in Fig.~\ref{fig:alpha_fit_eg}.  The fit is 
conducted over the region between 10 and 40 de~Broglie wavelengths. The short
length scale cutoff is to avoid the contribution of the non-universal normal
atoms, for which the thermal de~Broglie wavelength sets the appropriate decay
length. The long distance cutoff is chosen to be small compared to the
length scale $L$, to avoid the effect of periodic boundary
conditions on the long range correlations.

\begin{figure}[htbp]
  \begin{center}
  \includegraphics[width=12cm]{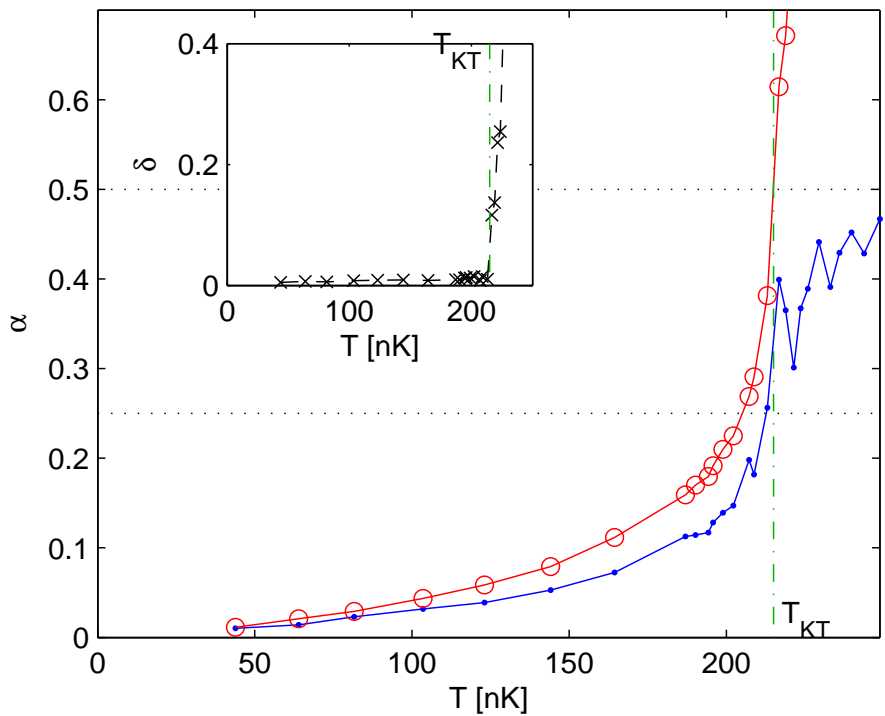}
  \end{center}
  \caption
  [Two methods for determining the algebraic decay coefficient for $g^{(1)}$]
  { \label{fig:alpha_comparison}
  Comparison of two methods for determining the algebraic decay coefficient
  $\alpha(T)$ for the first-order correlation function $g^{(1)}(\v{x},\v{x}')$.
  The line with circle markers represents direct fits to $g^{(1)}$.  These fits
  fail at the transition temperature as shown by the sharply diverging value 
  of $\abs{\delta(T)}$ in the inset.  The filled points represent the
  values $\alpha'(T)$ obtained from a simulation of the experimental analysis
  procedure of Ref.~\cite{Hadzibabic2006}, described in section~\ref{G1interfere}.
  Horizontal dotted lines at 0.25 and 0.5 correspond to the expected values
  of $\alpha'$ just below and above the transition, respectively
  \cite{Hadzibabic2006}.  The vertical line is the BKT transition temperature,
  as estimated from the superfluid fraction calculated in section~\ref{sec:fc_fs}.
  }
\end{figure}

The quality of the fitting procedure, and the breakdown of the expression in
Eq.~\eqref{eqn:algebraic_decay} at the BKT transition can be observed by adding an
additional degree of freedom to the fitting function.  In particular, at each 
temperature we fit the quadratic 
$\ln(g^{(1)}) = A - \tilde{\alpha} \ln(x) + \delta\ln^2(x)$  
and extract the parameter $\delta$ ($\tilde{\alpha} \approx \alpha$ is
discarded).  The abrupt failure of the fits can be observed in the inset of
Fig.~\ref{fig:alpha_comparison} as a sudden increase in the value of
$\abs{\delta(T)}$ --- an effect which is in excellent agreement with the value
of $\Tkt$ as estimated from the superfluid fraction.

\subsubsection{Calculation of $g^{(1)}$ via interference patterns}\label{G1interfere}

So far a direct probe of the \textit{in situ} spatial correlations has not been possible, although important progress has been made by the NIST group \cite{Clade2008}.
In the experiments of Hadzibabic \textit{et al.} \cite{Hadzibabic2006} a scheme
proposed by Polkovnikov  \textit{et al.} \cite{Polkovnikov2006} was used to
infer these correlations from the ``waviness'' of interference patterns produced
by pair of quasi-2D systems (see section~\ref{siminterference}). 
 In this section we simulate the experimental data analysis method, and compare inferred predictions for the correlation function against those we can directly calculate. This allows us to characterise the errors associated with this technique arising from finite-size effects and finite expansion time.

To make this analysis we follow the procedure outlined in
Ref.~\cite{Hadzibabic2006}. We fit our numerically generated interference
patterns (see section~\ref{siminterference})  to the function 
\begin{equation}
	F(x,z) = G(z) \Bktsq{1 + c(x) \cos\Bkt{\frac{2 \pi z}{D} + \theta(x)}},
\end{equation}
where $G(z)$ is a Gaussian envelope in the $z$-direction, $c(x)$ is the
interference fringe contrast, $D$ is the fringe spacing and $\theta(x)$ is the
phase of the interference pattern in the $z$-direction.

Defining the function
\begin{equation}
	C(L_x) = \frac{1}{L_x} \int_{-L_x/2}^{L_x/2}  dx\, c(x) e^{i\theta(x)},
\end{equation}
the nature of spatial correlations is then revealed by the manner in which $\exval[\big]{\abs{C(L_x)}^2}$ decays with $L_x$. In particular, we identify the parameter $\alpha'$, defined by $\exval[\big]{\abs{C(L_x)}^2}
\propto L_x^{-2\alpha'}$ \cite{Polkovnikov2006}. For an infinite 2D system in
the superfluid regime  ($T<\Tkt^{\infty}$)  $\alpha' = \alpha$ (i.e., $\alpha'$ corresponds to the algebraic decay of correlations). For $T>\Tkt^{\infty}$,
 where correlations decay exponentially, $\alpha'$ is equal to $0.5$.

Fitting $\exval[\big]{\abs{C(L_x)}^2}$ to the algebraic decay law
$A L_x^{\;-2\alpha'}$ we can determine $\alpha'$. A comparison between $\alpha'$
inferred from the interference pattern and $\alpha$ obtained directly from
$g^{(1)}$ is shown in Fig.~\ref{fig:alpha_comparison}.
Both methods give broadly consistent predictions for $\alpha$ when $T<\Tkt$,
however our results show that there is a clear quantitative difference between
the two schemes, and that $\alpha'$ underestimates the coefficient of algebraic
decay in the system (i.e., using $\alpha'$ in Eq.~(\ref{eqn:alpha_rhos_T}) would
overestimate the superfluid density). Near and above the transition temperature, where the
fits to $g^{(1)}$ fail,  we observe that $\alpha'$ converges toward $0.5$.  The
agreement between $\alpha$ and $\alpha'$ in the low temperature region improves as the size of the grid is increased. 


%

\subsection{Vortices and pairing}
\label{sec:vortices}

The simplest description of the BKT transition is that it occurs as a result of 
vortex pair unbinding: At $T<\Tkt$ vortices only exist in pairs of opposite 
circulation, which unbind at the transition point to produce free vortices that 
destroy the superfluidity of the system.  However, to date there are no direct 
experimental observations of this scenario, and theoretical studies of 2D Bose
gases have been limited to qualitative inspection of the vortex distributions.
In the c-field approach vortices and their dynamics are clearly revealed,
unlike other ensemble-based simulation techniques such as quantum Monte Carlo
where the vortices are obscured by averaging\footnote{
For example, \cite{Filinov2010} calculates the vortex density, but only 
indirectly via a relation with the quasiparticle density.
}. This gives us a unique
opportunity to investigate the role of vortices and pairing in a dilute Bose
gas.

\begin{figure}[htbp]
  \begin{center}
  \includegraphics[width=11cm]{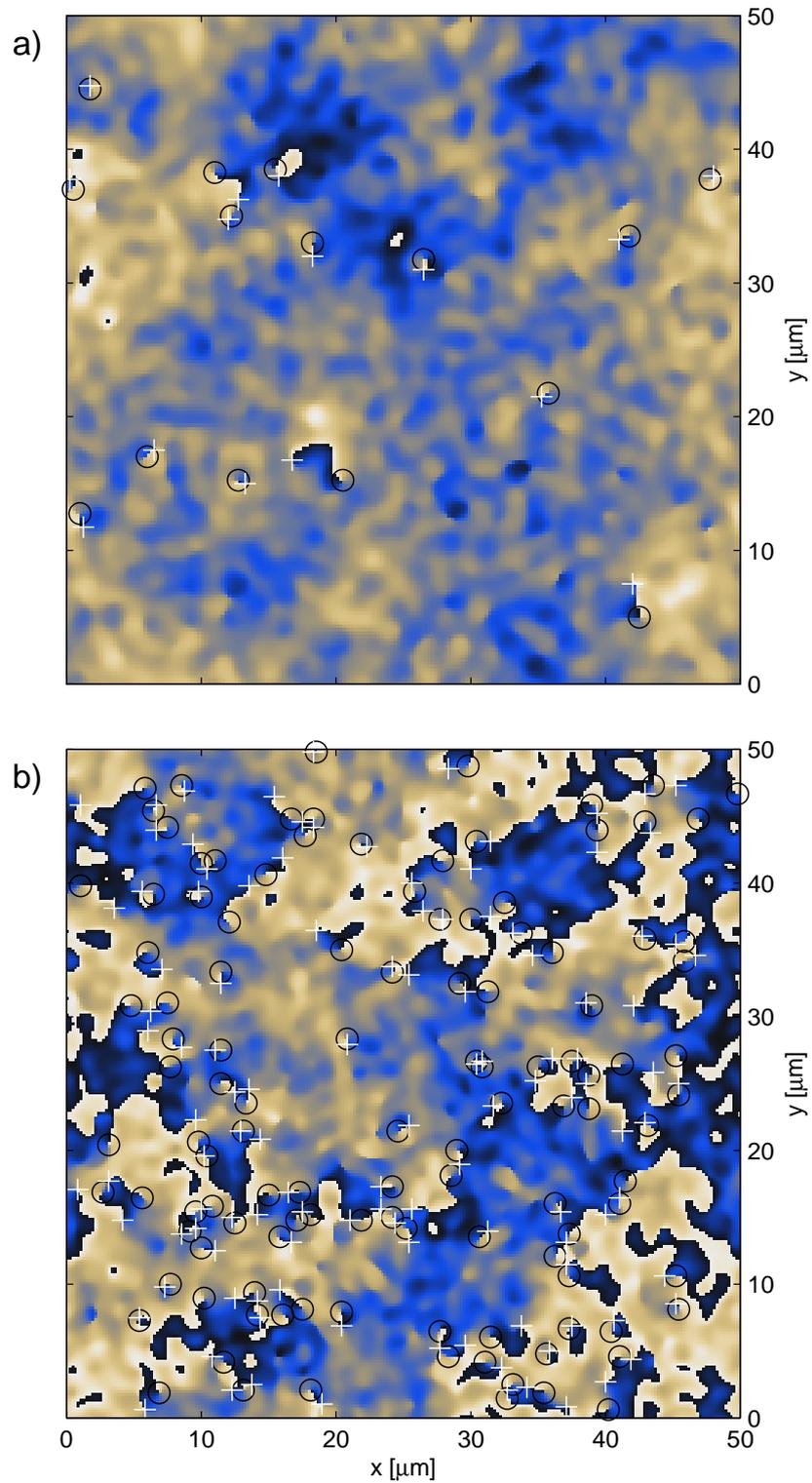}
  \end{center}
  \caption[Phase profile of a c-field showing vortex pairing at two temperatures]
  {\label{fig:vortex_pairing_eg}
	Phase profile of a c-field with vortices indicated. 
	 Vortices with clockwise (white $+$) and anticlockwise (black $\circ$) 
	 circulation.  The phase of the classical field is indicated by shading
	 the background between dark blue (phase 0) and light yellow (phase $2\pi$).
	 (a) Distinctive pairing below the transition at $T=207\text{nK}\approx0.93\Tkt$.
	 (b) A ``vortex plasma'' above the transition at $T=238\text{nK}\approx1.07\Tkt$.
  }
\end{figure}

We detect vortices in the c-field microstates by analysing the phase profile of 
the instantaneous field (see appendix \ref{sec:vortex_detection}). An example 
of a phase profile of a field for $T<\Tkt$ is shown in 
Fig.~\ref{fig:vortex_pairing_eg}(a). The vortex locations reveal a pairing 
character, that is, the close proximity of pairs of positive (clockwise) and 
negative (counterclockwise) vortices relative to the average vortex separation.
An important qualitative feature of our observed vortex distributions is that
at high temperatures, pairing does not disappear from the system entirely.
Indeed, most vortices at high temperature could be considered paired or grouped
in some manner, as shown in Fig.~\ref{fig:vortex_pairing_eg}(b).
Perhaps this is not surprising, since positive and negative vortices have a
logarithmic attraction, and we observe them to create and annihilate readily in
the c-field dynamics. However, this does indicate that the use of pairing to
locate the transition may be ambiguous, and we examine this aspect further
below.

\begin{figure}[htbp]
  \begin{center}
  \includegraphics[width=12cm]{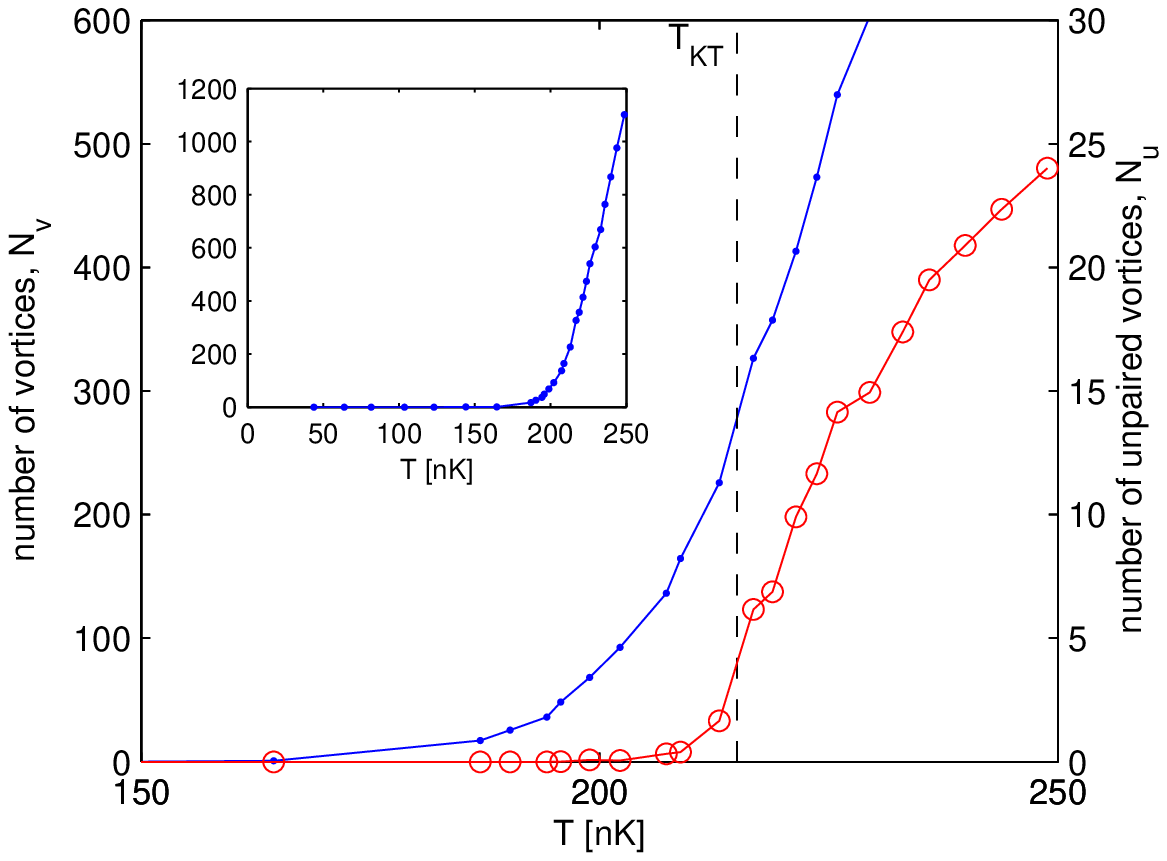}
  \end{center}
  \caption
  [Number of paired and unpaired vortices as a function of temperature]
  { \label{fig:nv_vs_T}
  Total number of vortices (dots) and number of unpaired vortices (circles) as
  a function of temperature near the transition.  While $N_v$ at the
  transition temperature is already very high, $N_u$ becomes nonzero only close
  to the transition, providing clear evidence of vortex unbinding at work.  The
  inset shows the variation in the total number over the full temperature range
  of the simulations.  Above the transition temperature the growth in the
  number of vortices becomes linear with temperature.
  }
\end{figure}

It is also of interest to measure the number of vortices, $N_v$, present in the
system as a function of temperature (see Fig.~\ref{fig:nv_vs_T}). At the lowest
temperatures the system is in an ordered state, and the energetic cost of
having a vortex is prohibitive. As the temperature increases there is a rapid
growth of vortex population leading up to the transition point followed by 
linear growth above $\Tkt$.

\subsubsection{Radial vortex density}
\begin{figure}[hptb]
  \centering
  \begin{center}
  \includegraphics[width=12cm]{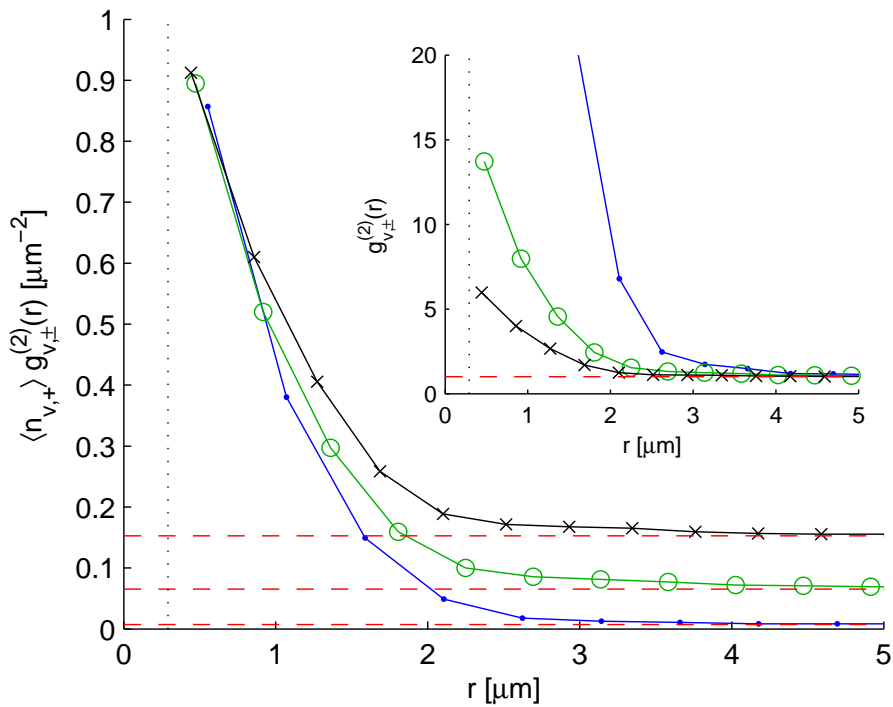}
  \end{center}
  \caption
  [Vortex pair distribution function at various temperatures]
  {\label{fig:pairing_pdfs}
  Angular average of the two-vortex pair distribution functions for vortices of 
  opposite sign.  Three temperatures centred about the transition are shown:
  dot markers $T=194$nK $\approx 0.9\Tkt$, $f_c = 0.34$; circle markers
  $T=217$nK $\approx 1.01\Tkt$, $f_c = 0.076$; cross markers $T=236$nK $\approx
  1.1\Tkt$, $f_c = 0.006$.  The vertical dotted line shows the value of the
  healing length at $T = 0$.  The main plot shows $g_{v,\pm}^{(2)}$ normalised
  by the positive vortex density; comparable magnitudes for the peaks near 
  $r = 0$ show that vortex pairing remains important over the range of 
  temperatures studied, not only below the transition.  The inset shows
  $g_{v,\pm}^{(2)}$ in the natural dimensionless units for which
  $g_{v,\pm}^{(2)}({r}) \to 1$ as $r \to \infty$.
  }
\end{figure}

The most obvious way to characterise vortex pairing is by defining a pair
distribution function for vortices of opposite sign.  Adopting the notation of
Ref.~\cite{Giorgetti2007}, this is
\begin{equation}
G_{v,\pm}^{(2)}(\v{r}) = \Exval{\rho_{v,+}(\v{0}) \rho_{v,-}(\v{r})},
\end{equation}
where $\rho_{v,+}$ is the vortex density function which consists of a sum of
delta spikes,
\begin{equation}
\rho_{v,+}(\v{r}) = \sum_{i=1}^{N_{v,+}} \delta(\v{r} - \v{r}^+_i)
\end{equation}
for positive vortices at positions $\Bktcl{\v{r}^+_i}$.  We use the analogous
definition for $\rho_{v,-}$.  The associated dimensionless two-vortex
correlation function is
\begin{equation} \label{eqn:g2vpm}
g_{v,\pm}^{(2)}(\v{r}) = \frac{G_{v,\pm}^{(2)}(\v{r})}{\exval{\rho_{v,+}(\v{0})} \exval{\rho_{v,-}(\v{r})}}.
\end{equation}
The angular average of $g_{v,\pm}^{(2)}$ can be calculated directly from
the detected vortex positions using a binning procedure on the pairwise distances
$\norm{\v{r}^+_i - \v{r}^-_j}$, and is shown in Fig.~\ref{fig:pairing_pdfs}.

These results quantify the effect discussed earlier:  Positive and negative vortices show a pairing correlation
that does not disappear above $\Tkt$. The characteristic size of this
correlation, given by twice the width of the peak feature in
Fig.~\ref{fig:pairing_pdfs}, is $l_{\rm{cor}}\sim3\mu$m (taking full width half
maximum).

The shape of our pairing peak is qualitatively similar to that described in
Ref.~\cite{Giorgetti2007}.  However, in contrast to their results the width does not
appear to change appreciably with temperature.  Additional simulations show
that increasing the interaction strength causes the peak to become squarer and
wider.  It is clear that while the pair size and strength revealed in
$g_{v,\pm}^{(2)}({r})$ does not change appreciably as the transition is
crossed, the amount of pairing relative to the background uncorrelated vortices
changes considerably.  This background of uncorrelated vortices is given by the
horizontal plateau $g_{v,\pm}^{(2)}({r}) \to 1$ at large $r$ as shown in the
inset.

\subsubsection{Revealing unpaired vortices with coarse-graining}


The function $G_{v,\pm}^{(2)}(r)$ clearly indicates the existence of vortex
pairing in the system.  However, it does not provide a convenient way to locate
the transition temperature, because a large amount of pairing exists both below
and above the transition:  The expected number of neighbours for any given
vortex --- roughly, the area of the pairing peak of
$\exval{n_{v,+}}G_{v,\pm}^{(2)}(r)$ shown in Fig.~\ref{fig:pairing_pdfs} ---
does not change dramatically across the transition.  $\exval{n_{v,+}} =
\exval{n_v}/2$ is the expected density of positive vortices.

We desire a quantitative observation of vortex unbinding at the transition and 
have therefore investigated several measures of vortex pairing\footnote{For
example, the Hausdorff distance (see, for example, \cite[p.~105]{Papadopoulos2005}) between
the set $\{\v{r}^+_i\}$ of positive vortices and the set $\{\v{r}^-_i\}$ of
negative vortices.}.  However, measures based directly on the full set of
vortex positions seem to suffer from the proliferation of vortices at high
temperature --- an effect that tends to wash out clear signs of vortex
unbinding.  With this in mind, we have developed a procedure for measuring the
number of \emph{unpaired} vortices in our simulations, starting from the
classical field rather than the full set of vortex positions.

\begin{figure}[htbp]
  \begin{center}
  \includegraphics[width=12cm]{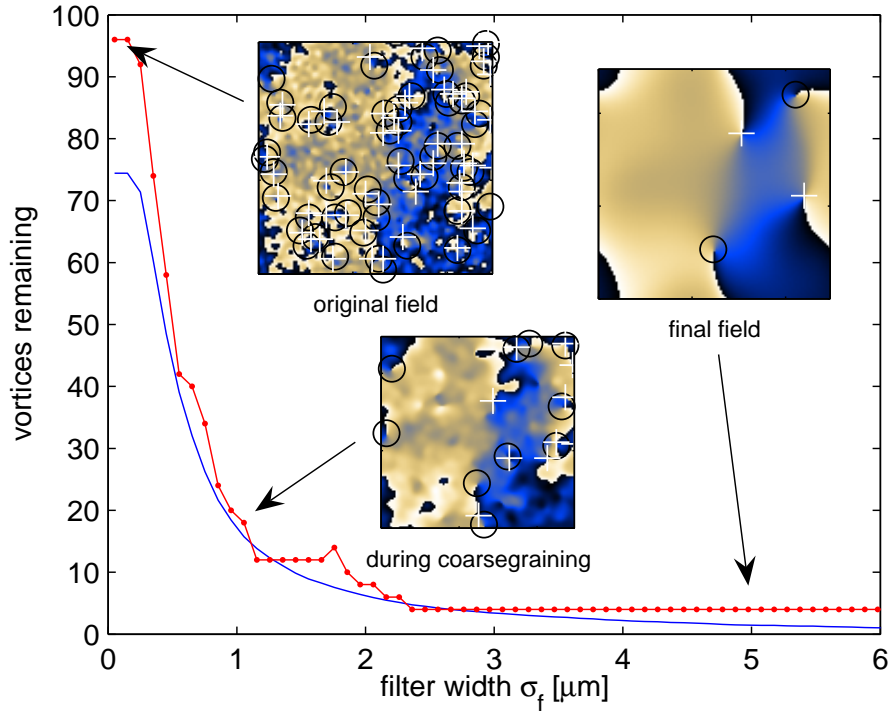}
  \end{center}
  \caption
  [Coarse-graining procedure for detecting unpaired vortices]
  { \label{fig:coarsegrain}
  The coarse-graining procedure: number of vortices as a function of filter
  width for a temperature near the transition.  The smooth curve is an average
  over many realisations of the field, whereas the stepped curve shows typical
  behaviour of the number for a single field.  
  Insets show the coarse-grained fields for various filter widths; the
  transformation removes vortex-antivortex pairs that are separated by
  approximately less than the standard deviation of the filter.  In this 
  example $N_u = 4$ unpaired vortices remain at $\sigma_f = 5$ $\mu$m.
  }
\end{figure}

The basis of our approach for detecting unpairing is to
coarse-grain the classical field by convolution with a Gaussian filter of
spatial width (standard deviation) $\sigma_f$.  This removes all vortex pairs
on length scales smaller than $\sigma_f$.  Figure \ref{fig:coarsegrain}
shows the count of remaining vortices as a function of filter width, along with
some examples of coarse-grained fields. For $\sigma_f\gtrsim l_{\rm{cor}}$, the
number of remaining vortices levels off and only decreases slowly with
increasing $\sigma_f$.  Ultimately the number of remaining vortices goes to
zero as $\sigma_f\to L$.

Setting the filter width to be  larger than the characteristic pairing
distance, $l_{\rm{cor}}$, yields a coarse-grained field from which the pairs
have been removed, but unpaired vortices remain.  In our simulations we have
$l_{\rm{cor}}\approx3$ $\mu$m; we take the vortices that remain after
coarse-graining with a Gaussian of standard deviation $\sigma_f=5$ $\mu$m to
give an estimate of the number of unpaired vortices, $N_u$.  Figure
\ref{fig:nv_vs_T} shows that $N_u$ becomes nonzero only near the transition, in 
contrast to $N_v$ which is nonzero well below $\Tkt$.  The sharp increase in
$N_u$ at $\Tkt$ is a quantitative demonstration of vortex unbinding at work.

In the experiment of Ref.~\cite{Hadzibabic2006}, the fraction of interference 
patterns with dislocations (see, for example, Figs.~\ref{fig:interference_eg}(c) and (d))
was measured.  While isolated vortices are clearly identified by interference
pattern dislocations, a lack of spatial resolution in experiments means that
this type of detection method obscures the observation of tightly bound vortex
pairs.  The experimental resolution of 3 $\mu$m is broadly consistent with the
scale of the coarse-graining filter (i.e., $\sigma_f = 5$ $\mu$m).  With this
in mind, we introduce the quantity $\punpair(T)$, defined as the probability
of observing an unpaired vortex in a $50 \times 50$ $\mu$m control volume at a
given temperature\footnote{We choose a fixed control volume with $L=50$
$\mu$m in order to compare results between simulations with different grid
sizes.}.  For the 50 $\mu$m grid we have simply $\punpair(T) = \Pr(N_u \ge 1)$.

\begin{figure}[htbp]
  \begin{center}
  \includegraphics[width=12cm]{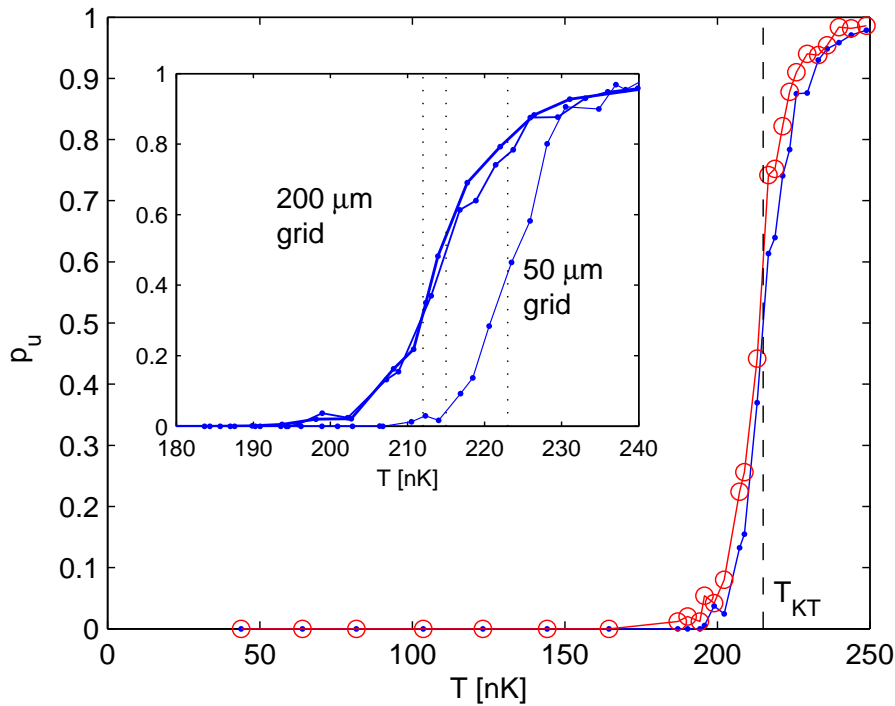}
  \end{center}
  \caption[Comparison of vortex unpairing measures]
  { \label{fig:pairing2}
  Comparison of vortex unpairing measures.  The dots are our pairing measure
  based on coarse-graining the field.  Circles represent the pairing as
  determined by the number of dislocations in the simulated interference
  patterns.  This was the same
  method used in the experimental analysis of Ref.~\cite{Hadzibabic2006} and
  coincides remarkably well with our coarse-graining based measure.  Both curves are consistent 
  with the vertical line showing the transition temperature $\Tkt$ as
  determined from the superfluid fraction calculation in section~\ref{sec:fc_fs}.
  The inset shows the calculated coarse-grained pairing measure for all three 
  grid sizes, along with vertical lines showing the estimates for $\Tkt$
  derived from the superfluid fraction calculations.
  }
\end{figure}

Computing $\punpair(T)$ from our simulations yields the results shown in
Fig.~\ref{fig:pairing2}.  Our results show a dramatic jump in $\punpair$
at a temperature that is consistent with the transition temperature $\Tkt$
determined from the superfluid fraction calculation presented in
section~\ref{sec:fc_fs}.

From the definition, we expect that $\punpair$ should be close to the
experimentally measured frequency of dislocations.  To demonstrate this
relationship, we have simulated interference patterns (as described in
section~\ref{siminterference}) and detected dislocations using the experimental
procedure of Ref.~\cite{Hadzibabic2006}: A phase gradient $d\theta/dx$ was
considered to mark a dislocation whenever $\Abs{d\theta/dx} > \pi/4$ rad$/\mu$m.
From this we can compute the probability of detecting at least one dislocation
as a function of temperature.  As shown in Fig.~\ref{fig:pairing2}, the results
of this procedure compare very favourably with our measure of pairing based on
$\punpair$.  We note that inhomogeneous effects in experiments probably
broaden the jump in $\punpair$ appreciably compared to our homogeneous
results.

\section{Conclusion}
\label{sec:conclusion}

In this chapter we have used c-field simulations of a finite-sized homogeneous 
system in order to investigate the physics of the 2D Bose gas in a regime 
corresponding to current experiments.  We have directly computed the condensate 
and superfluid fractions as a function of temperature, and made comparisons to 
the superfluid fraction inferred both from the first-order correlation
function, and the experimental interference scheme. Our results for these
quantities provide a quantitative test of the interference scheme for a finite
system. 

An intriguing possibility is the direct experimental observation of
vortex-antivortex pairs, their distribution in the system, and hence a
quantitative measurement of their unbinding at the BKT transition. We have
calculated the vortex correlation function across the transition and provided a
coarse-graining scheme for distinguishing unpaired vortices. These results
suggest that the dislocations observed in experiments, due to limited optical
resolution, provide an accurate measure of the unpaired vortex population and
accordingly are a strong indicator of the BKT transition.

We briefly discuss the effect that harmonic confinement (present in
experiments) would have on our predictions. The spatial inhomogeneity  will
cause the superfluid transition to be gradual, occurring first at the trap
centre where the density is highest, in contrast to our results where the
transition occurs in the bulk.

Bisset \textit{et al.}\ \cite{Bisset2009} used an extension of the c-field method for the trapped 2D gas to examine $g^{(1)}$ and found similar results for the onset of algebraic decay of correlations at the transition.  Their analysis was restricted to the small region near the trap centre where the density is approximately constant; we expect the results of our vortex correlation function and the coarse-graining scheme should similarly be applicable to the trapped system in the central region.  Except in very weak traps, the size of this region is relatively small and will likely prove challenging to measure experimentally.

Our results for the homogeneous gas emphasise the clarity with which \textit{ab initio} theoretical methods can calculate quantities directly observable in experiments, such as interference patterns. This should allow direct comparisons  with experiments, providing stringent tests of many-body theory.

%% file: sfrac/sfrac.tex
\chapter{Superfluid fraction and the PGPE}
\label{sfrac_chapter}

\begin{chap_desc}
In this chapter we describe a method for calculating the superfluid fraction
from a PGPE classical field simulation.  We first present the analytical
derivation, showing how the superfluid density arises in certain limits of the
momentum density autocorrelations.  This is followed by discussion of a
numerical implementation relevant to the 2D simulations of chapter
\ref{bkt_chapter}.
\end{chap_desc}

\section{Introduction}
\label{sfrac_introduction}

Superfluidity is a famous example of macroscopic quantum behaviour, and is
typically discussed in macroscopic terms.  In particular, one characterises a
superfluid by its zero viscosity; the ability to ``flow without friction'' 
through a narrow channel.  Although such macroscopic ideas are easily
expressed, it is not trivial to connect them to microscopic theories such as
the PGPE formalism in an efficient way.  In the following we briefly provide
some relevant background before following with the details of our derivation in
the next section.  For further background theory we refer the reader to chapter
6 of Ref.~\cite{Pitaevskii_Stringari2003} which provides an accessible overview
of superfluid theory as relevant to experiments on ultracold Bosons.

According to Landau's phenomenological model of superfluidity \cite{Landau1941},
it is possible to model a superfluid system as a ``mixture'' of two liquids: a
superfluid part without viscosity, and a normal part.  This idea was introduced
to describe the residual viscosity which remains in liquid helium, even below
the superfluid transition temperature.  One defines a \emph{superfluid
fraction} as the ratio $f_s \equiv \rho_s/\rho$ of the superfluid density
$\rho_s$ to the total density $\rho$ of the system\footnote{Note that we follow
convention and use the \emph{mass} density $\rho$ in the current chapter rather
than the number density $n$ that is used in the rest of the
thesis.\label{mass_density_footnote}
}.

The two-fluid model was put on firmer ground by Putterman and Roberts
in Ref.~\cite{Putterman1983}.  Starting from the equations for a single
nonlinear classical fluid, they considered the presence of small amplitude
excitations on top of a background fluid.  Using only a separation of scales
argument, they have derived kinetic equations for these thermal excitations.  In
the hydrodynamic (collision dominated) regime, the model then reduces to the
Landau two-fluid model of superfluidity.  The connection between this model and
the classical field methods is further described in \cite{Salman2013}.



In Landau's theory, superfluidity may be predicted from the form of the energy
spectrum of elementary excitations.  Let $\epsilon(\v{p})$ be the energy of an
excitation with momentum $\v{p}$ measured with respect to a stationary
background of fluid.  To understand the origin of superfluidity, we imagine that the fluid
occupies a narrow channel with walls moving at velocity $\v{u}$ with respect to
the fluid.  In the frame of reference where the walls are stationary, the
energy of the excitation is found to be
$\epsilon'(\v{p}) \equiv \epsilon(\v{p}) - \v{u}\cdot\v{p}$
after applying a Galilean transformation to the Hamiltonian\footnote{
Note that in our notation $\v{u}$ is the velocity of the walls with respect to
the fluid, so the superfluid velocity is $-\v{u}$ with respect to the walls.
This is opposite from the convention used in
Ref.~\cite{Pitaevskii_Stringari2003}.
}.
This is negative for sufficiently large $\v{u}$, making the formation of
excitations energetically favourable.
However, if $\v{u}$ is small enough, it may be the case that
$\epsilon'(\v{p}) > 0$ for all $\v{p}$ and excitations are energetically
forbidden.  This leads to Landau's criterion for superfluidity: $\norm{\v{u}}$
must be smaller than the critical velocity,
\begin{equation}
  v_c \equiv \min_\v{p} \frac{\epsilon(\v{p})}{p},
\end{equation}
where $p = \norm{\v{p}}$.

It is worth noting that any system with quadratic dispersion relation
$\epsilon(\v{p}) \propto p^2$ for small $p$ has a critical velocity of zero and
cannot be a superfluid.  In particular, this includes the ideal gas, where the
elementary excitations are simply particles and we have $\epsilon(\v{p}) =
p^2/2m$.  On the other hand, introducing interactions as in the Hamiltonian
Eq.~\eqref{swave_bose_gas_Hamiltonian} modifies the energy so that
$\epsilon(\v{p}) \propto p$ at small $p$, which allows the system to support
superfluidity.  For sufficiently weak interactions, Hamiltonian
\eqref{swave_bose_gas_Hamiltonian} may be approximately diagonalised via the
Bogoliubov transformation (see, for example, \cite[\S
4.3]{Pitaevskii_Stringari2003}), yielding the classic dispersion relation
\begin{equation}
  \epsilon(\v{p}) = \sqrt{\frac{U_0 n}{m} p^2 + \Bkt{\frac{p^2}{2m}}^2}
\end{equation}
for the energies of elementary excitations, known as Bogoliubov
quasiparticles.


The approximate diagonalisation discards terms corresponding to the interaction 
of quasiparticles, which is a good approximation for sufficiently weak
interactions and low temperatures.  Further, this leads to a well known method
for computing the superfluid fraction.  At nonzero temperature, the Bose
distribution gives the number of non-interacting quasiparticles at each energy,
\begin{equation}
  N_{\v{p}} = \Bktsq{\exp\Bkt{\frac{\epsilon(\v{p}) + \v{p}\cdot\v{u}}{kT}} - 1}^{-1}.
\end{equation}
This may be used to evaluate the expected momentum density
$\Exval{\hat{\v{p}}}_\v{u}$, which we attribute to the motion of the normal 
fraction, so that
\begin{equation}
  \rho_n \v{u} = \Exval{\hat{\v{p}}}_\v{u} = \int \v{p} N_\v{p} \frac{d\v{p}}{h^3},
\end{equation}
where we have used the phase space volume $1/h^3$ appropriate to three dimensions.
For consistency with the next section, we use the notation
$\exval{\cdot}_\v{u}$ to mean an expectation value with respect to the system
where the walls are moving.  After some manipulation, we arrive at an 
expression for the density of the normal fraction
\cite{Pitaevskii_Stringari2003}
\begin{equation}\label{normal_density_via_quasiparticles}
    \rho_n = -\frac{1}{3} \int \deriv{N_\v{p}(\epsilon)}{\epsilon} p^2 \frac{d\v{p}}{h^3}.
\end{equation}


This method is valid when the physical picture of non-interacting
quasiparticles is valid --- in particular, one requires weak interactions and
low temperatures.  In the context of the classical field method for ultracold
Bose gases, the method has previously been used to compute the superfluid
fraction, see for example Ref.~\cite{Zawitkowski2006}.  Unfortunately it is not
valid for the system considered in chapter \ref{bkt_chapter} for two reasons.
First, we wish to compute the superfluid fraction over a wide range of
temperatures, from zero to slightly above the transition and the
underlying assumptions are invalid near the transition
\cite[p.~66]{Pitaevskii_Stringari2003}.  Second, the weakly interacting limit is
especially difficult to reach in two dimensions because it requires the
inequality $\ln\ln(1/na^2) \ll 1$ to be satisfied as discussed in
Ref.~\cite{Fisher1988} (see also Ref.~\cite{Prokofev2002}).

Given our two-dimensional system at moderate temperatures, we require a 
non-perturbative alternative to Eq~\eqref{normal_density_via_quasiparticles}.
Starting from the macroscopic definition of superfluidity, one naturally
imagines performing a time-varying numerical experiment in order to determine 
the superfluid density within the PGPE formalism.  For example, in the
homogeneous case we might examine the drag force produced when a perturbing
potential is moved across the system; zero drag would imply a superfluid
fraction of 100\%.  Dynamical PGPE simulations were used in a similar way in
Ref.~\cite{Simula2008} to provide evidence for superfluidity in a 2D trapped
Bose gas, by analysing the dynamics of the ``scissors mode'' oscillation.

Deducing superfluidity from dynamical simulations is certainly possible, but is
far from ideal: For the work presented in chapter \ref{bkt_chapter} it would
mean performing an additional set of numerically expensive simulations at every
temperature of interest.  Our method address this problem by using linear
response theory to relate the superfluid fraction to the long wavelength limit
of the second order momentum density correlations.  The method is attractive
because the momentum correlations may be extracted directly from PGPE
simulations at thermal equilibrium.  This allows the superfluid fraction to be
computed from the same set of simulations as the temperature, chemical
potential, and other thermal properties; there is no need to perform an
expensive special purpose simulation for the sole purpose of calculating the
superfluid fraction.

\hfill

\section{Momentum density correlations and the superfluid fraction}

Our derivation is based on the microscopic theory presented in
Ref.~\cite[p.214]{Forster1975}, (see also \cite{Baym1968} and
\cite[p.96]{Pitaevskii_Stringari2003}).
The central idea is to establish a relationship between (i) the
autocorrelations of the momentum density in the simulated ensemble and (ii) the
linear response of the fluid to slowly moving solid boundaries; (i) is a
quantity we can calculate, while (ii) is related to the basic properties of a
superfluid via a simple thought experiment.

\begin{figure}[hptb]
	\centering
	\psfrag{velu}[][]{$\v{u}$}
	\psfrag{limy}[][]{$\displaystyle\lim_{L_y\to\infty}$}
	\psfrag{limx}[][]{$\displaystyle\lim_{L_x\to\infty}$}
	\psfrag{superflui}[][]{Superfluid}
	\psfrag{Ly}[][]{$L_y$}
	\psfrag{Lx}[][]{$L_x$}
	\psfrag{aa}[][]{a)}
	\psfrag{bb}[][]{b)}
	\psfrag{cc}[][]{c)}
	\psfrag{dd}[][]{d)}
    \begin{center}
	\includegraphics[width=12cm]{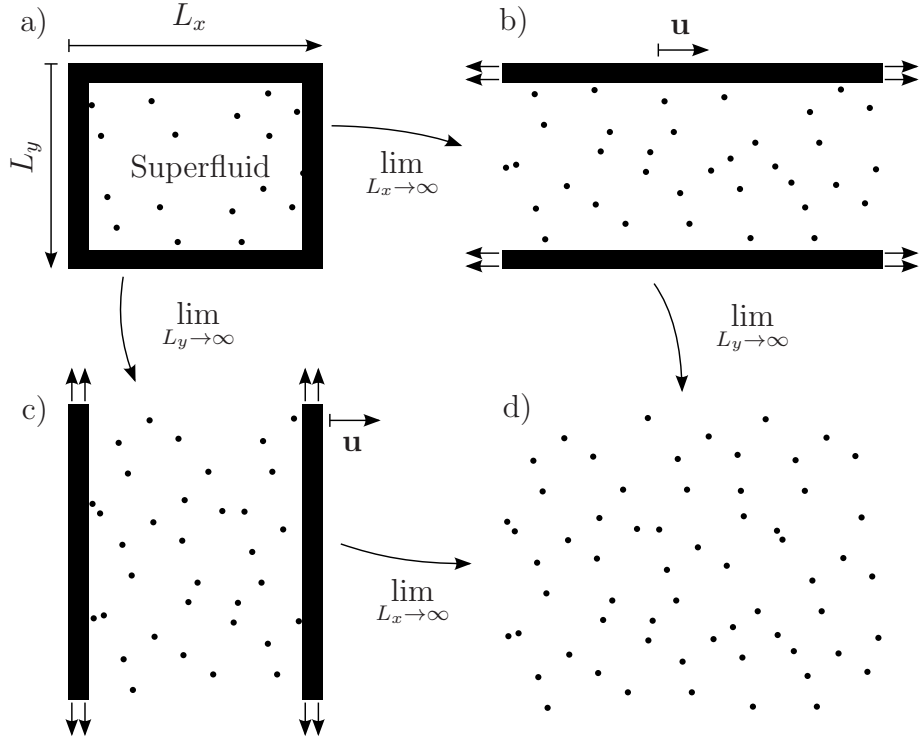}
    \end{center}
	\caption
	[Thought experiment used in deriving the superfluid density]
	{ \label{fig:superfluid_thought_expt}
	Thought experiment used in deriving the superfluid density.  The walls
	move with velocity $\v{u}$ in the $x$ direction.  To begin with, we imagine
	that the superfluid sits in a box of dimensions $L_x \times L_y$ as shown
	in (a).  We later take the limit as the box walls recede to infinity to get 
	the thermodynamic limit (d).  The order of the limits is critically 
	important: the path (b) leads to superflow while the path (c) results in the 
	entire fluid moving along with the walls.
	}
\end{figure}

To connect the macroscopic, phenomenological description of superfluidity with
our microscopic theory, we make use of the standard thought experiment shown 
schematically in Fig.~\ref{fig:superfluid_thought_expt}(b): Consider an 
infinitely long box, $B$ containing superfluid, and accelerate the box along
its long axis until it reaches a small velocity $\v{u}$.  Due to viscous
interactions with the walls, such a box filled with a normal fluid should have
a momentum density at equilibrium of $\exval{\md}_\v{u} = \rho \v{u}$.
As above, the notation $\exval{\cdot}_\v{u}$ denotes an expectation value in
the ensemble with walls moving with velocity $\v{u}$.

Because the superfluid part is nonviscous, the observed value for the momentum 
density in a superfluid is less than the value $\rho\v{u}$ expected for a
classical fluid.  As above, we attribute the observed momentum density, $\rho_n
\v{u}$, to the ``normal fraction'' where $\rho_n$ is the normal fluid density.
The superfluid fraction remains stationary in the lab frame, even at
equilibrium and makes up the remaining mass with density $\rho_s = \rho -
\rho_n$.

In order to apply the usual procedures of statistical mechanics to the thought
experiment, we consider two frames: the ``lab frame'' in which the walls move
with velocity $\v{u}$ in the $x$-direction and the ``wall frame'' in which the
walls are at rest.

Assuming that the fluid is in thermal equilibrium with the walls, the density matrix in
the grand canonical ensemble is given by the usual expression $\hat{\rho} =
e^{-\beta(\hat{H}_{\v{u}} - \mu \hat{N})} / \Tr\bkt[\big]{e^{-\beta(\hat{H}_{\v{u}} - \mu \hat{N})}}$ where
$\hat{H}_{\v{u}}$ is the Hamiltonian of the system in the wall frame and $\beta = 1/k_B T$.
A Galilean transformation relates $\hat{H}_{\v{u}}$ to the Hamiltonian in the lab
frame, $\hat{H}_{\v{u}} = \hat{H} - \v{u}\cdot\hat{\v{P}}+ \frac{1}{2}M u^2$, where $\hat{\v{P}} =
\int_B d^2\vx\; \md(\v{x}) $ is the total momentum, $M = mN$ is the total mass and
$\md(\v{x})$ is the momentum density operator at point $\v{x}$.  The
expectation value for the momentum density in the presence of moving walls is
then given by the expression
\begin{align}
\Exval{\md(\v{x})}_\v{u} 
  &= \Tr[\hat{\rho}\,\md(\v{x})], \\
  &= \frac{\Tr\bkt[\big]{e^{-\beta(\hat{H} - \v{u}\cdot\hat{\v{P}} + (m u^2 / 2 - \mu) \hat{N})}\md(\v{x})}}
  {\Tr\bkt[\big]{e^{-\beta(\hat{H} - \v{u}\cdot\hat{\v{P}} + (m u^2 / 2 - \mu) \hat{N}}}}.
\end{align}
Expanding this expression to first order in $\v{u}$ yields
\begin{equation}
  \Exval{\md(\v{x})}_\v{u} = \Exval{\md(\v{x})} +
  \beta\bkt[\big]{\exval{\md(\v{x}) \v{u}\cdot\hat{\v{P}}}
  - \Exval{\md(\v{x})}\exval{\v{u}\cdot\hat{\v{P}}}},
\end{equation}
where all the expectation values on the right hand side are now taken in the
\emph{equilibrium} ensemble with the walls at rest.  Since $\exval{\md(\v{x})}
= 0$ in our equilibrium ensemble, this simplifies to
\begin{align}
  \Exval{\md(\v{x})}_\v{u} &= \beta\exval[\big]{\md(\v{x}) \hat{\v{P}}} \cdot \v{u}, \\
  &= \beta\int_B d^2\v{x}' \, \Exval{\md(\v{x})\md(\v{x}')} \cdot \v{u}, \label{eqnMomdensRaw}
\end{align}
where $\md(\v{x})\md(\v{x}')$ is a rank-two tensor; the outer product of
$\md(\v{x})$ and $\md(\v{x}')$.  [In two dimensions this means
$\Exval{\md(\v{x})\md(\v{x}')}$ is a 2$\times$2 matrix for each pair of
coordinates $(\vx, \vx')$.]

To make further progress, we wish to take the limit as the system gets very 
large (we will notate this limit as $B \to \infty$).
To this end, we first consider some
properties of the correlation functions in the infinite system.  The infinite
system is homogeneous, which implies that $\Exval{\md(\v{x})\md(\v{x}')}_\infty
= \Exval{\md(\v{x}+\v{r})\md(\v{x}'+\v{r})}_\infty$ for any $\v{r}$, where
$\Exval{\cdot}_\infty$ indicates an average in the infinite system.  As
a consequence, we may express the correlations --- in the infinite system ---
in terms of the Fourier transform in the relative coordinate $\v{x}'-\v{x}$:
\begin{align}
\Exval{\md(\v{x})\md(\v{x}')}_\infty
	&= \Exval{\md(\v{0})\md(\v{x}'-\v{x})}_\infty \\
	&= \frac{1}{(2\pi)^2} \int d^2\v{k} \, e^{i\v{k}\cdot(\v{x}'-\v{x})} \chi(\v{k}),
\end{align}
where all the important features of the correlations are now captured by the tensor
\begin{equation}
  \chi(\v{k}) = \int d^2\v{r} \, e^{-i\v{k}\cdot\v{r}} \Exval{\md(\v{0})\md(\v{r})}_\infty.
  \label{eqn:chiDefinition}
\end{equation}
Because of the isotropy of the fluid in the infinite system, $\chi(\v{k})$
obeys the transformation law $\chi(O\v{k}) = O^{-1}\chi(\v{k})O$, for any
$2\times 2$ orthogonal matrix $O$.  This implies that $\chi$ may be decomposed
into the sum of longitudinal and transverse parts:
\begin{equation}
  \chi(\v{k})
    = \tilde{\v{k}}\tilde{\v{k}} \chi_l(k) + \bkt[\big]{I - \tilde{\v{k}}\tilde{\v{k}}} \chi_t(k)
  \label{eqn:chiDecomp}
\end{equation}
where $\tilde{\v{k}} = \v{k}/k$, $k = \norm{\v{k}}$, $I$ is the identity and 
the juxtaposition of vectors $\tilde{\v{k}}\tilde{\v{k}}$ represents the outer
product as above.  The transverse and longitudinal functions $\chi_t$ and
$\chi_l$ are scalars that depend \emph{only} on the length $k$.

We now return our attention to the finite system.  If the finite box $B$ is
large then the momentum correlations in the bulk will be very similar to the
values for the infinite system.  Therefore, when $\v{x}$ and $\v{x}'$ are far
from the boundaries, we may approximate
\begin{align}
\Exval{\md(\v{x})\md(\v{x}')}
	&\approx \Exval{\md(\v{x})\md(\v{x}')}_\infty \\
	&= \frac{1}{(2\pi)^2} \int d^2\v{k} \, e^{i\v{k}\cdot(\v{x}'-\v{x})} \chi(\v{k})
\end{align}
which in combination with Eq.~\eqref{eqnMomdensRaw} yields
\begin{align}
  \Exval{\md(\v{x})}_\v{u}
    &\approx \beta \int_B d^2\v{x}'\, \frac{1}{(2\pi)^2} \int d^2\v{k} \, 
      e^{i\v{k}\cdot(\v{x}' - \v{x})} \chi(\v{k}) \cdot \v{u} \\
    &= \beta \int d^2\v{k} \, \Delta_B(\v{k}) e^{-i\v{k}\cdot\v{x}} \chi(\v{k}) \cdot \v{u}.
\end{align}
Here we have defined the nascent delta function $\Delta_B(\v{k}) \equiv \frac{1}{(2\pi)^2}
\int_B d^2\v{x}' \, e^{i\v{k}\cdot\v{x}'}$ which has the property
$\Delta_B(\v{k}) \to \delta(\v{k})$ as $B \to \infty$.

We are now in a position to carry out the limiting procedure to increase the
box size to infinity.  However, care must be taken because the simple expression
$\lim_{B\to\infty} \Exval{\md(\v{x})}_\v{u}$ is not well defined without 
further qualification of the limiting process $B\to\infty$.

To resolve this subtlety we must insert a final vital piece of physical reasoning.
Let us assume for simplicity that $\v{u}$ is directed along the $x$-direction,
and the box $B$ is aligned with the $x$ and $y$ axes with dimensions
$L_x{\times}L_y$.  As shown in Fig.~\ref{fig:superfluid_thought_expt}, there
are two possibilities for taking the limits, representing different physical
situations.

On the one hand [Fig.~\ref{fig:superfluid_thought_expt}(b)], we may take the
limit $L_x\to\infty$ first, which gives us an infinitely long channel in which
superfluid can remain stationary while only the normal fraction moves with the
walls in the $x$-direction.  We have
\begin{align}
\rho_n\v{u}
	&= \lim_{L_y\to\infty} \lim_{L_x\to\infty} \Exval{\md(\v{x})}_\v{u} \\
	&= \lim_{L_y\to\infty} \lim_{L_x\to\infty} \beta \int d^2\v{k} \,
	    \Delta_B(\v{k}) e^{-i\v{k}\cdot\v{x}} \chi(\v{k}) \cdot \v{u} \label{sfrac_lim_with_integral} \\
    &= \beta \lim_{k_y\to 0} \lim_{k_x\to 0} \chi(\v{k}) \cdot \v{u}
\end{align}
where we use the fact that $\Delta_B(\v{k})$ can be decomposed into the product
$\Delta_{L_x}(k_x)\Delta_{L_y}(k_y)$ with $\Delta_{L}(k) \to \delta(k)$ as
$L\to\infty$.  Employing the decomposition of $\chi$ given in
Eq.~\eqref{eqn:chiDecomp} allows the density of the normal fraction to be
related to the transverse component of $\chi$ evaluated at zero:
\begin{equation}
	\rho_n = \beta\lim_{k\to 0}\chi_t(k) = \beta \chi_t(0).
\end{equation}

On the other hand [Fig.~\ref{fig:superfluid_thought_expt}(c)] we may take the
limit $L_y\to\infty$ first, resulting in an infinitely long channel --- with
velocity \emph{perpendicular} to the walls --- in which the entire body of the
fluid must move regardless of whether it is a superfluid or not.
In a similar way to the 
previous paragraph, $\rho\v{u} = \beta \lim_{k_x\to \v{0}} \lim_{k_y\to \v{0}}
\chi(\v{k}) \cdot \v{u}$, and making use of the decomposition in
Eq.~\eqref{eqn:chiDecomp} we find that the total density is related to the
longitudinal component of the correlations:
\begin{equation}
	\rho = \beta \lim_{k\to 0} \chi_l(k) = \beta \chi_l(0).
\end{equation}

With these expressions, the normal fraction $f_n$ may finally be expressed
directly as
\begin{equation} \label{eqn:normal_fraction}
  f_n = \rho_n/\rho = \lim_{k\to 0} \chi_t(k) \big/ \lim_{k\to 0} \chi_l(k)
\end{equation}
while the superfluid fraction is $f_s = 1 - f_n$.  Thus, we have expressed the 
superfluid and normal fractions in terms of a correlation function which can be
directly computed in the thermal ensemble; there is no need to deal with
difficult moving boundary conditions or other dynamical perturbations in the
simulation itself.

\section{Numerical procedure}
\label{superfluid_fraction_numerics}
To determine the superfluid fraction from a PGPE simulation, we need to
estimate the tensor of momentum density correlations $\chi$ using simulation
results.  For a finite system constrained to a periodic simulation box --- as
studied in chapter \ref{bkt_chapter} --- we may only compute the momentum
correlations at discrete grid points.  The discrete analogue of
Eq.~\eqref{eqn:chiDefinition} leads to the expression
\begin{equation}
  \chi(\v{k}) \propto \Exval{\v{p}_{\v{k}}\v{p}_{-\v{k}}}
  \label{eqn:chiFinite}
\end{equation}
where the constant of proportionality is not important to the final result, and
$\v{p}_{\v{k}}$ are the discrete Fourier coefficients of $\v{p}(\v{x})$ over 
our simulation box.

The momentum density operator is given by
\begin{equation}
  \md(\v{x}) = \frac{i\hbar}{2}\Bktsq{(\del \hat{\psi}^\dagger(\v{x}) ) \hat{\psi}(\v{x}) - 
  \hat{\psi}^\dagger(\v{x}) \del \hat{\psi}(\v{x}) }
\end{equation}
which may be derived by considering the continuity equation for the number 
density, $\exval[\big]{\hat{\psi}^\dagger(\v{x}) \hat{\psi}(\v{x})}$.  For a 
given classical field Eq.~\eqref{eqn:Cfield}, the Fourier coefficients of 
$\v{p}$ may be written as
\begin{equation}
  \v{p}_{\v{k}} = \frac{\hbar}{2\sqrt{A_B}} \sum_{\v{k}'} (2\v{k}'+\v{k})c^*_{\v{k}'}c^{\phantom{*}}_{\v{k}+\v{k}'},
\end{equation}
where $A_B$ is the area of the system.  Computing a value for all $\v{p}_\v{k}$
at each time step, we then evaluate $\chi(\v{k})$ via the usual ergodic
averaging procedure using Eq.~\eqref{eqn:chiFinite}.

Having evaluated $\chi(\v{k})$, we are left with performing the decomposition 
into longitudinal and transverse parts.  For this, simply note that
Eq.~\eqref{eqn:chiDecomp} implies $\chi_l(k) = \tilde{\v{k}}\cdot
\chi(\v{k}) \cdot \tilde{\v{k}}$, and $\chi_t(k) = \tilde{\v{w}}\cdot \chi(\v{k})
\cdot\tilde{\v{w}}$, where $\tilde{\v{w}}$ is a unit vector perpendicular to
$\tilde{\v{k}}$.

\begin{figure}[htbp]
  \begin{center}
  \includegraphics[width=12cm]{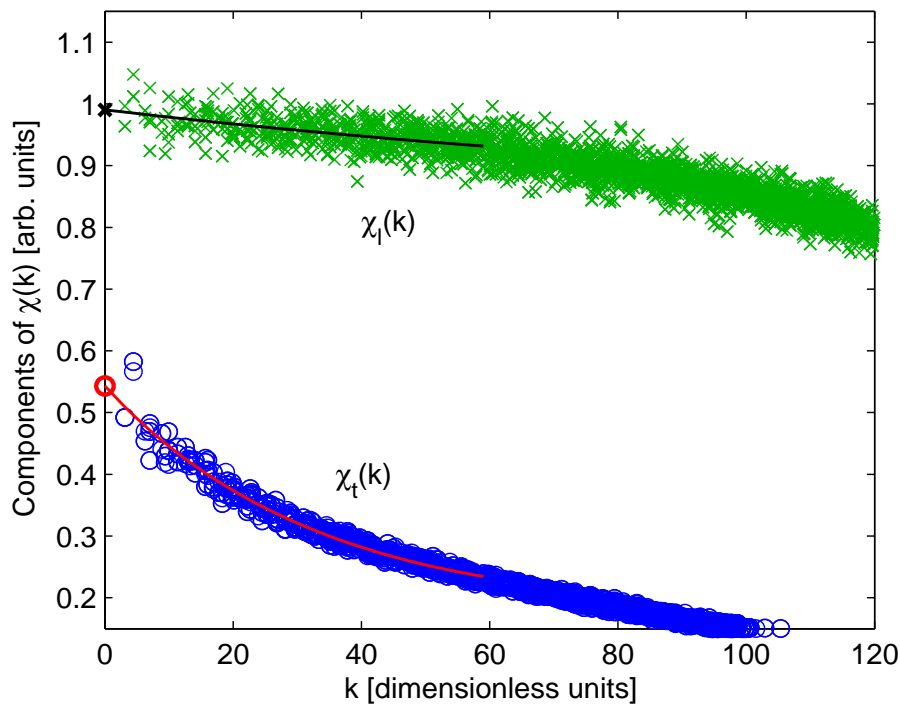}
  \end{center}
  \caption
  [Extrapolation of transverse and longitudinal momentum correlations]
  { \label{fig:sfl_lims_eg}
  Example fitting and extrapolation to $k=0$ for the transverse and longitudinal
  components of the momentum density autocorrelation tensor, $\chi$.  The 
  apparent functional form for $\chi_t$ and $\chi_l$ changes with temperature
  --- particularly near the transition --- which along with the sampling noise
  makes them difficult to fit reliably.  The data shown corresponds to a
  temperature $T \approx 0.99 \Tkt$ slightly below the transition.
  }
\end{figure}

Values for $\chi_t$ and $\chi_l$ may be collected for all angles as a function 
of $k$, and a fitting procedure used to perform the extrapolation $k \to 0$; 
this procedure is illustrated in Fig.~\ref{fig:sfl_lims_eg}.  At low 
temperatures, the extrapolation is quite reliable, but becomes more difficult 
near the superfluid transition where sampling noise increases and $\chi_t(k)$ changes
rapidly near $k=0$.  Without a known functional form, we settled for a
quadratic weighted least squares fit of $\ln(\chi_t)$ and $\ln(\chi_l)$ versus $k$.
A weighting of $1/k$ was used to account for the fact that the density of 
samples of $\chi$ vs $k$ scales proportionally with $k$ due to the square grid 
on which $\chi(\v{k})$ is evaluated.  The logarithm was used to improve the fits
of $\chi_t$ very near the transition where it varies non-quadratically near
$k=0$.  The fitting procedure and extrapolation to $k=0$ generally produces
reasonable results, but is somewhat sensitive to numerical noise.  For this
reason, the computed superfluid fraction at high temperatures is not exactly
zero (see Fig.~\ref{fig:fc_fs_vs_T} on page \pageref{fig:fc_fs_vs_T}).

\section{Discussion}

We have applied the technique described in this chapter to study the BKT phase
in chapter \ref{bkt_chapter}.  This included computing the superfluid fraction
as a function of temperature --- see Fig.~\ref{fig:fc_fs_vs_T} on page
\pageref{fig:fc_fs_vs_T} --- and the 
shape of the curve is consistent with the expectation of a universal jump\footnote{
Note that we do not expect an exact discontinuity in our calculated superfluid 
fraction due to the finite size of the system and the presence of statistical
noise.
}
in the superfluid density, as predicted by Nelson and Kosterlitz
\cite{Nelson1977}.   The location of the BKT transition --- as deduced from the
sharp disappearance of the superfluid fraction --- is also consistent with the
behaviour of the other physical quantities computed in chapter
\ref{bkt_chapter}, including the decay of spatial correlation functions and the
vortex unpairing.

Unlike explicitly constructing a dynamical simulation, our method is based 
on correlations that are calculated in the stationary thermal ensemble.
This is a great practical advantage because it avoids the need for additional 
numerically costly simulations.  It also avoids any concern that the dynamical
perturbation in such simulations might disrupt the thermal background,
resulting in a perturbed measurement of the superfluid fraction.  Having said
this, it would be interesting to compare our results to the superfluid fraction
as deduced from the drag force felt by a moving impurity potential.

We finish by discussing some limitations and possible extensions to our
technique.  The main limitation is that the method appears to rely critically on
translational symmetry; at the very least it is clear that the thought
experiment used in the derivation doesn't make sense for a trapped system.  This
is unfortunate, since all experimental systems are necessarily trapped in some
way or another and any quantitative comparisons with actual experiments must
take this inhomogeneity into account.  So far the superfluid fraction for the
trapped system has been determined by using the universality result for the
critical density in the homogeneous gas \cite{Prokofev2002} in combination with
the local density approximation \cite{Holzmann2008, Bisset2009}.  It would be
interesting to be able to compute the superfluid fraction independently as we
have done here, but it is not obvious how to generalise the derivation to this
case.

In this thesis we have calculated the superfluid fraction only for the 
two-dimensional case.  In 1D the tensor $\chi(\v{k})$ is degenerate so there is no
transverse component available, and the derivation fails to make sense.  There
is no such problem in 3D, and we expect our method to provide useful results in
this case.


%% file: var1d/var1d.tex
\newcommand{\wfa}{\psi}
\newcommand{\wffull}{\phi}

\newcommand{\ccs}{\chi + \chi^*}
\newcommand{\pdz}{\partial_z}

\chapter{Effective 1D equations for trapped Bose-Einstein condensates}
\label{var1d_chapter}

\begin{chap_desc}
  This chapter presents an ansatz for solutions to the 3D GPE in the quasi-1D
  regime.  Our ansatz expresses the full 3D wavefunction in terms of a pair of
  1D complex fields that describe the amplitude and width of an elongated BEC.
  We derive equations of motion for the 1D fields using the Lagrangian
  formalism and solve them numerically for several test scenarios.  We compare 
  with other 1D approaches, and the full 3D solution.
\end{chap_desc}

\section{Introduction}

The Gross-Pitaevskii equation (GPE) described in section \ref{GPE_derivation} 
[Eq.~\eqref{GPE}] is a
remarkably successful tool for describing experimental BEC dynamics
near $T=0$.  The reasons are simple: it is accurate for a large subset of
experiments, and efficient numerical schemes are fairly simple to implement.
Nevertheless, solving the GPE in three dimensions can be computationally
demanding because of the size of the spatial grid required.  In some cases it 
is possible to mitigate this problem by dimensional reduction; there are at 
least two possibilities:  First, we may make use of symmetries in the physical
situation to reduce the number of dimensions.  For example, experimental
systems commonly have cylindrically symmetric trapping potentials; if the
initial state also has cylindrical symmetry we may then eliminate the angle
variable, which reduces the simulation to two dimensions.  Second, there are
experimental situations where the system is strongly confined in one or two
``transverse'' dimensions so that the dynamics in those dimensions is
particularly simple.  We may then make an ansatz for the wavefunction that
allows us to integrate out the transverse directions to produce a lower
dimensional \emph{effective equation} which may be solved on a much smaller
numerical grid.

Elongated, cylindrically symmetric cigar-shaped BECs are routinely produced
in the laboratory using a variety of trapping techniques including optical
lattices \cite{Greiner2001}, atom chips \cite{Hansel2001,Leanhardt2002,Folman2002,Schneider2003} and
other types of magnetic traps as first achieved by Görlitz \emph{et
al.}~\cite{Gorlitz2001}.  In parallel, several studies
(see, for example, \cite{Dutton2001,Simula2005a,Hoefer2006,Chang2008,Meppelink2009}) have
investigated the dynamics of dispersive shock waves\footnote{
A \emph{dispersive} shock wave occurs when dispersion rather than dissipation
dominates the physics at small length scales.  Dispersive shock waves are
characterised by pulse trains as seen in the simulations of
Ref.~\cite{Meppelink2009}, for example.
} in BEC.  Refs.~\cite{Chang2008,Meppelink2009} combined both one dimensionality 
and shock wave generation by rapidly applying or removing a perturbing
potential at the centre of a cigar-shaped BEC.  In both of these studies the
resulting shock waves were modelled numerically using a GPE or GPE-like 
mean-field approach.  The calculations in Ref.~\cite{Meppelink2009} required a high
spatial resolution in the longitudinal direction due to both the large size of
the condensate and the need to resolve shock fronts at small length scales.  As
a result, these calculations were restricted to using a 1D effective equation
for efficiency.  The method chosen was the non-polynomial Schrödinger
equation (NPSE) of Salasnich \textit{et al.}~\cite{Salasnich2002}, one of
several methods that adiabatically eliminate the transverse motion and assume
the shape of the transverse profile varies slowly as a function of the
longitudinal coordinate.

While the NPSE is convenient, it is not obvious that the assumptions used in 
its derivation are valid when dealing with shocks.  In particular, the
existence of a shock implies rapid variation of the density and other system
parameters with the longitudinal coordinate, and we would expect to this to carry
over to the shape of the transverse profile.  In this chapter we relax both of
the assumptions which go into deriving the NPSE --- we include both the
variation in the longitudinal direction and avoid making the adiabatic
approximation.

An additional motivation for this work was to derive a 1D effective equation
capable of simulating the expansion dynamics after turning off the trapping
potential.  These type of expansions are the standard experimental tool for
imaging condensates, but simulating them directly is difficult due to the large
size of the spatial grid required.  The alternative 1D ansätze discussed in
the next section eliminate the transverse velocity either implicitly or
explicitly, and are therefore fundamentally unable to deal with expansion.

\subsection{Previous work}

We consider quasi-1D systems where the $x$ and $y$ coordinates correspond to
the tightly confined transverse directions; the $z$ coordinate is the
weakly-trapped longitudinal direction.  We assume that the transverse trapping
potential is harmonic and cylindrically symmetric, while the longitudinal 
potential $V_z$ is generic.  The full potential is
\begin{equation}
	V(\vx) = V_z(z) + \frac{m\omega_\perp^2}{2} r^2,
\end{equation}
where $r=\sqrt{x^2+y^2}$ is the radial coordinate, $\vx=(x,y,z)$, $m$ is the
atomic mass and $\omega_\perp$ the angular frequency of the transverse harmonic
potential.

In the quasi-1D regime, $\omega_\perp$ is large enough to prevent
significant excitation of the transverse degrees of freedom.  The simplest 
effective 1D equation may be derived by considering the case without
interactions, $U_0=0$.  In this 
case the full 3D wavefunction $\wffull$ factorises and the transverse component
has the functional form
$\frac{1}{\sqrt{\pi}\sigma}e^{-r^2 / 2\sigma^2}$,
where $\sigma$ is the transverse width.  In the perturbative regime\footnote{
The perturbative regime is defined by $a_s n_1 \ll 1$ where $n_1$ is the
integrated 1D density --- see, for example, Ref.~\cite{Mateo2008}.
}
where $U_0$ is small but nonzero, this suggests the simple ansatz
\begin{equation} \label{eq:simple_ansatz}
	\wffull(\vx,t) = \wfa_1(z,t) \frac{1}{\sqrt{\pi}\sigma} e^{-r^2 / 2\sigma^2},
\end{equation}
with $\sigma$ taken equal to the width of the non-interacting ground state.
This ansatz works fairly well when $U_0$ is small and leads to the 1D GPE 
which describes the evolution of the unknown function $\wfa_1$:
\begin{equation}\label{1D_GPE}
    i\hbar\partial_t\wfa_1 =
        -\frac{\hbar^2}{2m}\pdz^2 \wfa_1 + V\wfa_1 +
        U_\mathrm{1D}\abs{\wfa_1}^2\wfa_1,
\end{equation}
where $U_\mathrm{1D}$ is the effective 1D nonlinearity constant.  For
larger $U_0$ the transverse width increases substantially and the 1D GPE
becomes rather inaccurate unless the value of $\sigma$ is adjusted accordingly.
An appropriate value may be computed using a variational calculation, assuming
a constant density in $z$.

More sophisticated one-dimensional approximations have been investigated by
several authors; we review those that are relevant to the current work below.
An attempt has been made to keep to the notation used in the original papers,
with some modifications for consistency.

The first of the
more sophisticated approximations was the NPSE, as described by Salasnich
\textit{et~al.}\ \cite{Salasnich2002} in 2002.  They used the ansatz
\begin{equation}\label{NPSE_ansatz}
\wffull(\vx,t) = \frac{1}{\sqrt{\pi}\sigma(z,t)}
                e^{-r^2/2\sigma^2(z,t)} f(z,t),
\end{equation}
along with the assumption that the width $\sigma$ changes sufficiently slowly
in the longitudinal direction that $\pdz\sigma$ is negligible.  The equation
for $f$ which results is known as the nonpolynomial Schrödinger equation due
to the nonpolynomial nonlinear term that arises after eliminating the
transverse width $\sigma$:
\begin{equation}\label{NPSE}
    i\hbar \,\partial_t f =
        -\frac{\hbar^2}{2m}\pdz^2 f + V_z f +
        \Bktsq{
        \frac{U_0}{2\pi a_\perp^2} \frac{\abs{f}^2}{\sqrt{1 + 2a_s\abs{f}^2}} +
        \frac{\hbar\omega_\perp}{2} \Bkt{
        	\frac{1}{\sqrt{1 + 2a_s\abs{f}^2}} +
			\sqrt{1 + 2a_s\abs{f}^2}
		}
		} f,
\end{equation}
where $a_\perp=\sqrt{\hbar/m\omega_\perp}$ is the transverse length scale.
Implicit in the ansatz from Eq.~\eqref{NPSE_ansatz} is the assumption that transverse
dynamics given by changes in $\sigma$ are much faster than the dynamics of $f$
in which we are interested.  To see this, note that the phase of $\wffull$ does
not depend on the coordinate $r$, which implies the absence of a radial
superfluid velocity --- the ansatz does not support transverse dynamics
independently of the field $f$.

This lack of transverse dynamics was addressed more carefully by Muñoz Mateo
and Delgado \cite{Mateo2008,Mateo2009}, who showed that the general ansatz
\begin{equation}
\wffull(\vx,t) = \varphi(r,n_1(z,t)) \,\phi_1(z,t)
\end{equation}
may be used to derive an equation for $\phi_1$ via the adiabatic elimination of the 
transverse degrees of freedom.  Here $\varphi$ is some family of radial
wavefunctions, parametrised by the local 1D condensate density $n_1$.  By
assuming that $\pdz n_1 = 0$, making the adiabatic approximation,
and integrating out the transverse direction, they obtained the remarkably
simple equation
\begin{equation}\label{MDE}
	i\hbar \,\partial_t \phi_1 =
		-\frac{\hbar^2}{2m}\pdz^2 \phi_1 + V_z \phi_1 + \mu_\perp(n_1) \phi_1,
\end{equation}
where
\begin{equation}
	\mu_\perp(n_1) =
		\iint dx\,dy\; \varphi^* \Bkt{-\frac{\hbar^2}{2m}\del_\perp^2 + 
			\frac{m\omega_\perp^2}{2}r^2 + U_0 n_1 \abs{\varphi}^2} \varphi
\end{equation}
is the local chemical potential.  (We will refer to Eq.~\eqref{MDE} as the
Muñoz Mateo-Delgado equation (MDE) from now on.)
Choosing the formula $\mu_\perp(n_1) = \hbar\omega_\perp \sqrt{1 + 4a_s n_1}$
to interpolate between known limits at large and small $U_0$, they were able to
obtain very accurate predictions of in-trap oscillations with a range of smooth 
initial conditions, when benchmarked against a full 3D calculation
\cite{Mateo2008}.

When the transverse dynamics are important or the condensate width varies
rapidly with the longitudinal coordinate $z$, we expect approaches based on the
adiabatic approximation to become less accurate.  Kamchatnov \textit{et
al.}~\cite{Kamchatnov2004} investigated a different generalisation of the NPSE
ansatz,
\begin{equation}
\wffull(\vx,t) = \frac{1}{\sqrt{\pi}b(z,t)}
		e^{-r^2 / 2b^2(z,t)} e^{(i/2)\alpha(z,t)r^2} \tilde{\psi}(z,t),
\end{equation}
using a variational formalism.  The added generality and lack of assumptions
regarding the $z$ derivatives of the fields were motivated by the desire to 
describe solitons involving short length scales.  The Kamchatnov ansatz
results in coupled equations\footnote{
We refer the reader to Ref.~\cite{Kamchatnov2004} for details of the
equations --- they are not used further here, and are reasonably complex to
write down.
}
for the two real fields $b(z,t)$, $\alpha(z,t)$ and one complex field
$\tilde{\psi}(z,t)$, which were used to analytically investigate solitonic
solutions in various limits, in addition to small amplitude linear waves.  An
important feature is the inclusion of the phase factor $\alpha$, which allows
for ``breathing mode'' dynamics in the transverse direction.

In this chapter we introduce a variation of the Kamchatnov approach, which ---
while mathematically equivalent --- leads to more compact equations.  In
contrast to the earlier work, we investigate general numerical solutions.  We
show cases where the solutions are accurate, and highlight a number of generic
difficulties which will arise for any similar factorisation ansatz.

\section{Formalism}

\subsection{The Ansatz}

We consider the following approximation to the 3D wavefunction:
\begin{equation} \label{eq:DGA}
\wffull(\vx,t) = \wfa(z,t) e^{\chi(z,t) r^2}.
\end{equation}
Here both $\wfa$ and $\chi$ are complex time-varying fields in one spatial
dimension.  The field $\chi$ is related to the transverse Gaussian width and
transverse superfluid speed, respectively, via
\begin{equation}\label{chi_interpretation}
\sigma(z) = 1/\sqrt{-2\Re\chi(z)}
\quad\text{and}\quad
v_\perp(z,r) = 2\abs{\Im\chi(z)}r.
\end{equation}
The field $\wfa$ contains density and longitudinal phase information along with
a normalisation factor for $\chi$; with this ansatz, the 1D density has the
form
\begin{equation} \label{eq:1Ddens}
n(z,t) = \iint dx\; dy\; \abs{\wffull(\vx,t)}^2 =
	\frac{\pi\Abs{\wfa(z,t)}^2}{-[\chi(z,t)+\chi^*(z,t)]}.
\end{equation}
In what follows we will often suppress time and space arguments to avoid 
excessive notational clutter.

\subsection{Equations of motion}
\label{var1d_eqns_of_motion}

To derive the equations of motion we use a time dependent variational 
formalism essentially the same as that described in section
\ref{DF_variational_principle}.  We begin by computing a Lagrangian for
the one dimensional effective theory by integrating out the transverse
dimensions.  Taking derivatives of the effective Lagrangian and putting these
into the Euler-Lagrange equations then yields the equations of motion in the
usual manner.

One possible Lagrangian\footnote{
This form of the Lagrangian is manifestly symmetrical with respect to $\wffull$ 
and $\wffull^*$ but we could equally well have chosen a form more similar to
that used in section \ref{DF_variational_principle}.
}
for the three-dimensional classical $\Abs{\wffull}^4$ field theory is
\begin{equation}\label{full_GPE_lagrangian}
\Lcal[\wffull] = \int d\vx\; \frac{i\hbar}{2} \Bkt{\wffull^*\dot{\wffull} - \dot{\wffull}^*\wffull} - \Hcal[\wffull],
\end{equation}
where the Hamiltonian functional is given by
\begin{equation}
\Hcal[\wffull] = \int d\vx\; \Bkt{\frac{\hbar^2}{2m} \Abs{\del \wffull}^2 + V\Abs{\wffull}^2 + \frac{U_0}{2} \Abs{\wffull}^4}.
\end{equation}
Differentiating $\Lcal$ with respect to $\wffull^*$ yields the 3D GPE via
the Euler-Lagrange equations.  If instead we substitute our ansatz from
Eq.~\eqref{eq:DGA} into Eq.~\eqref{full_GPE_lagrangian} we may perform the Gaussian
integrals over the transverse coordinates, $x$ and $y$.  We then obtain a
reduced Lagrangian which is a function of the fields $\wfa$ and $\chi$:
\begin{multline} \label{eq:DGA_L}
\Lcal[\wfa,\chi] = \pi \int dz\; \Bigg\{
\frac{i\hbar}{2} \bktsq[\bigg]{
	\frac{\dot{\chi} - \dot{\chi}^*}{(\ccs)^2} \Abs{\wfa}^2
	- \frac{\wfa^*\dot{\wfa} - \wfa\dot{\wfa}^*}{\ccs}
	}
	\\
- \frac{\hbar^2}{2m}\bigg[
	\frac{4 \abs{\chi}^2 \abs{\wfa}^2}{(\ccs)^2}
	- \frac{2 \abs{\pdz\chi}^2 \abs{\wfa}^2}{(\ccs)^3}
	- \frac{\abs{\pdz\wfa}^2}{\ccs}
	+ \frac{\wfa\pdz\wfa^*\pdz\chi + \wfa^*\pdz\wfa\pdz\chi^*}{(\ccs)^2}
	\bigg]
	\\
+ \frac{V_z\abs{\wfa}^2}{\ccs}
- \frac{m\omega_\perp^2\abs{\wfa}^2}{2(\ccs)}
+ \frac{U_0\abs{\wfa}^4}{4(\ccs)} \Bigg\}.
\end{multline}

As usual, stationarity of the action implies the Euler-Lagrange equations for
our two fields, which are
\begin{equation}
	\deriv{}{t}\Bkt{\fderiv{\Lcal}{\dot{\chi}^*}} = \fderiv{\Lcal}{\chi^*} \quad\text{and}\quad
	\deriv{}{t}\Bkt{\fderiv{\Lcal}{\dot{\wfa}^*}} = \fderiv{\Lcal}{\wfa^*}.
\end{equation}
Performing the functional derivatives (see appendix \ref{funcderivs_appendix}),
rearranging and simplifying the results gives the equations of motion for
$\chi$ and $\wfa$:
\begin{align} \begin{split} \label{eq:DGA_dynamics}
i\hbar \dot{\chi} &= \frac{\hbar^2}{2m} \Bktsq{
	-\pdz^2\chi
	- 2\frac{\pdz\wfa}{\wfa}\pdz\chi
	+ 4\frac{(\pdz\chi)^2}{\ccs}
	- 4\chi^2
}
+ \frac{m\omega_\perp^2}{2}
+ \frac{U_0}{4}\abs{\wfa}^2(\ccs), \\
i\hbar \dot{\wfa} &= \frac{\hbar^2}{2m} \Bktsq{
	- \pdz^2\wfa
	+ \frac{2(\pdz\chi)^2\wfa}{(\ccs)^2}
	- 4\wfa\chi
}
+ V_z \wfa
+ \frac{3U_0}{4} \abs{\wfa}^2 \wfa.
\end{split}\end{align}

We highlight two relevant features of the equations above that may not be
immediately obvious: First, Eqs.~\eqref{eq:DGA_dynamics} are energy conserving
due to the time independence of the Lagrangian\footnote{
The sophisticated way to say this is that the time translation symmetry of the
Lagrangian implies energy conservation via Noether's theorem.
}.  Second, our choice of generalised coordinates in Eq.~\eqref{eq:DGA} forces 
us to give up the Hamiltonian structure of the phase space in exchange for a
compact parametrisation of the wavefunction.  This is not physically 
problematic, but does affect our choice of numerical methods.

\subsection{Conservation of normalisation and conserved current}
An important property of any low-dimensional effective equation arising from
the GPE is the conservation of normalisation of the wavefunction; this
corresponds to conservation of the total number of atoms during time evolution.
The total normalisation is
\begin{equation}
N = \int d\vx\; \Abs{\wffull(\vx)}^2 = \int dz \; \frac{\pi\abs{\wfa}^2}{-(\ccs)}.
\end{equation}

In principle, $N$ is a function of time; to prove that it is not and the total
number is conserved, we show that $\dot{N} = 0$.  Expanding $\dot{N}$ and
inserting the equations of motion to remove the resulting factors of
$\dot{\wfa}$ and $\dot{\chi}$, we have:
\begin{align}
\dot{N} &= -\pi \int dz\; \Bktsq{\frac{\dot{\wfa}\wfa^* + \dot{\wfa}^*\wfa}{\ccs}
	- \frac{\abs{\wfa}^2(\dot{\chi} + \dot{\chi}^*)}{(\ccs)^2}} \\
&= \frac{i\hbar\pi}{2m} \int dz \Bktsq{ \Bkt{
	- \frac{\wfa^*\pdz^2\wfa}{\ccs}
	+ \frac{\abs{\wfa}^2\pdz^2\chi}{(\ccs)^2}
	+ \frac{2\wfa^*\pdz\wfa\pdz\chi}{(\ccs)^2}
	- \frac{2\abs{\wfa}^2(\pdz\chi)^2}{(\ccs)^3}
} - \text{c.c.} }
\end{align}
where c.c.\ stands for the complex conjugate of the preceding bracketed term.
Finally, we may remove all second order derivatives using integration by parts,
assuming that the boundary terms are zero\footnote{In a periodic system the
individual boundary terms are generally nonzero but nevertheless correctly
cancel due to periodicity.
}.
The resulting terms cancel out, implying that $\dot{N} = 0$ and normalisation
is conserved.

Conservation of normalisation suggests that there should also be a conserved 
one-dimensional current $j(z,t)$, obeying the equation
\begin{equation}
	\dot{n}(z,t) = -\pdz j(z,t).
\end{equation}
Computing $\dot{n}$ directly from the definition in Eq.~\eqref{eq:1Ddens}
and rearranging shows that the appropriate conserved current may be written
\begin{equation}
j = \frac{\hbar}{2mi}\Bktsq{
    \pi\Bkt{\frac{\abs{\wfa}^2\pdz\chi}{(\ccs)^2} - \frac{\wfa^*\pdz\wfa}{\ccs}}
    - \text{c.c.} }.
\end{equation}
If we define the 1D wavefunction $\eta = \wfa \sqrt{-\pi/(\ccs)}$ with the
properties $n(z,t) = \abs{\eta(z,t)}^2$ and $\arg\Bktsq{\wffull(0,0,z,t)} =
\arg\Bktsq{\eta(z,t)}$, the current takes on a somewhat more familiar form
\begin{equation}
j = \frac{\hbar}{2mi}\Bkt{
	\eta^* \pdz\eta - \eta\pdz\eta^*
	+ \abs{\eta}^2 \frac{\pdz(\chi^*-\chi)}{\ccs} }.
\end{equation}
The usual conserved probability current for quantum mechanics consists of the
first two terms in the expression for $j$.  The extra term is a consequence of 
integrating out the radial structure.

\subsection{Ground states}

We make use of ground states as physically reasonable initial conditions for
the numerical simulations in the next section.  Here we describe the ground 
state equations to be solved, along with the simplest approximate solution.

Ground states of Eqs.~\eqref{eq:DGA_dynamics} have no transverse
dynamics, that is, $\dot{\chi} = 0$.  Further, the transverse speed $v_\perp$
is zero which implies $\Im\chi = 0$.  The time evolution of $\wfa$ is the
simple phase rotation $\dot{\wfa} = (-i\mu/\hbar) \wfa$ for some chemical 
potential $\mu$, and we may assume for simplicity that $\wfa$ is real at $t=0$.
With these observations, we see that the ground state obeys the
time-independent system of equations
\begin{align} \begin{split} \label{eq:DGA_gs}
0 &= \frac{\hbar^2}{2m} \Bktsq{
	-\pdz^2\chi
	- 2\frac{\pdz\wfa}{\wfa}\pdz\chi
	+ 2\frac{(\pdz\chi)^2}{\chi}
	- 4\chi^2
}
+ \frac{A}{2}
+ \frac{U_0}{2}\abs{\wfa}^2\chi, \\
0 &= \frac{\hbar^2}{2m} \Bktsq{
	- \pdz^2\wfa
	+ \frac{(\pdz\chi)^2\wfa}{2\chi^2}
	- 4\wfa\chi
}
+ (V_z - \mu) \wfa
+ \frac{3U_0}{4} \abs{\wfa}^2 \wfa,
\end{split}\end{align}
where both $\chi$ and $\wfa$ are real.

In cases where the local terms dominate in Eqs.~\eqref{eq:DGA_gs} --- slow 
spatial variation, large interactions, or large densities --- we may ignore the
$z$ derivative terms.  This is equivalent to using $\chi$ and $\wfa$ as
determined from a spatially homogeneous system with the same local density, and
is therefore known as a local density approximation (LDA).

Setting $\pdz\wfa = 0$ and $\pdz\chi = 0$, Eqs.~\eqref{eq:DGA_gs} reduce
to
\begin{equation}\label{LDA_state}\begin{split}
    -\frac{2\hbar^2}{3m} \chi^2 + \frac{2}{3}(\mu-V)\chi + \frac{A}{2} = 0, \\
    \abs{\wfa}^2 = \frac{4}{3U_0} \Bktsq{\frac{2\hbar^2}{m}\chi + (\mu - V)}.
\end{split}\end{equation}
These algebraic equations are easily solved for $\chi$ and $\wfa$.  We take the
branch of the square root such that $\chi<0$ so that the transverse width 
in Eq.~\eqref{chi_interpretation} is a real number.  We note that
Eqs.~\eqref{LDA_state} are analogous to the well known Thomas-Fermi
approximation for the GPE ground state.

%

\section{Numerical Simulations}


In this section we present numerical results for the simulation of 
Eqs.~\eqref{eq:DGA_dynamics} along with comparisons to full 3D simulations and other
proposed effective 1D equations.  Our 1D equations are discretised on a uniform 
grid with periodic boundary conditions; derivatives are calculated spectrally
via the Fourier transform.  We use the standard fourth order Runge-Kutta
integrator for time stepping.  For the numerical simulations we use units
such that $\hbar = m = 1$, and have arbitrarily chosen $\omega_\perp = 10$ and a
periodic domain $-8 \le z < 8$ with wavefunction normalisation $N = 1000$.
With these choices, nonlinear behaviour becomes strongly apparent at values of 
the interaction strength $U_0\sim 1$.


The full 3D system was assumed to be cylindrically symmetric, allowing a
reduction to a two-dimensional equation.  We simulated this 2D equation using
the XMDS2 software \cite{XMDS2} with a Bessel Fourier basis.  Time stepping was 
achieved using an adaptive fourth/fifth order Runge-Kutta solver\footnote{named
\texttt{ARK45} in the XMDS software} for time stepping.

\subsection{Ground states}


A common approach for finding the ground state of the GPE is to minimise the
Hamiltonian via continuous-time steepest-descent optimisation, the so-called
``imaginary time'' method\footnote{We note that for the equations
presented here, steepest-descent minimisation is \emph{not} equivalent to
replacing the time $t$ with $i\tau$ and evolving in $\tau$.  This is because
the equations of motion are not Hamiltonian after the transformation
in Eq.~\eqref{eq:DGA}.\label{Hamiltonian_footnote}
}.  Unfortunately, this method does not always work well for the
equations presented here, because the energy depends only weakly on the 
transverse degrees of freedom $\chi$ in regions of low density.  This means
that convergence of $\chi$ to the ground state value can be extremely slow in 
regions of high potential $V_z$.

To avoid this problem, we solve a discretised version of the ground state
differential equations \eqref{eq:DGA_gs} directly using Newton's method.  The
spatial discretisation and derivative calculation used here is the same as for
the dynamical equations.  The desired normalisation for the wavefunction is 
obtained by treating the chemical potential as one of the unknowns and
adjoining the normalisation condition to the set of equations.

Convergence of Newton's method for this system requires a starting point which
is quite close to the true solution.  For this we make use of the LDA solution
Eq.~\eqref{LDA_state}.  We find
that the presence of low density regions where the LDA gives $\wfa = 0$ results
in Newton's method failing with a singular Jacobian.  A simple method to avoid
this problem is to smooth the LDA solution with a spatial filter with
exponentially decaying tails.

We also needed to find ground states for the NPSE, MDE and 3D equations for
comparison purposes.  The imaginary time method is sufficient in these cases
because the equations arise directly from 3D or effective 1D Hamiltonians,
albeit with sometimes unusual nonlinearity terms.  To satisfy a given
normalisation we use a slightly unusual version of the imaginary time method
where the chemical potential $\mu$ is treated as an unknown and adjusted
continuously along with the wavefunction (see appendix \ref{itime_mu_method} 
for details).

\subsection{Test case: soliton formation}


\label{waveguide_sim}
As a test case, we consider evolution in a circular waveguide ($V_z = 0$) after
releasing the ground state of the barrier potential $V_z =
C\exp(-10 z^2)$ at time $t = 0$.  The barrier intensity $C=50$ is chosen
such that the density inside the barrier is depressed to few percent of the
background density to induce strongly nonlinear evolution.

\begin{figure}[p]
\centering
\includegraphics[height=20cm]{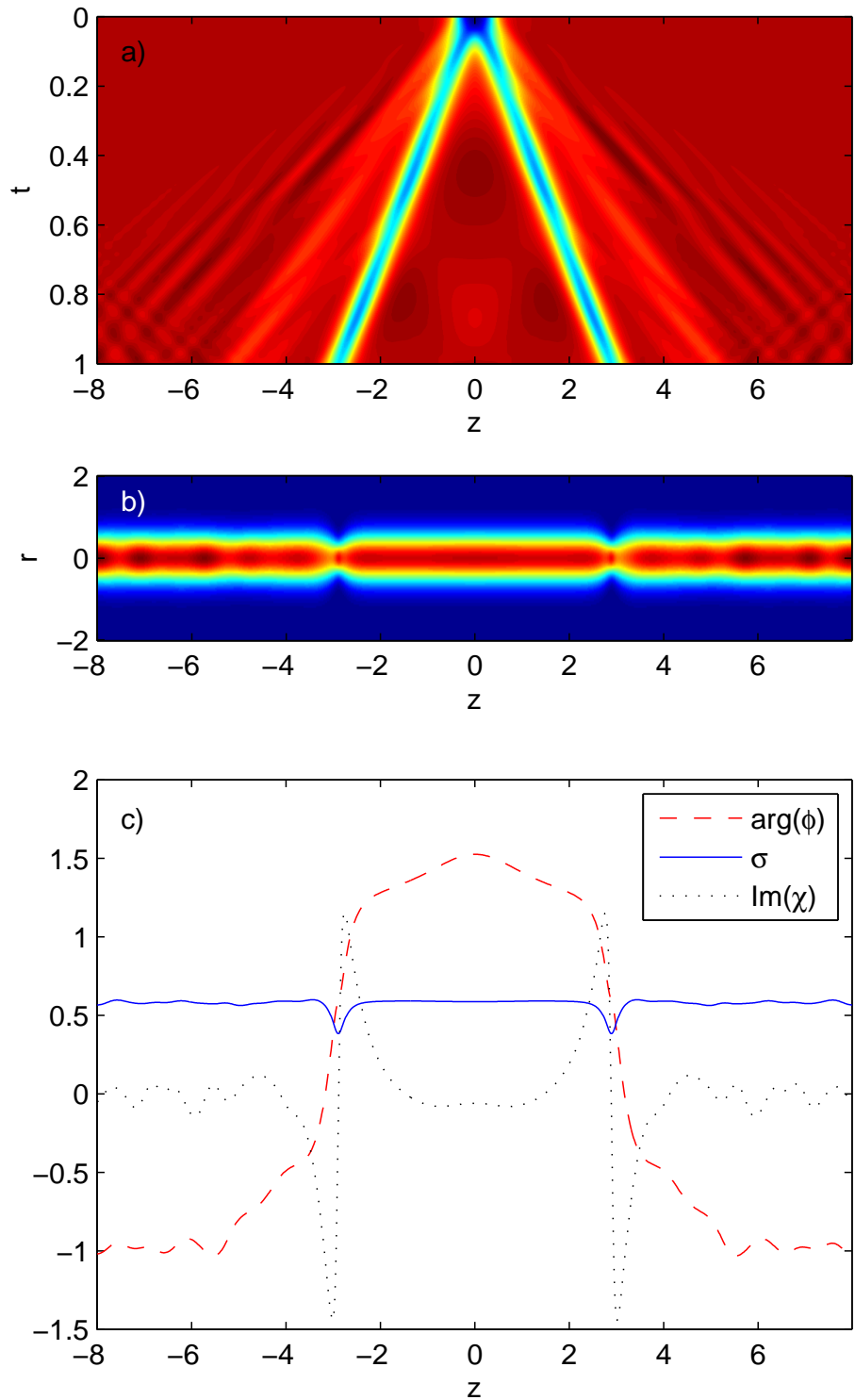}
\caption[Solitons on a constant density background]
{Solitons on a constant density background.  The initial condition is 
the ground state of a Gaussian barrier potential.  (a) Spatial density
as a function of time. (b) Density slice through the centre of the inferred 3D 
cloud at $t=1$. (c) Detail of the fields at $t=1$ --- dashes indicate the phase 
profile, $\arg(\wfa)$, the solid line is the width of the Gaussian, $\sigma$, and
the dotted line shows the scaling factor for the radial velocity, $\Im\chi$.
}
\label{dga_solitons}
\end{figure}

After finding the ground state, the barrier is turned off and the atoms fill 
the resulting hole.  For sufficiently large interaction strength $U_0$, the
excitations are carried away as a combination of sound waves and a pair of grey
solitons form, as shown in Fig.~\ref{dga_solitons}(a).  The associated phase
profile presented in Fig.~\ref{dga_solitons}(c) shows the expected jump in
phase for a soliton across the density depression at $t=1$.  In addition there
is a decrease in the width and associated radial flow.  


\begin{figure}[p]
\centering
\includegraphics[height=21cm]{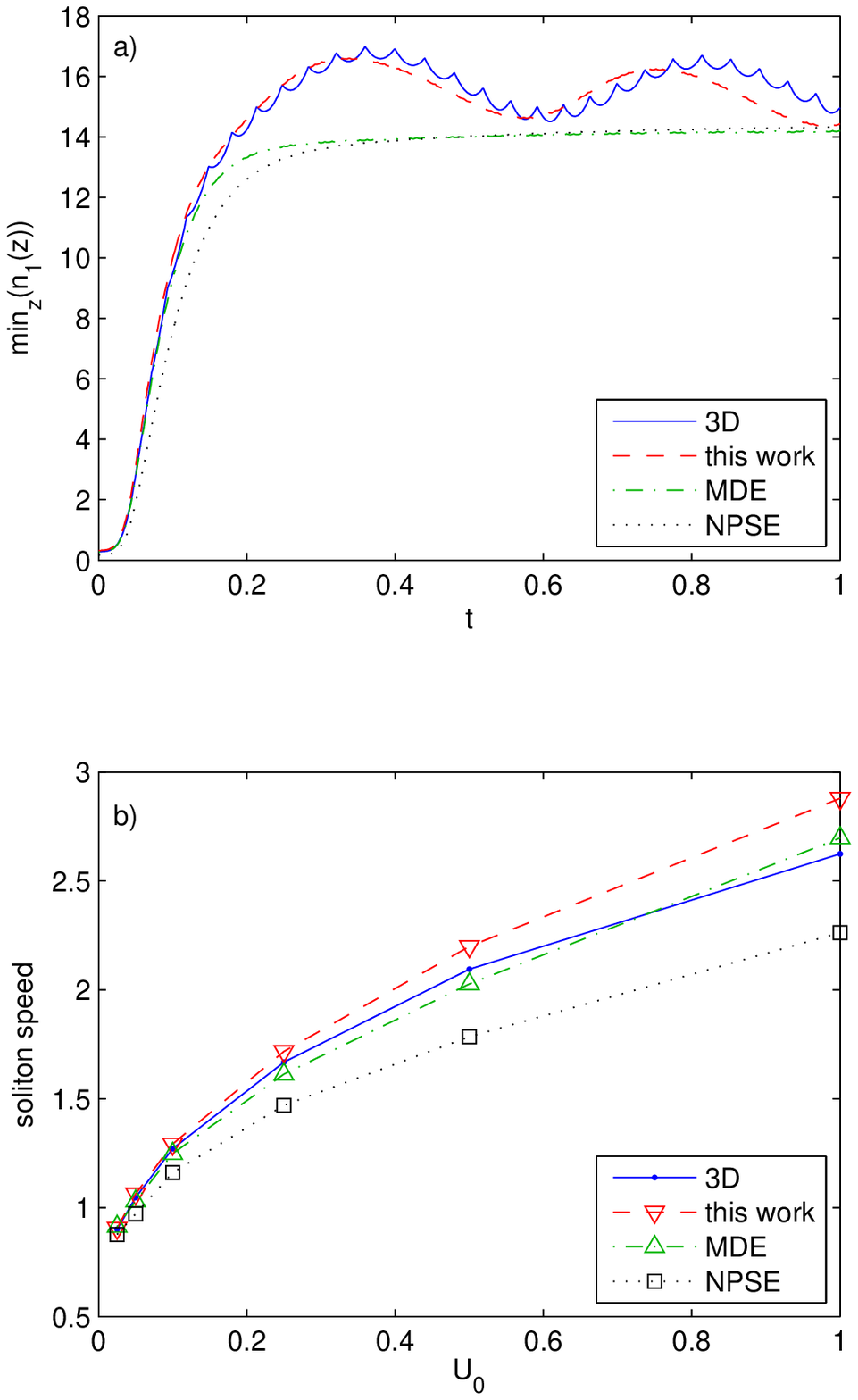}
\caption[Comparison of simulation methods for the soliton waveguide problem]
{
Comparison between various simulation methods for the soliton
waveguide problem of section \ref{waveguide_sim}.  (a) The soliton depth as a 
function of time for $U_0=0.25$.  The high frequency oscillations in the 3D 
solution correspond to a small excitation of the higher order transverse modes.
(b) The soliton speed as a function of the interaction strength $U_0$.
}
\label{soliton_comparison}
\end{figure}

To evaluate the 1D model we compare with the full 3D results, and also to two
alternative effective 1D equations: the NPSE and the MDE.  A comparison of the 
soliton speed as a function of interaction strength is shown in 
Fig.~\ref{soliton_comparison}(b).  We see that our equations and the MDE are
competitive over most of the range, however the MDE wins out at larger
interaction strengths.  Of note is the NPSE results that are significantly
less accurate than our equations, even though the same transverse Gaussian
ansatz is used.

We note that the MDE has a particular advantage in the reproduction of accurate
ground states: the interpolating form used for the local chemical potential
accurately represents the important properties of the transverse wavefunction.
In contrast, both methods which assume a transverse Gaussian profile are less
accurate when the transverse wavefunction is deformed due to the interaction
energy.  Nevertheless, our equations are significantly more accurate than the
NPSE, which indicates that derivatives of the transverse wavefunction
parameters have a measurable effect for these initial conditions.

Further examination of Fig.~\ref{dga_solitons}(a) shows an apparent
oscillation in the soliton depth coupled with the emission of sound waves.
This feature is not present in the MDE or NPSE simulations but is present
in the full 3D results.  To quantify the effect,
Fig.~\ref{soliton_comparison}(a) shows the soliton depth as a function of
time.  We see that the initial condition causes an oscillation in the
transverse width which feeds back into the soliton depth in this case.  While
the width cannot be observed after integrating out the transverse direction,
the soliton depth can.

\subsection{Transverse shock wave formation}


We have presented a case above where our equations are competitive with other
effective 1D equations for modelling the full 3D dynamics.
Nevertheless, we have found recurring stability problems over the course of
many numerical experiments with varying initial conditions.  Such instability
is typically characterised by sharp shock-like features that initially develop
in the transverse width $\chi$.  We observed that the shocks are often
associated with the density passing close to zero, and inspecting the equation
of motion for $\chi$ in Eqs.~\eqref{eq:DGA_dynamics} suggests that the term
containing $1/\wfa$ might be responsible: When $\wfa$ is very small this term
becomes very large, potentially resulting in stiff equations.  Explicit schemes
such as the classic fourth order Runge-Kutta method become unstable in the
presence of stiffness unless an unmanageably small time step is used.

We tested this hypothesis by making use of a Matlab implementation
\cite{Engstler2010} of the RADAU5 algorithm \cite{Hairer1996}.  RADAU5 is a
fifth order instance of the Radau IIA class of implicit Runge-Kutta algorithms,
and is known for its exceptional stability properties \cite{Hairer1996}.
Nevertheless we found the results to be inconclusive: while it seemed to help
for some initial conditions, there were certainly cases where the solutions
diverged regardless of the type of numerical integrator used.

After further investigation, we believe we have identified the underlying
cause of these problems: The ideal transverse width is not always a continuous
function of the transverse slices of the 3D wavefunction.  To make matters
worse, discontinuities develop dynamically and it is not generally obvious
which initial conditions will lead to problems.

\begin{figure}[htbp]
\centering
\includegraphics[width=14cm]{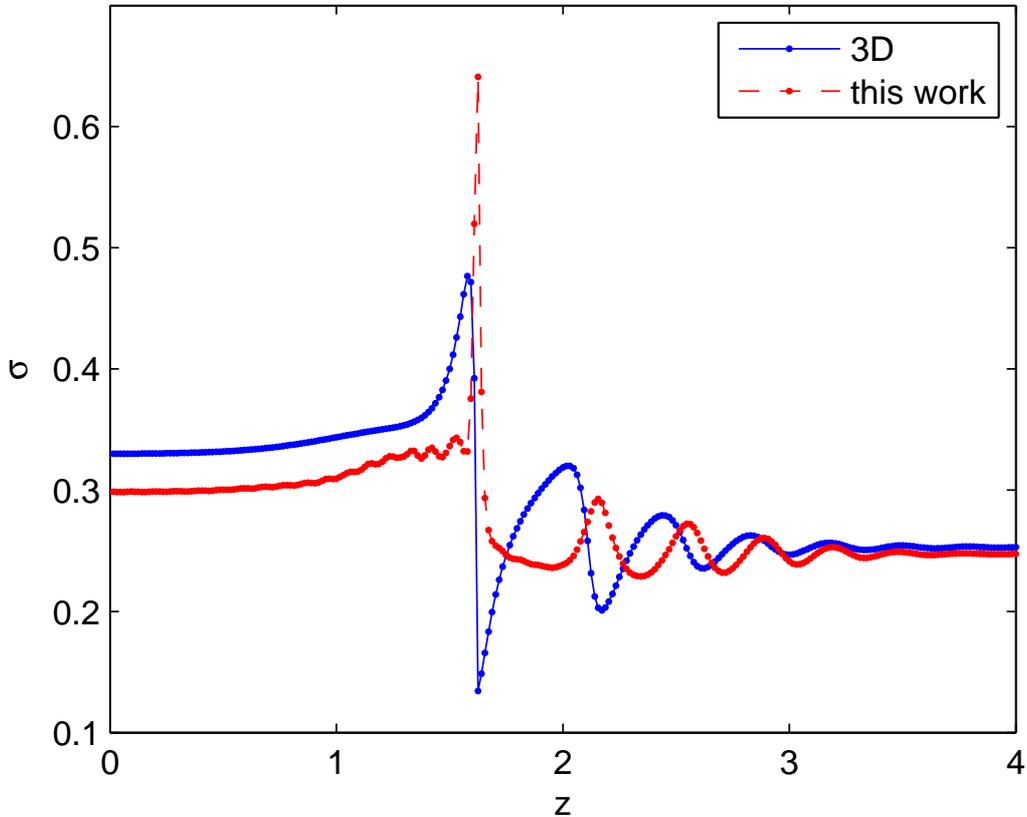}
\caption[Discontinuities in the transverse Gaussian width]
{Transverse Gaussian width (standard deviation $\sigma$) of the wavefunction
density from our ansatz, compared to a 3D simulation at time $t=0.17$.  
This is a zoomed view in the $z$ axis, clearly showing the discontinuities at 
$z\approx 1.6$.
}
\label{waveguide_sigma_blowup}
\end{figure}

\begin{figure}[htbp]
\centering
\includegraphics[width=14cm]{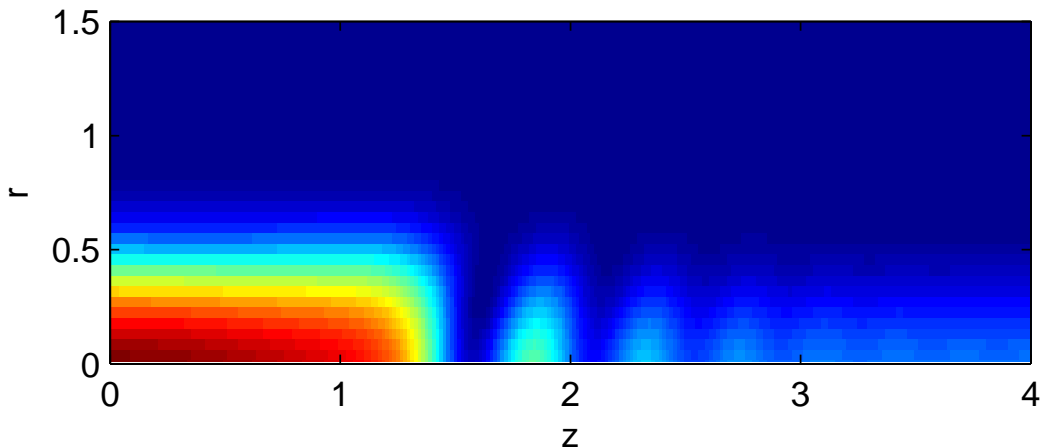}
\caption[Radial structure of 3D waveguide simulation]
{Density snapshot from a 3D simulation showing radial structure at time $t=0.17$,
as in Fig.~\ref{waveguide_sigma_blowup}.
}
\label{waveguide_discont_3d_U01}
\end{figure}

To demonstrate the issue, we fix $U_0=0.1$ and consider initial conditions that
are the ground state of the trapping potential
\begin{equation}\label{dga_Gaussian_well_ICs}
    V(z) = V_0\Bktsq{1 - e^{-10 z^2}}.
\end{equation}
By changing $V_0$, the nonuniformity of the initial conditions can be increased
from a completely homogeneous system ($V_0 = 0$) to one in which all the density
is concentrated near $z=0$.  For sufficiently strong initial nonuniformity
($V_0 = 50$) we observe the formation of discontinuities in the transverse
width of the 3D system at a finite time $t\approx 0.17$ after the pulse is
released, as shown in Fig.~\ref{waveguide_sigma_blowup}.  We computed the 3D
width by nonlinear least squares fitting of a Gaussian to the 3D transverse
density profile shown in Fig.~\ref{waveguide_discont_3d_U01}.  Also shown in
Fig.~\ref{waveguide_sigma_blowup} is the width arising from our ansatz,
according to Eq.~\eqref{chi_interpretation}.  
While the width oscillations are not reproduced accurately, the existence and
position of the discontinuity is correct.  As a result, numerical methods that
assume $\chi$ is smooth will generally fail within the next few time steps 
after the snapshot shown in Fig.~\ref{waveguide_sigma_blowup}.

\begin{figure}[htbp]
\centering
\includegraphics[width=11cm]{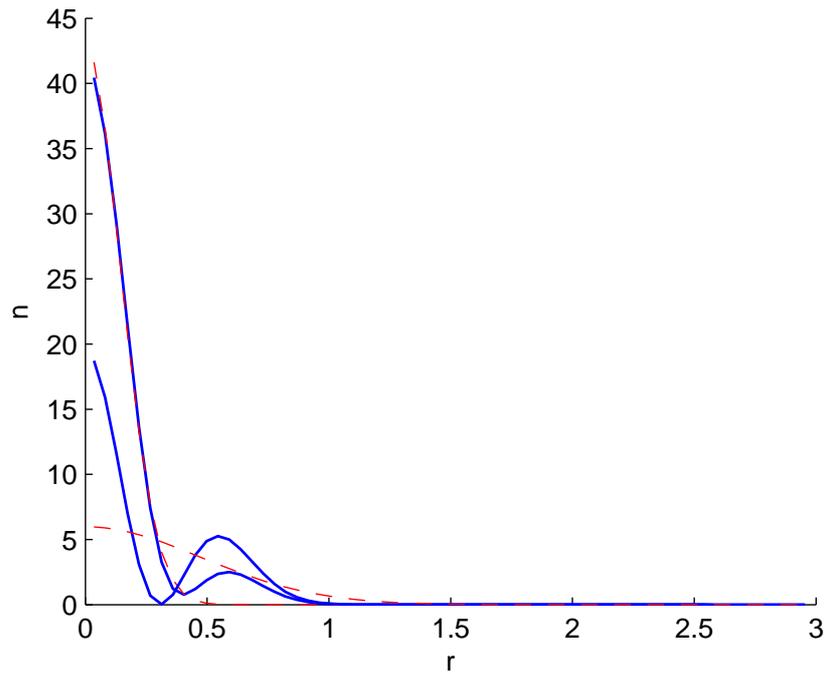}
\caption[Fits to the 3D wavefunction on both sides of a width discontinuity]
{Two transverse Gaussian fits (red) to the radial profile of the 3D
wavefunction (blue) at adjacent grid points in the $z$ direction.  This example
shows how a discontinuity in the fitting parameters may arise even when the
underlying function to be fitted is continuous.  The grid points in $z$ span
the position of the discontinuity in $\sigma$ shown in
Fig.~\ref{waveguide_sigma_blowup}.
}
\label{waveguide_discont_3d_fits}
\end{figure}

The existence of discontinuities comes as a surprise; one might expect that
continuous evolution of a smooth underlying field should lead to smooth fitting
parameters.  The 3D transverse profile in Fig.~\ref{waveguide_discont_3d_U01}
shows that the underlying density is smooth as expected, and that the
transverse profile is well behaved along most of the $z$ coordinate.  However,
in the region of very low density near the discontinuity ($z\approx 1.6$), the
transverse profile has two peaks as shown in
Fig.~\ref{waveguide_discont_3d_fits}.  The discontinuity comes about because
the least squares solution jumps suddenly from fitting the thin central peak to
fitting a wider combination of the central and secondary peaks.  Even though
the density is low in the problematic region, the discontinuity in width is
enough to destabilise the numerical solution.

We expect that \emph{any} ansatz for the 3D wavefunction containing a single
transverse width parameter will suffer from similar problems.  It is somewhat
ironic that the NPSE, MDE and similar approaches based on the adiabatic
approximation remain stable and somewhat accurate even in regimes where our
equations fail: Such methods avoid the problem by assuming that the transverse
profile depends only weakly on $z$!

%
%

\section{Conclusion}

In this chapter we have examined the general Gaussian ansatz for the quasi-1D Bose gas
tightly confined in a transverse 2D harmonic trap.  We have compared the ansatz to 
two alternative 1D effective equations, the NPSE and MDE.  We have found that some 
transient behaviours --- such as an oscillation in soliton depth after release
from a trap --- can only be captured with an ansatz that allows transverse
breathing motion.  Such motion is prohibited by the NPSE and MDE due to an 
assumption that the transverse degrees of freedom adiabatically follow the 
local density.  Further, we have shown that the method is more accurate than
the NPSE due to the inclusion of derivatives of the transverse width.

However, we have found that our general Gaussian ansatz has several important
failings that make it impractical for general use.  First and most severely,
discontinuities appear in the Gaussian width parameter for a wide range of
sufficiently excited initial conditions.  Second, finding ground states
numerically is significantly more difficult than with competing 1D
approximations.  Third, the assumption of a Gaussian profile implies less
accurate ground states than the MDE when the interaction strength is
sufficiently strong.

The appearance of discontinuities in the width at finite time is an interesting
issue which we trace to the appearance of two peaks in the transverse structure.
Although this problem typically occurs in regions of low density and energy, it
spells disaster for numerical solutions.  We expect it is a generic feature of
any similar ansatz which fits the transverse wavefunction using a single width
parameter.  One could consider additional parameters to get a better fit to the
transverse wavefunction, such as taking the next one or two harmonic oscillator
basis states, scaled by a Gaussian width parameter.  However, the resulting
equations would clearly be more complex to derive --- and more unwieldy to use 
--- than those presented here.

Our results show that out of the three effective 1D equations considered, the
MDE \cite{Mateo2008,Mateo2009} has
the best combination of reliability, simplicity and relative accuracy.

%% file: conclusion.tex
\chapter{Conclusion}
\label{conclusion_chapter}

In this thesis we have used numerical simulations to investigate the physics of
ultracold Bose gases confined to one- and two-dimensional geometries.

Our two-dimensional simulations were carried out using the PGPE classical field 
technique.  The PGPE allowed us to investigate the finite temperature regime
surrounding the BKT superfluid phase transition, in a parameter regime 
that is relevant to current experiments.  We directly computed the superfluid
and condensate fractions and found that these drop to zero at the same
temperature: in our finite-sized system there is no clear separation between
BEC and BKT transitions.  This is in contrast to what is expected based on the
theory of the infinite-sized system, emphasising the importance of finite-size
effects.

The unbinding of vortex pairs above the BKT transition is one of the defining
features of the transition, and we developed a coarse-graining technique to
quantify the number of unpaired vortices in the system.  Results proved
consistent with several other characteristic features, including the decay of 
the first-order correlation function and the sharp drop of the superfluid 
fraction to zero.

With a detailed microscopic theory at hand, we were able to simulate and 
validate the experimental measurement technique of Ref.~\cite{Hadzibabic2006},
including the method used to deduce the first-order correlations from
interference experiments.  In addition, our simulations suggest that direct 
observation of vortex pairs is not possible due to limited imaging resolution.
However, for this same reason it is likely that the dislocations observed in
experimental interference patterns correspond to free vortices, and accordingly 
are a strong indicator of the BKT transition.

To investigate the BKT transition we required a method for calculating the
superfluid fraction from a PGPE classical field simulation.  Ideally such a 
method would be based on properties of the equilibrium thermal ensemble as 
ergodically sampled by the PGPE.  We were able to derive such a method, using 
linear response theory to relate the superfluid fraction to the 
autocorrelations of the momentum density operator.  Our method provided
physically reasonable results that were consistent with the other physical
properties of the 2D gas.

One-dimensional effective equations allow for efficient simulation of
elongated cigar-shaped BECs.  We derived equations of motion using the 
Lagrangian formalism with a Gaussian ansatz, and solved these numerically for
several test cases.  We found that in certain cases our equations are more 
accurate than effective equations based on the adiabatic approximation.
However, they are numerically unstable for a wide range of initial conditions.
We investigated this instability and found that it is an inherent weakness of
any similar ansatz which relies on a single variational width parameter.

In summary, this thesis contributes original insight into the physics of
ultracold Bose gases in low dimensions, shedding light on recent experiments
and supplying several new theoretical and numerical tools.

%% file: funcderivs.tex
\chapter{Mathematical techniques for classical field theory}
\label{funcderivs_appendix}

\section{The Wirtinger calculus: derivatives of nonholomorphic functions}
\label{sec_wirtinger_calculus}

The Wirtinger derivative is the suitable generalisation of the real derivative to
general complex valued functions for the purposes of making linear approximations.
As we will see below, the usual derivative of complex analysis is much too
strict for these uses.

The derivative in complex analysis is defined using the familiar limit
definition used in the real calculus, only with complex variables.  Consider a
function $f\colon\Cfield\to\Cfield$ and some point $z_0\in\Cfield$; the
derivative of $f$ at $z_0$ is
\begin{equation} \label{complex_analysis_deriv_def}
    \Evalat{\deriv{f}{z}}{z_0} \equiv \lim_{z\to z_0}\frac{f(z) - f(z_0)}{z-z_0}
\end{equation}
(see, for example, \cite[page 30]{Churchill1960}).
If the limit converges to the same value irrespective of the way in which 
$z\to z_0$ we say that $f$ is differentiable at $z_0$.  A function is 
called \emph{holomorphic} on some domain $D$ if it is differentiable at every
point in an open set containing $D$.  The existence conditions for complex
derivatives defined like this are \emph{much} stricter than for the real
derivative.  In particular, a necessary condition for the existence of the
limit in Eq.~\eqref{complex_analysis_deriv_def} is the Cauchy Riemann conditions:
If $f(x+iy) = u(x,y) + i v(x,y)$ with $u,v\colon \Rfield^2 \to \Rfield$, the
Cauchy-Riemann conditions are $\pderiv{u}{x} = \pderiv{v}{y}$, and
$\pderiv{v}{x} = -\pderiv{u}{y}$.

Deriving the equations of motion for a classical field theory from an action
principle depends on taking derivatives of the action functional with respect 
to the fields and setting these to zero.  That is, for an action $S[\psi]$ we 
would like to do something like set
\begin{equation}
    \fderiv{S}{\psi} = 0 \, ,
\end{equation}
which should give us the equations of motion.  However, the action $S$ is
generally a \emph{real} function for any complex value of the field $\psi$, and
as such cannot possibly satisfy the Cauchy-Riemann conditions\footnote{At least
not in a nontrivial way: for a real function $f$ we have $v(x,y) = 0$ and the
derivative conditions immediately show that $u(x,y)$ must be a constant.}.
From the point of view of traditional complex analysis this immediately leads
to a problem: how do we make use of the derivative $\delta S/\delta\psi$ when
this derivative isn't even defined according to (some functional generalisation
of) the definition in Eq.~\eqref{complex_analysis_deriv_def}?  This is the main problem
that we tackle in this section.

One complicating factor for our particular case is that the derivatives of interest
are functional derivatives with respect to the field $\psi$.  Luckily, the
central problem remains intact if we forget about the fields and consider a 
nonholomorphic function $f$ of a single complex variable $z$.  This
simplification is made throughout the rest of the section.

If we can't use the usual complex derivative for common problems, what is the
alternative?  Derivatives are all about linear approximations\footnote{In
fact, the Fréchet derivative --- an elegant generalisation of the derivative
to Banach spaces --- is by definition exactly the linear part of the affine
approximation to a function.},
so to understand what to do about the peculiar strictness of the
Cauchy-Riemann conditions we consider a linear expansion of $f$ about the point
$z_0 = x_0+iy_0$ in terms of the real variables $x$ and $y$.  Assuming that $u$
and $v$ are differentiable, this is simply
\begin{align}
    f(x+iy) &\approx
        f(z_0) +
        \Bkt{\Evalat{\pderiv{u}{x}}{z_0} + i\Evalat{\pderiv{v}{x}}{z_0}} \Delta x +
        \Bkt{\Evalat{\pderiv{u}{y}}{z_0} + i\Evalat{\pderiv{v}{y}}{z_0}} \Delta y \\
        &= f(z_0) + \Evalat{\pderiv{f}{x}}{z_0} \Delta x +
                    \Evalat{\pderiv{f}{y}}{z_0} \Delta y
\end{align}
where $\Delta x=x-x_0$ and $\Delta y=y-y_0$.
This is a straightforward expansion using the real calculus so it's clear that 
this linear approximation is valid regardless of the Cauchy Riemann conditions.
Using $\Delta z = z - z_0$, $\Delta z^* = z^* - z_0^*$ for the conjugate and
suppressing the subscripts for brevity (all derivatives being evaluated at
$z_0$), we have
\begin{align}
    f(z)
    &\approx f(z_0) + \pderiv{f}{x} \Delta x + \pderiv{f}{y} \Delta y \\
    &= f(z_0) + \pderiv{f}{x}\; \frac{1}{2}(\Delta z + \Delta z^*) +
                \pderiv{f}{y}\; \frac{1}{2i}(\Delta z - \Delta z^*) \\
    &= f(z_0) +
        \frac{1}{2}\Bkt{\pderiv{f}{x} - i\pderiv{f}{y}} \Delta z + 
        \frac{1}{2}\Bkt{\pderiv{f}{x} + i\pderiv{f}{y}} \Delta z^*.
        \label{complex_linear_expansion}
\end{align}
This is an elementary but interesting result: the general linear approximation
to any real-differentiable complex-valued function may be expressed as a linear
combination of $\Delta z$ and $\Delta z^*$ with the coefficients given by a
simple combination of the real derivatives.  Furthermore, we can now express
the ratio inside the limit in Eq.~\eqref{complex_analysis_deriv_def} as
\begin{equation}
    \frac{f(z) - f(z_0)}{\Delta z} = 
        \frac{1}{2}\Bkt{\pderiv{f}{x} - i\pderiv{f}{y}} + 
        \frac{1}{2}\Bkt{\pderiv{f}{x} + i\pderiv{f}{y}} e^{-2i\theta}
\end{equation}
where $\theta = \arg\Delta z$ which reveals the general form of the directional
dependence of the limit.  If $f$ is to be complex differentiable, the value of
the right hand side must be independent of $z$ which implies $\pderiv{f}{x} +
i\pderiv{f}{y} = 0$; rearranging we see that this is exactly an expression of
the Cauchy Riemann conditions.

Coming back to Eq.~\eqref{complex_linear_expansion}, we \emph{define} two
new derivative operators
\begin{equation}
    \label{wirtinger_deriv_defs}
    \pderiv{f}{z} \equiv \frac{1}{2}\Bkt{\pderiv{f}{x} - i\pderiv{f}{y}}
    \quad\text{and}\quad
    \pderiv{f}{{z^*}} \equiv \frac{1}{2}\Bkt{\pderiv{f}{x} + i\pderiv{f}{y}},
\end{equation}
which allow us to write the linear approximation as
\begin{equation}
    \label{wirtinger_linear_expansion}
    f(z) \approx f(z_0) + \pderiv{f}{z} \Delta z + \pderiv{f}{{z^*}} \Delta z^*.
\end{equation}
At first sight this expression may seem like a mere curiosity that has been
manufactured to resemble the real calculus of functions of two variables.
On the contrary, it turns out that there is a powerful and general calculus for
the operators $\pderiv{}{z}$ and $\pderiv{}{{z^*}}$.  Most importantly, given a
function $f(z)$ written in terms of $z$ and $z^*$, it is possible to compute
derivatives directly with the new operators using the familiar rules of
calculus for two \emph{real} variables, and \emph{without} decomposing $f$ into
real and imaginary parts.  This calculus is known as the Wirtinger calculus
\cite{Remmert1991} after the Austrian mathematician Wilhelm Wirtinger\footnote{
It has also been called the $\Cfield\Rfield$-calculus by Kreutz-Delgado
\cite{Kreutz2009}.}.

\subsection{Properties of the Wirtinger derivative}

We now discuss the general properties of the Wirtinger derivative as discussed
in Ref.\ \cite[\S 1.4]{Remmert1991}.  Several properties are immediate from the
definitions in Eq.~\eqref{wirtinger_deriv_defs}:
\begin{itemize}
    \item The derivative operators are linear, due to the linearity of the real
        partial derivatives.
    \item They obey the usual product rule.
    \item The behaviour under complex conjugation is given by
        \begin{equation}
            \Bkt{\pderiv{f}{z}}^* = \pderiv{f^*}{{z^*}}.
        \end{equation}
    \item The derivatives satisfy the ``independence of $z$ and
        $z^*$'' property,
        \begin{equation}
            \label{z_zstar_independence}
            \pderiv{z^*}{z} = 0 \quad\text{and}\quad \pderiv{z}{{z^*}} = 0.
        \end{equation}
\end{itemize}
The last property has great importance from a practical computational
viewpoint, and we return to it after discussing the chain rule.

Using the conjugation property and the linear approximation formula in
Eq.~\eqref{wirtinger_linear_expansion} it is straightforward to show that the
Wirtinger derivatives obey a familiar looking chain rule.  We have
\begin{align}
    f(g(z))
    &\approx f\Bkt{g(z_0) + \pderiv{g}{z}\Delta z + \pderiv{g}{{z^*}} \Delta z^*} \\
    &\approx f(g(z_0)) +
        \Bkt{\pderiv{f}{g}\pderiv{g}{z} + \pderiv{f}{{g^*}}\pderiv{g^*}{z}} \Delta z +
        \Bkt{\pderiv{f}{g}\pderiv{g}{{z^*}} + \pderiv{f}{{g^*}}\pderiv{g^*}{{z^*}}} \Delta z^*
\end{align}
where $\pderiv{f}{g} \equiv \Evalat{\pderiv{f}{z}}{g(z)}$ and we have used the 
linear approximation formula twice, first for $g$ and then for $f$.  Identifying
the coefficient in front of the $\Delta z$ using the linear approximation
formula, we have shown that the chain rule for
Wirtinger derivatives is
\begin{equation}
    \pderiv{}{z}f(g(z)) = \pderiv{f}{g} \pderiv{g}{z} +
                    \pderiv{f}{{g^*}} \pderiv{g^*}{z},
\end{equation}
with the analogous expression for $\pderiv{}{{z^*}}f(g(z))$.  It is worth 
noting the close resemblance of this chain rule with the formula from the
more familiar calculus on $\Rfield^2$,
\begin{equation}
    \pderiv{}{x} f(g(x,y)) = \pderiv{f}{g_1}\pderiv{g_1}{x} + \pderiv{f}{g_2}\pderiv{g_2}{x}
\end{equation}
where $f,g\colon \Rfield^2\to\Rfield^2$ and $g(x,y) = [g_1(x,y),g_2(x,y)]$.

The relationship with the complex derivative in Eq.~\eqref{complex_analysis_deriv_def}
of a holomorphic function is simply that
\begin{equation}
    \deriv{f}{z} = \pderiv{f}{z};
\end{equation}
the definitions coincide for this important special case, as expected from the 
linear approximation formula.  For the same reason, it's possible to see that
the Cauchy-Riemann conditions have a particularly nice form in this formalism:
\begin{equation}
    \pderiv{f}{{z^*}} = 0.
\end{equation}

The generalisation of the Wirtinger calculus to higher dimensions is completely
straightforward from a calculational viewpoint, because the variables $z_k$ are
genuinely independent for distinct $k$.  We summarise this here by asserting 
that everything works just as in the real multivariable calculus, with the main 
difference being that variables always occur in conjugate pairs.

\subsection{Computing with Wirtinger derivatives}

The independence property $\pderiv{z^*}{z} = 0$ --- along with the other 
basic properties --- leads to the commonly repeated\footnote{See, for example, 
Ref.~\cite[page 598]{Goldstein1980}.} mnemonic device 
\begin{quote}
    \em
    To compute with the Wirtinger derivatives, treat $z$ and $z^*$ as
    ``independent variables'' and proceed exactly as for the calculus of two
    real variables.
\end{quote}
This ``independence'' is clearly problematic in a purely algebraic sense, but
is extremely useful for the purposes of computing derivatives because it results
in all the right rules when interpreted correctly.  It is often useful to write
functions $f(z)$ redundantly as $f(z,z^*)$ to aid in applying the mnemonic.
For example, $f(g(z))$ would become $f\bkt[\big]{g(z,z^*), g^*(z,z^*)}$ and the
chain rule follows after remembering the version for two independent real
variables.

We now present two examples that illustrate how computing with the Wirtinger
calculus avoids the need to break a function into real and imaginary parts.
A simple but useful example is the derivative of the absolute value:
\begin{align}
    \pderiv{}{z}\abs{z} &= \pderiv{}{z} (zz^*)^{1/2} \\
                  &= \frac{1}{2} (zz^*)^{-1/2} \pderiv{}{z} (zz^*) \\
                  &= \frac{z^*}{2\abs{z}} \,
\end{align}
where the second line is an application of the chain rule with the square root 
as the outer function and noting that the second term that appears in the chain
rule vanishes because $\pderiv{}{{z^*}} z^{1/2} = 0$.  In a
similar way, $\pderiv{}{{z^*}} \abs{z} = \frac{z}{2\abs{z}}$.

As a more complicated example, suppose we wanted to find the stationary points
of the real valued function
\begin{equation}
    f(z) = \Abs{(z+1)^{10} + z^*}.
\end{equation}
(We present this example because it is closely related to finding the 
stationary points of an action which is generally also a real valued function 
of complex arguments.)
For general complex-valued functions, the stationary points are the points at
which $\pderiv{f}{z} = 0$ and $\pderiv{f}{{z^*}} = 0$ simultaneously, but for
real valued functions we have $\pderiv{f}{{z^*}} = \Bkt{\pderiv{f}{{z}}}^*$ and
it suffices to satisfy the first condition.  Computing the derivative is an 
exercise in applying the chain rule to $h(g(z))$ with functions $h(z) = \abs{z}$ 
and $g(z) = (z+1)^{10} + z^*$:
\begin{align}
    \pderiv{f}{z}
        &= \pderiv{h}{g}\pderiv{g}{z} + \pderiv{h}{{g^*}}\pderiv{g^*}{z} \\
        &= \Bkt{\frac{[(z+1)^{10} + z^*]^*}{2\abs[\big]{(z+1)^{10}+z^*}}}
           \Bkt{10(z+1)^9}
           +
           \Bkt{\frac{(z+1)^{10} + z^*}{2\abs[\big]{(z+1)^{10}+z^*}}}
           \Bkt{1} \\
        &= \frac{10\Bktsq{(z^*+1)^{10} + z}(z+1)^9 + (z+1)^{10} + z^*}
                {2\abs[\big]{(z+1)^{10}+z^*}}.
\end{align}
Therefore, finding the stationary points in this case involves solving the high
order polynomial
\begin{equation}
    10\Bktsq{(z^*+1)^{10} + z}(z+1)^9 + (z+1)^{10} + z^* = 0.
\end{equation}
This could also have been derived by finding and setting \emph{both} real
derivatives to zero, but doing so using the formalism of the Wirtinger calculus
is more convenient in this case.

\section{Functional derivatives}
\label{sec_functional_derivatives}

%
%
%
%
%
%

In the Lagrangian formulation of dynamics the action $S$ is a \emph{functional}
--- that is, a function taking other functions as arguments\footnote{This is
obviously true for field theories, but remains true when we have a finite number
of dynamical variables because such variables are functions of time.}.
Stationarity of the action means that the derivatives with respect to
the fields are zero; this section investigates what we mean by such derivatives
and how to calculate them.

Throughout the section we make two simplifying assumptions to streamline the
presentation: First, the functionals under consideration and their arguments
are real.  Second, the argument fields are functions of a single independent
variable so that all integrals are one-dimensional.  Lifting these assumptions
presents no particular difficulty and is covered very briefly in the next
section.

\subsection{An example}

As in the previous section, we start by looking at linear approximations.
Consider for example the simple functional
\begin{equation}
    S[f] \equiv \int dx\; g(x) f^2(x),
\end{equation}
where $g$ is some fixed function, and $f$ may be anything that is well behaved
enough so that the integral is finite.  Evaluating $S$ at $f(x) + \eta(x)$ for
some small test function $\eta$ and expanding around $f$, we have
\begin{align}
    S[f+\eta]
        &= \int dx\;  g(x) \bktsq[\big]{f(x) + \eta(x)}^2 \\
        &= \int dx\; g(x) \bktsq[\big]{f^2(x) + 2f(x)\eta(x) + \eta^2(x)} \\
        &\approx \int dx\; g(x) f^2(x) + \int dx\; 2g(x)f(x)\eta(x) ,
\end{align}
where we have discarded the quadratic terms in $\eta$.  We see that the change
in $S$ which results from changing $f$ to $f + \eta$ is approximately a
linear functional:
\begin{equation} \label{S_variation_L_def}
    S[f + \eta] - S[f] \approx L_{f}[\eta] \equiv \int dx\; 2g(x)f(x)\eta(x) .
\end{equation}
In some sense the linear functional $L$ defined here actually \emph{is} the
derivative of $S$ --- known as the Fréchet derivative --- but a
slightly different definition is usually favoured by physicists, based on an
analogy with the gradient.
\vspace{0.5cm}

To motivate the definition of functional derivative commonly used in
physics we compare to the discrete setting.  For a finite number of variables
the analogue of $S$ would be some function $R(\v{f}) \equiv \sum_{i=1}^N g_i
f_i^2$ for $\v{f},\v{g} \in \Rfield^N$.  The linear approximation to $R$ about
$\v{f}$ is
\begin{equation}
    R(\v{f} + \v{e}) - R(\v{f})
        \approx \sum_{i=1}^N 2 g_i f_i e_i
        = \sum_{i=1}^N \pderiv{R}{f_i} e_i
        = \grad R \cdot \v{e}.
\end{equation}
The correspondences with the continuous case are $f \leftrightarrow
\v{f}$, $\eta \leftrightarrow \v{e}$, $g \leftrightarrow \v{g}$ and
$x\leftrightarrow i$.  With this in mind, it's clear that the function
$2g\cdot f$ in Eq.~\eqref{S_variation_L_def} is acting in an analogous role to the
gradient $\grad R = \bktsq[\big]{ \pderiv{R}{f_1}, \dotsc , \pderiv{R}{f_N} }
= [ 2 g_1 f_1, \dotsc , 2g_N f_N ]$.  The \emph{functional derivative} is
therefore written using similar notation to the partial derivative:
\begin{equation}
    \fderiv{S}{f(x)} = 2g(x) f(x).
\end{equation}
\vfill

\subsection{General definition}

%
%
%

Roughly speaking, 
the \emph{functional derivative} of a functional $S$ evaluated at $f$ is a
distribution\footnote{A \emph{distribution} is also known as a generalised
function and includes such objects as the Dirac delta ``function''; see, for
example, Ref.~\cite{Hoskins1999}.
} $\fderiv{S}{f}$ such that
\begin{equation}
    S[f + \eta] - S[f] = \int dx\; \fderiv{S}{f(x)} \eta(x) + O(\norm{\eta}^2)
\end{equation}
for all test functions $\eta$.  (Note that a more formal and general definition
may be given in terms of the Fréchet derivative\footnote{Borrowing the
definition of the Fréchet derivative, we might define $\fderiv{S}{f}$ as the
distribution that satisfies
\begin{equation}
    \lim_{\norm{\eta}\to 0} \frac{1}{\norm{\eta}}
    \Norm{S[f + \eta] - S[f] - \int dx\; \fderiv{S}{f(x)} \eta(x) } = 0.
\nonumber
\end{equation}
}.)  As indicated in the previous section, the distribution $\fderiv{S}{f}$ is
a generalisation of the gradient vector from a finite to an infinite number of
independent variables.  In the case that the distribution is simply a function,
we may evaluate it at $x$ to give the \emph{number} $\fderiv{S}{f(x)}$ which is
intuitively the answer to the question ``how much does $S$ change if we change
$f$ by a small amount at the point $x$?''.

This intuition leads to the ``physicist's definition'' of functional derivative:
\begin{equation} \label{func_deriv_physicists_def}
    \fderiv{S}{f(x)} = \lim_{\epsilon\to 0}
        \frac{S[f + \epsilon\delta_x] - S[f]}{\epsilon}
        \qquad\text{(dubious!)}
\end{equation}
where $\delta_x(y) = \delta(x-y)$.
The notion of poking $f$ exactly at $x$ using a delta spike is intuitive but
this definition doesn't make much mathematical --- or even calculational ---
sense if taken literally.  For example, consider attempting to calculate the
derivative of the functional $\int dx\; g(x) f^2(x) $ from the previous section.
Trying to compute $S[f+\epsilon\delta_x]$ then results in a term containing the
integral of $\delta_x^2$ which has no well-defined value.  It is possible to
repair these problems by considering a sequence $\{\Delta_{x,n}\}$ of nascent
delta functions such that $\Delta_{x,n}\to\delta_x$ as $n\to\infty$, and taking
the limit $\epsilon\to 0$ \emph{before} the limit $n\to\infty$.  However, it
seems better to simply regard Eq.~\eqref{func_deriv_physicists_def} as an aid to
the intuition, and fall back to the idea of linear approximations for 
computational purposes.


%
%

\subsection{Calculating with functional derivatives}

Functional derivatives may usually be calculated directly by considering the
linear expansion of the functional, but in practice this can be tedious.  
Therefore it is useful to derive a few general formulae that can be applied to
a wide range of functionals.  Here we present a few such formulae that are
relevant to this thesis, along with an indication of how to derive them.
We direct the reader to Ref.\ \cite[\S 2.3]{Greiner1996} for additional
exposition.

The functionals of interest are generally of the form
\begin{equation}
    S[f] = \int dy\; g\bkt[\big]{f(y), f'(y)} 
\end{equation}
for some differentiable function $g$ of two variables.  The task is to create a
linear approximation for
\begin{equation}
    S[f+\eta] = \int dy\; g\bkt[\big]{f(y) + \eta(y), f'(y) + \eta'(y)} .
\end{equation}
Writing $g^{(1,0)}$ and $g^{(0,1)}$ for the first derivatives with respect to
the first and second arguments respectively, and briefly dropping the explicit
$y$ dependence of all functions for succinctness, the linear approximation is
\begin{align}
    S[f+\eta] 
      &= \int dy\; \bktsq[\Big]{ g(f,f') + g^{(1,0)}(f,f') \eta + g^{(0,1)}(f,f') \eta'} + O(\eta^2) \\
      &\approx S[f] + \int dy\; g^{(1,0)}(f,f') \eta + \int dy\; g^{(0,1)}(f,f') \eta'.
\end{align}
This is almost in the right form, but the last term has the derivative $\eta'$ 
of the test function rather than the test function itself.  The derivative may
be removed by integrating by parts and assuming that the boundary terms are
zero.  This is true for a periodic domain, an infinite domain where the fields
and their derivatives decay at infinity, or any domain where the test function 
$\eta$ is constrained to be zero at the boundary.
Integrating by parts leads to
\begin{equation}
    S[f+\eta] 
      \approx S[f] + \int dy\; \Bktsq{g^{(1,0)}\bkt[\big]{f(y),f'(y)} - \pderiv{}{y} \Bktsq{g^{(0,1)}\bkt[\big]{f(y),f'(y)}}} \eta(y)
\end{equation}
and as a result,
\begin{equation}
    \label{general_func_deriv_f_fprime}
    \fderiv{S}{f(x)}
        = g^{(1,0)}\bkt[\big]{f(x),f'(x)} - \pderiv{}{x} \Bktsq{g^{(0,1)}\bkt[\big]{f(x),f'(x)}}.
\end{equation}

It is interesting to note that this is almost an expression of the
Euler-Lagrange equation for a single dynamical variable.  To see this, note
that the derivative may more succinctly --- though less explicitly --- be
written as
\begin{equation}
    \fderiv{S}{f}
        = \pderiv{g}{f} - \pderiv{}{x} \Bktsq{\pderiv{g}{f'}}
\end{equation}
where by $\pderiv{g}{f}$ and $\pderiv{g}{f'}$ we mean the derivatives of $g$
with respect to the first and second arguments respectively, evaluated at
$(f(x), f'(x))$.  Making the suggestive replacements $g \to L$, $x \to t$,
$f\to q$, and $q' \equiv \dot{q}$, we have
\begin{equation}
    \fderiv{S}{q}
        = \pderiv{L}{q} - \pderiv{}{t} \Bktsq{\pderiv{L}{\dot{q}}},
\end{equation}
and setting $\fderiv{S}{q} = 0$ yields the Euler-Lagrange equation in its usual
form.

Table \ref{func_deriv_table} presents a summary of several useful functional
derivatives, most arising as special cases of 
Eq.~\eqref{general_func_deriv_f_fprime}.

\begin{table}
    \caption{\label{func_deriv_table}
    Some useful functional derivatives.  The notations $g^{(1,0)}$ and 
    $g^{(0,1)}$ mean derivatives of $g$ with respect to its first and second 
    arguments respectively.
    }
    \begin{center}
    \begin{tabular}{ll}
    \toprule
    Functional $S$ & Functional derivative $\fderiv{S}{f(x)}$ \\
    \midrule
    $\displaystyle \int dy\; g(f(y)) $   & $g'(f(x))$
    \\[1em]
    $\displaystyle \int dy\; g(y)f'(y) $ & $-g'(x)f(x)$
    \\[1em]
    $\displaystyle \int dy\; g(f(y), f'(y)) $ &
                        $g^{(1,0)}\bkt[\big]{f(x),f'(x)}
                         - \pderiv{}{x} \Bktsq{g^{(0,1)}\bkt[\big]{f(x),f'(x)}}$
    \\[1em]
    $\displaystyle f(y)$                & $\delta(x-y)$
    \\
    \bottomrule
    \end{tabular}
    \end{center}
\end{table}

%
%
%

%% file: bkt/bkt_appendix.tex
\chapter{Details of two-dimensional simulations}
\label{bkt_appendix}


\section{Simulation using the PGPE} \label{sec:sim_details}

Here we outline our procedure for determining the properties of 
the $\rC$ region and the steps used to create initial states for the PGPE
solver.  The $\rC$ region itself is characterised by the cutoff momentum $K$,
while the initial states are characterised by the energy $E_\rC$ and number
$N_\rC$.  We want to obtain values of these three properties that are 
consistent with a specified temperature $T$ and total number of atoms $N$.



\subsection{Hartree-Fock-Bogoliubov analysis} \label{sec:HFB}
%


To generate an initial estimate of the $\rC$ region parameters we
solve the self-consistent Hartree-Fock-Bogoliubov (HFB) equations in the 
so-called Popov approximation\footnote{
It is amusing to note the commentary in Ref.~\cite{Yukalov2006} that maintains that
Popov himself never suggested this trick, and would only have deemed it valid
very near the transition temperature!
} \cite{Griffin1996} to find an approximate thermal
state for the system at a temperature $T$.  The resulting state is a Bose
Einstein distribution of quasiparticles interacting only via the mean-field,
expressed in terms of the quasiparticle amplitudes $u_{\v{k}}$ and $v_{\v{k}}$.

Occupations for the $\rC$ region field may be computed directly from the
quasiparticle occupations via
\begin{equation}
	n_{\v{k}} = \Bkt{u_{\v{k}}^2 + v_{\v{k}}^2} N_B(E_\v{k}) + v_{\v{k}}^2,
\end{equation}
where $N_B$ is the Bose-Einstein distribution and $E_\v{k}$ is the
quasiparticle energy that is obtained by solving the Bogoliubov-de~Gennes
equations self-consistently \cite{Griffin1996}.
This allows us to compute the cutoff as
the maximum value of $\norm{\v{k}}$ consistent with sufficient modal
occupation:
\begin{equation}
K = \max \bktcl{\norm{\v{k}} \colon n_{\v{k}} \ge n_{\text{cut}} }.
\end{equation}
We choose $n_{\text{cut}} = 5$ for
the sufficient occupation condition on the $\rC$ region modes.

The number of atoms below the cutoff may be computed directly from the sum of
the condensate number $N_0$ and the number of $\rC$ region excited state atoms,
$N_{1\rC}$:
\begin{align}
N_\rC = N_0 + N_{1\rC},\quad\text{where}\quad
N_{1\rC} = \sum_{\v{k} \in \rC \backslash \{\v{0}\}} n_{\v{k}}.
\end{align}

For the total energy below the cutoff, we use the expression
\begin{equation}
E_\rC = \frac{\hbar^2}{mL^2} \Bkt{ \frac{g N_0^2}{2} + \lambda N_{1\rC} - g N_{1\rC}^2 } 
        + \sum_{\v{k} \in \rC \backslash \{\v{0}\}} E_{\v{k}} \Bktsq{N_B(E_{\v{k}}) - v_{\v{k}}^2}
\end{equation}
where $\lambda = g(N_0 + 2 N_{1\rC})$.  Rearranging, this is
\begin{align} \label{EC_hfb}
E_\rC = \frac{\hbar^2 g}{2mL^2}\Bkt{N_\rC^2 + N_{1\rC}^2}
    + \sum_{\v{k} \in \rC \backslash \{\v{0}\}} E_\v{k} \Bktsq{N_B(E_\v{k}) - v_{\v{k}}^2}.
\end{align}
The expression in Eq.~\eqref{EC_hfb} differs from  Eq.~(22) of Ref.~\cite{Griffin1996} as
we have retained the zeroth order (constant) terms that are required to match
the energy scale of the HFB analysis to the zero point of energy in the
classical field simulations.

\subsection{Initial conditions for fixed total number}


A simple comparison between simulations at varying temperatures can only be
carried out if the total number of atoms is fixed.  This presents a problem in 
our simulations: although the number of atoms and energy of the $\rC$ region
can be directly specified (see section \ref{sec:initial_given_Ec_Nc}), we may only
determine the total number after performing a simulation.  This is because the
number of atoms in the $\rI$ region depends on the temperature and chemical
potential that are calculated by ergodic averaging of the $\rC$ region
simulations.

Formally, this may be stated as a root finding problem: solve
\begin{equation}
	N(N_\rC, E_\rC) = N_\text{tot}
\end{equation}
with initial guess provided by the solution to the HFB analysis in
section \ref{sec:HFB}.  Although both $N_\rC$ and $E_\rC$ affect the total number
$N$, we choose to fix $N_\rC$ to the initial guess and to vary $E_\rC$ until the
desired total number is found.

We note that evaluating the function $N(N_\rC, E_\rC)$ is very computationally
expensive and difficult to fully automate because it involves a simulation and
several steps of analysis.  For this reason we use a nonstandard root finding
procedure:  For the first iteration we simulate three energies about the
initial guess $E_\rC$ such that the results crudely span $N_\text{tot}$; these
three simulations can be performed in parallel which significantly reduces the
time to a solution.  A second guess was obtained by quadratic fitting of
$E_\rC$ as a function of $N$ which gives $N$ accurate to within about 5\% of
$N_\text{tot}$.  An additional iteration using the same interpolation method
takes $N$ to within 0.3\%, which we consider sufficient.

We note that changing $E_\rC$ during the root finding procedure means we have
no direct control over the final temperature of each specific simulation.  In
our case this is not a problem because we only require a range of temperatures
spanning the transition.  In principle one could solve for a given temperature
by allowing $N_\rC$ to vary in addition to $E_\rC$.

\subsection{Initial conditions for given $E_\rC$ and $N_\rC$} \label{sec:initial_given_Ec_Nc}


We compute initial conditions for the $\rC$ region field in a similar way to
Ref.~\cite{Davis2002}.  Using the representation for the $\rC$ region given by
Eq.~\eqref{eqn:Cfield}, the task is to choose appropriate values for the
$\bktcl{c_\v{n}}$.  As a first approximation, choose the smallest value for a
momentum cutoff $K'$ such that the field with coefficients
\begin{equation} \label{eqn:pgpe_initial1}
c_{\v{n}}
= \begin{cases}
    A e^{i\theta_{\v{n}}} \qquad &\text{for } 0 < \norm{\v{k}} \leq K', \\
    0   \qquad &\text{for } \abs{\v{k}} > K',
  \end{cases}
\end{equation}
has energy greater than $E_\rC$.  Here $A$ is chosen so that the field has
normalisation corresponding to $N_\rC$ atoms, and $\theta_{\v{n}}$ is a
randomly chosen phase which is fixed for each mode at the start of the
procedure.  The random phases allow us to generate many unique random initial
states at the same energy.

By definition, the field defined by Eq.~\eqref{eqn:pgpe_initial1} has energy
slightly above the desired energy.  This problem is solved by mixing it with
the lowest energy state:
\begin{equation}
c_{\v{n}}
= \begin{cases}
    A' e^{i\theta_{\v{0}}} \qquad &\text{for } \v{n} = \v{0}, \\
    0   \qquad &\text{elsewhere},
  \end{cases}
\end{equation}
using a root finding procedure to converge on the desired energy $E_\rC$.
The scheme generates random realisations of a non-equilibrium
field with given $E_\rC$ and $N_\rC$ which are then simulated to equilibrium
before using ergodic averaging for computing statistics.

\section{$\rI$ region integrals} \label{sec:above_cutoff_integrals}

Our assumed self-consistent Wigner function (section \ref{sec:above_cutoff}) for
the $\rI$ region atoms takes a particularly simple form in the homogeneous
case:
\begin{equation}
W(\v{k},\v{x})
  = \frac{1}{(2\pi)^2} \frac{1}{e^{(\hbar^2\v{k}^2/2m + 2\hbar^2 gn_\rC/m - \mu_\rC)/k_B T} - 1 }.
\end{equation}
The above-cutoff density may then be found by direct integration:
\begin{align}
n_\rI(\v{x}) &= \int_{\norm{\v{k}} \ge K} d^2\v{k} \; W_\rI(\v{k},\v{x}), \\
           &= -\frac{1}{\lambda_\text{dB}^2} \ln \Bktsq{1 - e^{-(\hbar^2K^2/2m + 2\hbar^2 gn_\rC/m - \mu_\rC)/k_B T}}.
\end{align}

In a similar way, the assumed Wigner function allows any
desired physical quantity to be estimated via a suitable integral.  A particular
quantity of interest in the current work is the first-order correlation
function,  which can be obtained from the Wigner function as \cite{Naraschewski1999}
\begin{equation}
  G^{(1)}_\rI(\v{x}, \v{x}') = \int_{\norm{\v{k}} \ge K} d^2\v{k}\; e^{-i\v{k}\cdot(\v{x}-\v{x}')} \; W_\rI\bkt[\big]{\v{k},\tfrac{\v{x}+\v{x}'}{2}}.
\end{equation}

This integral is of the general form
\begin{equation}
  I_1(\v{r}) \equiv \int_{\norm{\v{k}} > K} d^2\v{k}\; \frac{e^{-i\v{k}\cdot\v{r}}}{e^{A\v{k}^2 + B} - 1},
\end{equation}
for constants $A$ and $B$.  Noting that $I_1$ depends only on the length
$r$ of $\norm{\v{r}}$, and transforming $k$ to polar coordinates $(\kappa, \theta)$, 
we have
\begin{align}
  I_1(\v{r}) &= \int_{K}^\infty d\kappa\;  \frac{\kappa}{e^{A\kappa^2 + B} - 1} \int_0^{2\pi} d\theta\; e^{-ir\kappa\cos\theta},\\
  &=  \int_{K}^\infty d\kappa\;  \frac{\kappa}{e^{A\kappa^2 + B} - 1} \,2 \Bktsq{\Gamma(\tfrac{1}{2})}^2 J_0(r\kappa),
\end{align}
(see Ref.~\cite[p902]{Gradshteyn2000} for the Bessel function identity).

Thus we obtain $G_\rI^{(1)}(\v{x}, \v{x}')$ in terms of 
a one-dimensional integral which may be performed numerically:
\begin{equation}
  G_\rI^{(1)}(\v{x}, \v{x}')
  = \frac{1}{2\pi} \int_{K}^\infty  d\kappa\;
    \frac{\kappa J_0(\kappa\norm{\v{x}-\v{x}'})}{e^{(\hbar^2 \kappa^2/2m + 2\hbar^2 gn_\rC/m - \mu_\rC)/k_B T} - 1}.
  \label{eqn:GI1}
\end{equation}

\section{Vortex detection} \label{sec:vortex_detection}
The defining feature of a ``charge-$m$'' vortex is that the phase $\theta$ of
the complex field $\psi(\v{x}) = \abs{\psi{(\v{x})}} e^{i\theta(\v{x})}$
changes continuously from $0$ to $2m\pi$ around any closed curve that circles
the vortex core.  We express our field $\psi$ on a discrete grid in position
space; the aim of vortex detection is then to determine which grid plaquettes
(that is, sets of four adjacent grid points) contain vortex cores.

To obtain the phase winding about a plaquette, first consider the phase at two
neighbouring grid points A and B.  We are interested in the unwrapped phase
difference $\Delta\theta_{\text{AB}}$ between the grid points; unwrapping
ensures that the phase is \emph{continuous} between A and B.  (In the discrete
setting such continuity is poorly defined; the best we can do is to correct for
the possibility of $2\pi$ phase jumps by adding or subtracting factors of
$2\pi$ so that $\Abs{\Delta\theta_{\text{AB}}} < \pi$.)  The unwrapped phase
differences around a grid plaquette tell us a total phase change $\theta_\text{wrap}
= \sum_i \Delta\theta_{i,i+1} = 2m\pi$ where $m\in\mathbb{Z}$ is the winding
number or ``topological charge''.

Due to the necessity of unwrapping the phase, a four-point grid plaquette cannot 
unambiguously support vortices with charge larger than one.  Luckily, such 
vortices are energetically unfavourable in 2D Bose gases
\cite[page 83]{Pitaevskii_Stringari2003} so we need only concern ourselves with
detecting vortices with winding number $\pm1$ in this work.  The positions
obtained from a given run of our vortex detection algorithm are the labelled
$\{\v{r}^+_i\}$ and $\{\v{r}^-_i\}$ for winding numbers $+1$ and $-1$,
respectively.

%% file: var1d/var1d_appendix.tex
\chapter{An imaginary time method with fixed normalisation}
\label{var1d_appendix}

%

\label{itime_mu_method}


Ground states for the GPE are often found using the so-called ``imaginary 
time'' method \cite{chiofalo2000}.  The prescription is to replace the time $t$
with the ``imaginary time'' $\tau = it$ and the potential $V$ with $V-\mu$ in
the equation of motion.  One then selects a desired chemical potential $\mu$,
and evolves the new equations in $\tau$ until a stationary state $\wffull_0$ is
reached.  This stationary state corresponds to a ground state of the GPE.

A primary advantage of the imaginary time technique is the extreme simplicity
of implementation, given the existence of a numerical solver for the real-time
version of the equation.  An important disadvantage is that the normalisation
is controlled only indirectly via the chemical potential, and it is common to
desire a fixed total number of atoms in a simulation instead.  In the following
we describe a method of achieving this by evolving $\mu$ along with the imaginary
time evolution of the fields.

\section{The usual imaginary time method}

We start by investigating the reasons why the traditional imaginary time method
works.  In the case that the interaction strength is zero, we have the linear
Schrödinger equation and the reasoning behind the method is straightforward:
The coefficient of the lowest energy eigenstate evolves as $e^{-E_0\tau/\hbar}$
and therefore becomes large compared to coefficients $e^{-E_k\tau/\hbar}$ as
$\tau\to\infty$, where $E_k$ is the energy of the $k$th eigenstate.  The
wavefunction converges exponentially to the ground state as a result, provided 
that it has nonzero overlap with the initial state.

In the case that the Hamiltonian is nonlinear there is no simple formal 
solution to rely on, so the justification for the imaginary time method must
be modified.  The appropriate starting point is to realise that we seek to
minimise the Hamiltonian under the constraint that the number of atoms, $N$ is
fixed.  That is, we wish to solve the constrained minimisation problem,
\begin{equation}\begin{split}
    \text{minimise}   \:\:&\Hcal[\wffull] \\
    \text{subject to} \:\:&\Ncal[\wffull] = N
\end{split}\end{equation}
where $\Ncal[\wffull] = \int dV\, \Abs{\wffull}^2$ is the normalisation functional and $N$ 
is the desired normalisation.  Proceeding via Lagrange multipliers, we may
define
\begin{equation}
    \Kcal[\wffull] \equiv \Hcal[\wffull] - \mu\Bkt{\Ncal[\wffull] - N},
\end{equation}
and solve for the critical point of $\Kcal$.  It is important to note that the
critical point of the Lagrange function is \emph{not} a minimum with respect to 
the variables $\{\wffull, \mu\}$; it is a saddle point.  To find the critical
point we must solve the two equations
\begin{equation}\label{K_critical_point}
    \fderiv{\Kcal}{\wffull^*} = 0
    \quad\text{and}\quad
    \pderiv{\Kcal}{\mu} = 0.
\end{equation}
The first of these is the condition that the right hand side of the GPE with an
extra energy offset $\mu$ is zero, while the second recovers the number
constraint:
\begin{equation}
    (\Hsp - \mu)\wffull + U\abs{\wffull}^2\wffull = 0
    \quad\text{and}\quad
    \Ncal[\wffull] - N = 0.
\end{equation}

%

In the conventional imaginary time method, we solve a restricted version of
this problem by fixing $\mu$ and \emph{minimising} $\Kcal$.  This works because 
the critical point of $\Kcal$ with respect to $\wffull$ is a minimum for any fixed
$\mu$, even though it is not a minimum when the combined variables
$\{\wffull, \mu\}$ are considered.  (To see this, note that $\Kcal$ is also a
valid Hamiltonian with an energy offset $\mu$.)

Minimisation of $\Kcal$ with fixed $\mu$ may be achieved by continuous-time
steepest-descent optimisation and this corresponds to the imaginary time
evolution.  This may seem like a coincidence but the fact that imaginary time 
evolution minimises the energy is a generic property of Hamiltonian systems in
complex phase space.




To see this, note that the GPE may be written
\begin{equation} \label{GPE_hamiltonian_formalism}
    i\hbar \pderiv{\wffull}{t} = \fderiv{\Hcal}{\wffull^*}.
\end{equation}
On the other hand, the steepest descent direction for $\Kcal$ is
$-\fderiv{\Kcal}{\wffull^*}$, so an evolution equation which takes us toward the
ground state is
\begin{equation}\label{Psi_itime_evolution}
    \hbar \pderiv{\wffull}{\tau} = -\fderiv{\Kcal}{\wffull^*}.
\end{equation}
The factor of $\hbar$ on the left is included only to emphasise the
similarities with Eq.~\eqref{GPE_hamiltonian_formalism}.  Taken together, 
this shows why the imaginary time method works: it corresponds to a steepest-descent 
minimisation of the Hamiltonian, with chemical potential added as a 
Lagrange multiplier for the number constraint.


\section{The imaginary time method with varying $\mu$}

The conventional imaginary time method as described above does not allow us to 
satisfy the normalisation constraint directly; to do this we must modify $\mu$ 
during the evolution.  Surprisingly, the appropriate evolution law turns out to 
be the \emph{uphill} evolution
\begin{equation}
    \pderiv{\mu}{t} = +\pderiv{\Kcal}{\mu}.
\end{equation}
Combined with Eq.~\eqref{Psi_itime_evolution} this takes us to the
appropriate saddle point of $\Kcal$ representing the solution of
Eqs.~\eqref{K_critical_point}.

A discrete-time version of this saddle point finding algorithm is described in
\cite[\S 14.2~p.429]{Luenberger1984} and the local convergence of the
continuous version proven in Ref.~\cite[Theorem 1]{Arrow1958}.  For convergence to
hold, the Hessian of $\Kcal$ with respect to $\wffull$ must be positive definite
at the critical point.  We note that this is the case for our system, because
for every fixed $\mu$ the critical point is in fact a minimum with respect to
variations of $\wffull$.  While the convergence theorem holds only locally, we
find that the Thomas-Fermi solution is a sufficiently accurate starting point
for reliable convergence in practice.


We finish by giving some intuitive reasoning behind the saddle point algorithm.
To simplify the exposition, we work with an analogous minimisation problem over
$\Rfield^n$:
\begin{equation}\begin{split}
    \text{minimise}\:\:& f(\v{x}) \\
    \text{subject to}\:\:& g(\v{x}) = 0
\end{split}\end{equation}
where $f,g\colon \Rfield^n \to \Rfield$.  The Lagrangian function for this 
minimisation is
\begin{equation}
    \Lcal(\v{x},\mu) = f(\v{x}) - \mu g(\v{x}),
\end{equation}
and we wish to solve for a critical point $(\v{x}_0, \mu_0)$ of
$\Lcal$.  In nearly all cases $(\v{x}_0, \mu_0)$ is a saddle point because to 
first order near the critical point,
\begin{align}
\mu g(\v{x}) &= \mu \Bktsq{g(\v{x}_0) + \left.\pderiv{g}{\v{x}}\right|_{\v{x}_0} \cdot (\v{x}-\v{x}_0) + \dotsc} \\
             &\approx \mu\grad g(\v{x}_0) \cdot (\v{x}-\v{x}_0),
\end{align}
where the constant term disappears because the constraint $g$ is satisfied at
the critical point.  This first-order approximation therefore amounts to a sum
of saddle-like terms of the form $\mu (x_i - x_{0,i})$.

In the case that the principle directions of the saddle point correspond to the
coordinate axes, for example
\begin{equation}
    h_1(x,y) = x^2 - y^2,
\end{equation}
one can intuitively expect that a steepest descent in $x$ and steepest
ascent in $y$ would converge to the saddle point. 
In the Lagrangian case, the principle directions do not line up 
with the coordinate axes, and instead we have a saddle point intuitively similar
to the function
\begin{equation}
    h_2(x,y) = x^2 - xy.
\end{equation}

Nevertheless, the evolution
\begin{align}
    \dot{\v{x}} &= -\pderiv{\Lcal}{\v{x}} \\
    \dot{\mu}   &= +\pderiv{\Lcal}{\mu}
\end{align}
still converges to the saddle point in many cases, provided we start
sufficiently close.  As noted above, this happens when the Hessian
of $\Lcal$ with respect to $\v{x}$ is positive definite \cite[Theorem
1]{Arrow1958}. This is the case when finding ground states of the GPE using the
saddle point algorithm.

%% file: other_papers.tex
\chapter{Additional work}
\label{additional_work_appendix}

The following two papers contain additional work completed during the course of
the PhD, but which do not tie into the main theme of the thesis.

In the first paper we report the first discovery of a Bell inequality for
observables with continuous spectra.  In this paper the current author
contributed the no-go proof on pages three and four, in collaboration with
E.~G.~Cavalcanti: There are no local hidden variable inequalities possible if
one considers only the first-moment correlations between continuous variables
at different sites.  The current author also helped with the final stages of
writing the paper.

The second paper proposes a computationally efficient alternative to the
traditional Uhlmann-Jozsa fidelity measure between two quantum states.  The
current author contributed section IV that discusses computational
efficiency, including the optimised C implementations, and additional
optimisation of the slower Matlab codes.  The current author also helped with 
polishing of the manuscript as a whole.

\textbf{Note for the arXiv version:}
Verbatim copies of the two papers, Refs.~\cite{Cavalcanti2007} and
\cite{Mendonca2008} were attached here in the version of this thesis submitted
for examination.  These have been omitted from the arXiv version due to
technical constraints.